\documentclass[12pt,preprint]{aastex}
\usepackage{graphics,natbib,verbatim}

\shorttitle{ATCA Survey of UCHIIs in the Magellanic Clouds}

\begin{document}

\newcommand{\ts}{\textsuperscript}
\newcommand{\dg}{$^\circ$\ }
\newcommand{\msun}{$M_{\sun}$\ }
\newcommand{\up}[1]{$\times$10\ts{#1}}
\newcommand{\uchii}{{UC\ion{H}{2} }}
\newcommand{\hii}{{\ion{H}{2} }}

\newcommand{\figsizetwo}{2.75in}
\renewcommand{\plottwo}[2]{\centerline{\resizebox{3.15in}{!}{\includegraphics{#1}}\resizebox{3.15in}{!}{\includegraphics{#2}}}}

\title{Australia Telescope Compact Array Survey of Candidate
Ultra-Compact and Buried HII Regions in the Magellanic Clouds}

\author{R\'{e}my Indebetouw, Kelsey E. Johnson}  
\affil{University of Wisconsin Astronomy Department, 475 N. Charter
St., Madison, WI 53705 (remy@astro.wisc.edu, kjohnson@astro.wisc.edu)}

\and
\author{Peter Conti}  
\affil{JILA, University of Colorado, 440-UCB, Boulder, CO, 80309 
(pconti@jila.colorado.edu)}

\begin{abstract}
We present a systematic survey for ultracompact \hii (\uchii) regions
in the Magellanic Clouds. Understanding the physics of massive star
formation (MSF) is a critical astrophysical problem.  The study of MSF
began in our galaxy with surveys of \uchii regions, but before now
this has not been done for other galaxies.
We selected candidates based on their Infrared Astronomical Satellite
(IRAS) colors and imaged them at 3 and 6 cm with the Australia
Telescope Compact Array (ATCA). Nearly all of the observed regions
contain compact radio sources consistent with thermal emission. Many
of the sources are related to optically visible \hii regions, and
often the radio emission traces the youngest and densest part of the
\hii region.  The luminosity function and number distribution of Lyman
continuum fluxes of the compact radio sources are consistent with
standard stellar and cluster initial mass functions.
This type of systematic assessment of IRAS diagnostics is important
for interpreting {\it Spitzer} Space Telescope data, which will probe
similar physical scales in nearby galaxies as IRAS did in the
Magellanic Clouds.
\end{abstract}

\keywords{stars: formation --- HII regions --- Magellanic Clouds}

\section{Introduction}
Massive stars play a major role in the dynamical evolution of
galaxies: they are responsible for the ionization of the interstellar
medium, their stellar winds and supernovae are dominant sources of
mechanical energy, their ultraviolet radiation powers far-infrared
luminosities through the heating of dust, and they are a main driver
of chemical evolution in the universe via supernova explosions at the
end of their lives.  However, despite the significant role of massive
stars throughout the universe, their birth is not well understood, and
we are only beginning to piece together a scenario for the youngest
stages of massive star evolution.  We have made some progress
understanding the early stages of massive star formation in the
Galaxy, but the current knowledge about the early stages of massive
star evolution in other environments is mediocre at best.

The reasons for this dearth of information about extremely young
massive stars in other galaxies are predominantly twofold: (1) the
earliest stages of massive star evolution are deeply enshrouded and
inaccessible in the optical and near-infrared regimes; and (2) high
spatial resolutions are necessary to disentangle the individual
massive stars from their surrounding environment and background
contamination.  Radio observations using the Australia Telescope
Compact Array (ATCA) are uniquely capable of overcoming these
obstacles for the galaxies nearest to our own, the Large and Small
Magellanic Clouds.  Synthesis imaging observations at centimeter
wavelengths are sensitive to the free-free emission emitted by the
densest compact ionized regions, and the centimeter spectral index can
be used to differentiate those extremely dense regions suffering from
self-absorption.

We have conducted a survey of candidate \uchii regions in the
Magellanic Clouds with high (1--2\arcsec) spatial resolution at 3 and
6~cm using the ATCA.  Section~\ref{observations} contains a
description of the FIR-based candidate selection, the radio observing
strategy, data reduction, and lists the detected sources.  Individual
compact \hii regions with interesting radio morphologies are described
in the Appendix, and the statistical properties of the population as a
whole are analyzed in section~\ref{population}.

\section{Observations}
\label{observations}

\subsection{Far Infrared Target Selection}

One goal of this project was to test the effectiveness of using
far-infrared (FIR) colors at fairly low spatial resolution to discover
massive protostellar objects.  This exercise using IRAS data in the
Magellanic Clouds is particularly interesting because the {\it
Spitzer} Space Telescope probes similar spatial scales in nearby
galaxies as IRAS did in the Magellanic Clouds.

The primary target selection criterion for this survey was thus FIR
color.  \citet[][hereafter WC89]{wc89b} determined that the majority
of Galactic ultra-compact \hii (\uchii) regions have particular mid-
to far-infrared colors which distinguish them from other astronomical
objects.  They indicated (their Figure~1b) that most sources with IRAS
colors $F_\nu(60\mu m) / F_\nu(12\mu m) \ge 20. $ and $F_\nu(25\mu m)
/ F_\nu(12\mu m) \ge 3.7$ are {\uchii}s (these colors correspond to
black-body temperatures $\lesssim$105 and $\lesssim$160~K,
respectively).  More importantly in the context of this work, they
found that there was little color contamination -- compared to a
random selection of Galactic IRAS sources, most of the objects in that
region of color-color space were \uchii regions.

\begin{figure}
\plotone{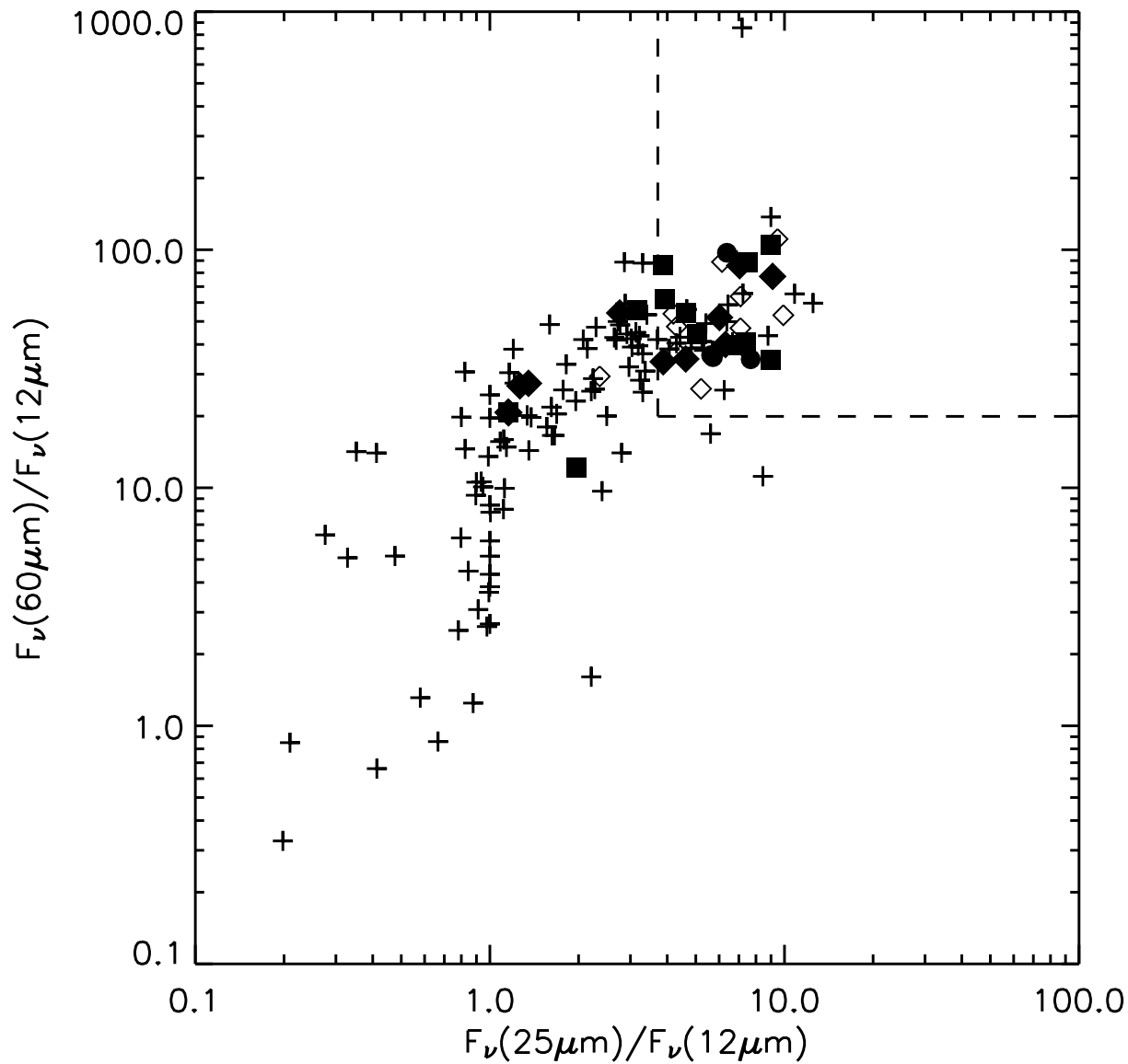}
\caption{\label{targets} Candidate \uchii regions in the Magellanic
Clouds.  All IRAS sources with Parkes radio detections in the Clouds
are marked (crosses), our candidates in bold symbols (filled diamonds
observed 5/2001, squares observed 5/2002, filled circles existing
archival observations).  The color selection box of WC89 is marked; we
used a slightly relaxed color selection criterion in order to test the
validity of that box at the lower spatial resolution available in the
Magellanic Clouds.}
\end{figure}

To exploit this fact, we first combined the IRAS point source catalog
with a multiwavelength Parkes radio continuum survey \citep[][and
references therein]{filip98a}.  The Parkes surveys have a resolution
of 2.8~arcmin at 8.55~Ghz (which corresponds to a linear distance of
$\sim 49$~pc in the SMC and $\sim 41$~pc in the LMC), so we took as a
match to each radio source the brightest IRAS source within 2.5
arcminutes.  We then examined the FIR to radio spectral energy
distribution of all the objects with slightly broader colors than the
WC89 criteria: F$_{24\mu m}$/F$_{12\mu m}~>$~1 and F$_{60\mu
m}$/F$_{12\mu m}~>$~10.  This sample contained 98 objects in the LMC
(47 with the stricter WC89 colors) and 22 in the SMC (8 with WC89
colors).  Finally, we chose those targets which were consistent with
having flat or inverted radio emission at $\lambda\sim$1--5~cm.  Thus
our sample is FIR-color selected, but objects which are highly
nonthermal at cm wavelengths at the resolution of the Parkes survey
are excluded.  This last criterion was imposed to bias against
supernova remnants and other sources of nonthermal emission.  Sources
consistent with optically thin thermal emission (slightly negative
spectral index $\alpha\sim-0.1$ where $F_\nu\propto\nu^\alpha$) were
kept as candidates under the presumption that \hii regions observed at
low resolution will be dominated by the optically thin envelopes, but
may still contain compact, more optically thick sources.
Figure~\ref{targets} shows our target objects in an IRAS color-color
space identical to that of WC89's Figure~1b.  The two targets which
fall farthest outside of the color selection box were cases for which
the IRAS flux densities or identification were ambiguous.

\subsection{Radio Observations and Data Reduction}

Candidate fields were observed with the ATCA during two runs in May
2001 and May 2002.  Simultaneous observations were obtained at 3~cm
(8.6~GHz) and 6~cm (4.8~GHz) in the highest resolution configurations
(6F in 2001, 6A in 2002).  Each field was observed for roughly 3
hours, with some time added for a few of the faintest fields.  We
observed 19 of the 29 selected fields in the LMC, and 7 of the 8
selected fields in the SMC.  For some of the fields, archival data was
available, which allowed compact sources to be tabulated for 3
additional candidates.  Our observations are summarized in
Table~\ref{obs.tab}, and the archival observations which we completely
re-reduced for consistency, are summarized in Table~\ref{arch.tab}.


\begin{deluxetable}{ccclrcl}
\tabletypesize{\scriptsize}
\tablecaption{\label{obs.tab} New Observations}
\tablehead{
\colhead{Target} & \colhead{RA} & \colhead{Dec} &
\colhead{Date} & \colhead{Exp.} & 
\colhead{Config} & \colhead{Phase}\\
&(J2000)&(J2000)&&\colhead{Time}&&\colhead{Calibrator(s)}
}
\startdata
0452-6927 & 4h 51m 53.00s & -69d 23' 29.99'' & 2002/5/6     & 169m & 6A & 0530-727\\
0452-6722 & 4h 52m 31.64s & -67d 17' 00.14'' & 2001/5/10-15 & 262m & 6F & 0355-669,0252-712,0407-658\\
0454-6716 & 4h 54m 49.92s & -67d 11' 58.80'' & 2001/5/11-15 & 249m & 6F & 0355-669,0252-712,0407-658\\
0456-6636 & 4h 56m 47.63s & -66d 32' 18.39'' & 2001/5/9     & 190m & 6F & 0515-674,0355-669 \\
0457-6632 & 4h 57m 45.00s & -66d 27' 29.99'' & 2002/5/6     & 175m & 6A & 0530-727\\
0505-6807 & 5h 04m 59.76s & -68d 03' 40.03'' & 2001/5/9     & 190m & 6F & 0515-674,0355-669 \\
0510-6857 & 5h 09m 50.00s & -68d 52' 59.99'' & 2002/5/6     & 166m & 6A & 0530-727\\
0519-6916 & 5h 18m 45.00s & -69d 14' 29.99'' & 2002/5/4     & 178m & 6A & 0530-727\\
0523-6806 & 5h 22m 55.00s & -68d 04' 29.99'' & 2002/5/4     & 174m & 6A & 0530-727\\
0525-6831 & 5h 24m 41.80s & -68d 29' 23.23'' & 2001/5/10    & 184m & 6F & 0515-674,0355-669 \\
0531-7106 & 5h 31m 20.00s & -71d 04' 29.99'' & 2002/5/3     & 193m & 6A & 0530-727\\
0532-6629 & 5h 32m 35.00s & -66d 27' 19.99'' & 2002/5/4     & 174m & 6A & 0530-727\\
0538-7042 & 5h 38m 22.01s & -70d 41' 08.34'' & 2001/5/11-17 & 238m & 6F & 0530-727\\
0539-6931 & 5h 39m 11.36s & -69d 30' 04.51'' & 2001/5/11    & 186m & 6F & 0530-727\\
0540-6940 & 5h 39m 45.00s & -69d 38' 39.99'' & 2002/5/3     & 164m & 6A & 0530-727\\
0542-7121 & 5h 41m 30.00s & -71d 18' 59.99'' & 2002/5/3     & 164m & 6A & 0530-727\\
0545-6947 & 5h 45m 24.57s & -69d 46' 33.73'' & 2001/5/11    & 192m & 6F & 0530-727\\
\tableline
0043-7321 & 0h 45m 29.90s & -73d 04' 56.63'' & 2001/5/14    & 220m & 6F & 0230-790\\
0046-7333 & 0h 48m 01.85s & -73d 16' 02.86'' & 2001/5/14    & 201m & 6F & 0230-790\\
0047-7343 & 0h 49m 31.70s & -73d 26' 50.26'' & 2001/5/14    & 200m & 6F & 0230-790\\
0057-7226 & 0h 59m 50.00s & -72d 10' 59.99'' & 2002/5/5     & 169m & 6A & 0230-790\\
0103-7216 & 1h 05m 16.67s & -72d 00' 04.64'' & 2001/5/10    & 212m & 6F & 0355-669,0252-712\\
0107-7327 & 1h 09m 12.00s & -73d 11' 41.99'' & 2002/5/6     & 159m & 6A & 0230-790\\
0122-7324 & 1h 24m 10.00s & -73d 08' 59.99'' & 2002/5/5     & 169m & 6A & 0230-790\\
\enddata
\tablecomments{LMC targets are above the solid line, SMC below. ``Target'' 
refers to the Filipovic Parkes catalog name. }
\end{deluxetable}

\begin{deluxetable}{ccclrclcl}
\tabletypesize{\scriptsize}
\tablecaption{\label{arch.tab} Archival Observations}
\tablehead{
\colhead{Target} & \colhead{RA} & \colhead{Dec} &
\colhead{Date} & \colhead{Exp.} & 
\colhead{Config} & \colhead{Phase} &
\colhead{Program} & \colhead{Name} \\
&(J2000)&(J2000)&&\colhead{Time}&&\colhead{Calibrator}&
} 
\startdata
0452-6927 & 4h 51m 48.00s & -69d 23' 48.00'' & 2000/4/8   & 47m & 6D & 0454-810 & C868 & N79A      \\
0452-6700 & 4h 52m 06.00s & -66d 55' 24.00'' & 2000/4/9   & 47m & 6D & 0454-810 & C868 & N4A       \\
0454-6916 & 4h 54m 24.00s & -69d 10' 60.00'' & 2000/4/9   & 47m & 6D & 0454-810 & C868 & N83B-1    \\
0456-6629 & 4h 57m 18.00s & -66d 23' 16.00'' & 2000/4/9   & 59m & 6D & 0454-810 & C868 & N11A      \\
0540-6940 & 5h 39m 42.00s & -69d 38' 54.00'' & 2000/4/9   & 47m & 6D & 0454-810 & C868 & N160A     \\
0540-6946 & 5h 40m 00.00s & -69o 44' 36.00'' & 2000/4/9   & 47m & 6D & 0454-810 & C868 & N159-5    \\
0540-6946 & 5h 39m 39.50s & -69d 44' 26.00'' & 1996/7/11  & 75m & 6C & 0515-674 & C520 & lmcx1     \\
0540-6946 & 5h 39m 38.72s & -69d 44' 45.97'' & 1995/8/30  & 70m & 6D & 0522-611 & C462 & 0540-697  \\
\hline
0043-7321 & 0h 45m 30.00s & -73d 04' 48.00'' & 2000/4/8   & 50m & 6D & 2353-686 & C868 & N12       \\
0107-7327 & 1h 09m 12.00s & -73d 11' 42.00'' & 2000/4/8   & 47m & 6D & 2353-686 & C868 & N81       \\
0122-7324 & 1h 24m 06.00s & -73d 08' 59.99'' & 2000/4/8   & 52m & 6D & 2353-686 & C868 & N88A      \\
\enddata
\tablecomments{
LMC targets are above the solid line, SMC below.  ``Target'' refers to
the Filipovic Parkes catalog name, as for our observations, while
``Name'' refers to what the source was called by the archival
observer.  }
\end{deluxetable}

The receivers simultaneously detect 4800~MHz ($\sim$6~cm) and 8640~MHz
($\sim$3~cm), with primary beams (field of view) of 10 and 5
arcminutes, respectively.  Each frequency had a bandwidth of 128~MHz,
and the resulting theoretical RMS noise for these observations is
$\approx 0.07$~mJy/beam. The absolute flux density scale was
determined by assuming a flux density for PKSB1934-638 of 5.83~Jy at
4.8~GHz and 2.84 at 8.6~GHz, and we estimate that the resulting flux
density scale is uncertain by less than $\sim 5\%$.

Calibration was carried out using the Miriad data reduction package,
including gain, phase, and polarization calibrations of the flux
density calibrator and phase calibrators.  Following the calibration,
data were loaded into the Astronomical Image Processing System (AIPS)
where they were inverted and cleaned using the task {\sc imagr}.  In
order to optimize the synthesized beam at 4.8~GHz, a fairly uniform
weighting was used (robust=0), while a more natural weighting was
using at 8.6~GHz (robust=3) in order to increase the sensitivity.
In addition, the $uv$ range of the 4.8~GHz
data was limited in order to better match the shortest $uv$ coverage of
the 8.6~GHz data. It should be noted that this does not create matched
beams at 3cm and 6cm, but simply mitigates the influence of extended
structure at 6cm; the 6cm observations still have denser $uv$ coverage
at short spacings, and the 3cm observations still have $uv$ coverage at
long spacings not available in the 6cm observations.  Based on the
resulting $uv$ coverage, the largest angular scales to which these data
are sensitive are $\lesssim 30''$ at 6~cm and $\lesssim 20''$ at 3~cm.
The images at both frequencies were then corrected for the primary
beam sensitivity.

Radio flux densities were measured in AIPS$++$ using two methods.  The
first method used the {\sc viewer} program to place identical
apertures around the 3~cm and 6~cm images, and background levels were
estimated using surrounding annuli.  Several combinations of apertures
and annuli were used in order to estimate the uncertainty in the flux
density.  The second method used the {\sc imagefitter} to fit
two-dimensional Gaussians to the sources.  Final flux densities were
determined by averaging these results, and the uncertainties were
determined from the scatter of these results in combination with the
uncertainty in the absolute flux density scale.  Because the quoted
uncertainties reflect the scatter from determining flux densities in
different ways, in some cases the relative uncertainties between the
flux densities at 3~cm and 6~cm are smaller than the absolute
uncertainty for either determination.  The final flux densities and
spectral indexes are listed in Tables \ref{lmcflux} and \ref{smcflux}.
It is important to note that for resolved objects, the
calculated spectral indexes may be lower limits (in the
sense of objects appearing less optically thick than they truly are)
depending on the geometry of the source; the observations are more
sensitive to large-scale optically thin emission at 6~cm than at 3~cm,
and the synthesized beams were $\sim$1''.5 at 3~cm and $\sim$2''.0 at
6~cm.

High-resolution (6~km configuration) archival data were available for
a few additional fields, and the sources detected in these data are
included in Tables \ref{lmcflux} and \ref{smcflux}, identified with
their ATNF program number.  In some cases, the archival data were
centered a few arcminutes (a significant fraction of the primary beam)
away from our pointing, so the same sources are detected, but the
errors are significantly lower in the pointing for which the source in
question is nearer the center of the primary beam.  The quality of the
archival data vary, and in some cases is inferior to our data due to
an insufficient frequency of observing the secondary phase calibrator
during periods of varying atmospheric conditions.  The data reduction
and error assessment was performed identically to our data; cases in which
the calibration was suspect are omitted, and cases in which we feel
less accurate calibration was obtained are reflected by larger error
bars.

\section{The Population of Detected Sources}

Many of the compact \hii regions detected in this survey are
associated with optically identified \hii regions.  This fact is not
surprising as the \hii regions observed in the radio are probably
compact sites of embedded star formation associated with the less
extincted (and possibly more evolved) regions ionized by young stars
and observable in optical light.  Some of our radio sources have
particularly interesting morphology, especially as compared to their
infrared or H$\alpha$ emission, and we highlight these in detail in
the appendix (\ref{individuals}).  

Tables \ref{lmcflux} and \ref{smcflux} list all the detected compact
radio continuum sources. 
Sources are named according the nearest radio continuum source in the
(low-resolution) Parkes surveys \citep{filip98a}.  If there is
more than one compact source in the area, the compact source most
clearly associated with the \hii region as seen in the mid-IR is
given the main designation, and other sources are designated with
direction relative to the first one (e.g. ``N'').  Also listed are
the nearest associated IRAS source, and the familiar shortened form
(e.g. N23A for LHA 120-N 23A) of the Lamont-Hussey optical designation
\citep{henize56}.  It is important to note, as mentioned above, that
because the synthesized beam is smaller at 3~cm than at 6~cm, the
spectral index is probably a lower limit that depends on the geometry
of the source.


\newcommand{\p}{&$\pm$&} 
\newcommand{\m}{&$<$&}
\newcommand{\n}{\multicolumn{3}{l}{---}}

\begin{deluxetable}{llccr@{}l@{}c@{}r@{}lr@{}l@{}c@{}r@{}lcrlcl}
\rotate
\tabletypesize{\scriptsize}
\tablecaption{Flux densities and spectral indexes of compact objects observed
in the SMC\label{smcflux}.}
\tablewidth{0pt}
\tablehead{
\colhead{radio source\tablenotemark{a}} & \colhead{IRAS source\tablenotemark{b}} & \colhead{RA\tablenotemark{c}}& \colhead{Dec}& 
\multicolumn{5}{c}{F$_3$\tablenotemark{d}} & 
\multicolumn{5}{c}{F$_6$} & 
\colhead{$\alpha$\tablenotemark{e}} & 
\colhead{log($Q$)\tablenotemark{f}} & \colhead{Sp.T.\tablenotemark{g}} &
\colhead{pgm\tablenotemark{h}} &
\colhead{notes, other IDs}\\
&&(J2000)&(J2000)&\multicolumn{10}{c}{}&&&&&}
\startdata
B0043-7321    & 00436-7321& 0 45 15.7& -73 ~6 ~7&    &  \m  3&  &  0&.8\p  &.3&     & 47.4&B0V  &    & \\
              &           &          &          &   2&  \p  1&  &  0&.6\p  &.4& +1.5& 47.8&O9.5V&C868& \\
B0043-7321(N) & 00436-7321& 0 46 13.1& -73 ~0 45&    &  \m 15&  &  1&.1\p  &.5&     & 47.5&B0V  &    & N12B?\\
	      &           &          &          &    &  \m 10&  &  1&.6\p  &.7&     & 47.7&B0V  &C868& \\\hline
B0046-7333    & 00462-7331& 0 48 ~8.5& -73 14 55&  17&  \p   &.5& 17&.4\p  &.3& -0.0& 48.8&O7.5V&    & N26\\\hline
B0047-7343    & 00477-7343& 0 49 29.0& -73 26 33&   7&.9\p   &.4&  8&.4\p  &.3& -0.1& 48.4&O8.5V&    & N33 \\
B0047-7343(N) & 00469-7341& 0 48 53.4& -73 24 57&   1&.2\p  1&.2&  1&  \p  &.3& +0.3& 47.5&B0V  &    & N33?\\\hline
B0057-7226    & 00574-7226& 0 59 14.9& -72 11 ~3&   7&  \p  2&  &  2&  \p 2&  & +1.8& 48.4&O8.5V&    & N66, em$\star$ Lin~350 \\
B0057-7226(S) & 00574-7226& 0 59 11~~& -72 11 40&   5&  \p  5&  &  7&  \p 7&  & -0.5& 48.2&O9V  &    & diffuse, N66 \\\hline
B0103-7216    & 01035-7215& 1 ~5 ~4.1& -71 59 25&  14&.1\p   &.4& 14&.1\p  &.4& +0.0& 48.7&B0V  &    & N78B\\
B0103-7216(N) & 01035-7215& 1 ~5 ~5.2& -71 59 ~1&   2&.4\p  1&  &  2&  \p 1&  & +0.3& 47.9&O9.5V&    & N78A\\\hline
B0107-7327    & 01077-7327& 1 ~9 12.9& -71 11 39&  34&  \p  1&  & 33&  \p 1&  & +0.0& 49.1&O6.5V&    & N81\\
B0107-7327(N) & 01077-7327& 1 ~9 20.6& -73 10 51&   1&.0\p   &.6&  1&.5\p  &.5& -0.6& 47.5&B0V  &    & N81\\
B0107-7327(W) & 01077-7327& 1 ~8 32.4& -72 11 19&    &.6\p   &.4&   &.8\p  &.3& -0.4& 47.3&B0V  &    & N81?\\\hline
B0122-7324    & 01228-7324& 1 24 ~7.9& -73 ~9 ~4&  85&  \p  1&  & 91&  \p  &.5& -0.1& 49.5&O5V  &    & N88 \\
	      &           &	    &          &  87&  \p  5&  & 90&  \p 4&  & -0.0& 49.5&O5V  &C868& \\
\enddata
\tablecomments{\ts{a}Radio designation is from the Parkes survey of \cite{filip98a}.
\ts{b}IRAS source which was used to identify each object as a candidate
\uchii region is also identified by name, and names are given for
those objects which could be identified with optical HII regions.
\ts{c}Coordinates are J2000. \ts{d}Flux densities are given in mJy.
\ts{e}$\alpha$ is the spectral index assuming $F_\nu\propto\nu^\alpha$. 
\ts{f}$Q_{49}$ is the Lyman continuum flux assuming that all 3~cm emission is from an 
ionization-bounded optically thin HII region.  For sources with no 3~cm detection, 
$Q$ is calculated from the 6~cm flux density. Assumed distance to the SMC is 60~Mpc 
\citep[][and references therein]{graczyk}.
\ts{g}The spectral type of a single star required to produce the given $Q$.
\ts{h}The ATCA program number is given for archival data taken by other investigators.
}
\end{deluxetable}

\begin{deluxetable}{llccr@{}l@{}c@{}r@{}lr@{}l@{}c@{}r@{}lcrlcl}
\rotate
\tabletypesize{\scriptsize}
\tablecaption{Flux densities and spectral indexes of compact objects observed
in the LMC\label{lmcflux}.}
\tablewidth{0pt}
\tablehead{
\colhead{radio source} & \colhead{IRAS source} & \colhead{RA}& \colhead{Dec}& 
\multicolumn{5}{c}{F$_3$} & 
\multicolumn{5}{c}{F$_6$} & 
\colhead{$\alpha$} & \colhead{log($Q$)} & \colhead{Sp.T.} &
\colhead{pgm} &
\colhead{notes, other IDs}\\
&&(J2000)&(J2000)&\multicolumn{10}{c}{}&&&&&}
\startdata
B0452-6927    & 04521-6928 & 4 51 53.3& -69 23 29& 160&  \p  5&  &138&  \p  3&  & +0.2&49.6&O5V  &    & N79A \\
              &            &          &          & 169&  \p  5&  &133&  \p  5&  & +0.3&49.6&O5V  &C868& \\
B0452-6927(E) & 04521-6928 & 4 51 53~~& -69 23 27&  30&  &&   &  & 25&  &&   &  & +0.3&48.8&O7.5V&    & 4 weak sources on E edge of main source\\\hline
B0452-6700(SW)& 04520-6700 & 4 52 ~9~~& -66 55 23&  45&  \p 15&  & 27&  \p 10&  & +0.7&48.7&O8V  &C868& elongated, N4A\\
B0452-6700(NE)& 04520-6700 & 4 52 12~~& -66 55 15&   5&  \p 10&  &  8&  \p  8&  & -0.7&48.1&09V  &C868& N4A\\\hline
B0452-6722(NW)& 04524-6721 & 4 52 12.4& -67 12 52&   3&.5\p  2&  &  2&.3\p   &.3& +0.6&47.9&O9.5V&    & N5?\\
B0452-6722(NE)& 04524-6721 & 4 52 33.4& -67 13 28&   2&.4\p   &.5&  0&.9\p   &.3& +1.4&47.7&B0V  &    & N5?\\\hline
B0454-6916(E) & 04546-6915 & 4 54 26.1& -66 11 ~2&  39&.3\p  2&  & 39&.1\p  1&  & +0.0&49.0&06.5V&C868& N83B\\
B0454-6916(W) & 04542-6916 & 4 53 58.6& -66 11 ~6&   2&.8\p  1&  &  3&.6\p   &.5& +0.4&47.8&O9.5V&C868& N83B\\\hline
B0456-6629(E) & 04571-6627 & 4 57 16.2& -66 23 21&  30&  \p  5&  & 31&  \p  5&  & +0.0&48.8&07.5V&    & diffuse, N11A\\
              &            &          &          &  32&  \p  5&  & 32&  \p  5&  & +0.0&48.9&07V  &C868& \\
B0456-6629(W1)& 04566-6629 & 4 56 57.3& -66 25 13&    &  \m 10&  &  3&.3\p   &.4&     &47.9&O9.5V&    & N11B\\
              &            &          &          &   1&.5\p   &.8&  3&.1\p  1&.7& -1.0&47.5&B0V  &C868& \\
B0456-6629(W2)& 04566-6629 & 4 56 47.8& -66 24 34&    &  \n   &  &  9&  \p  2&  &     &48.3&O9V  &    & diffuse, N11B\\
              &            &          &          &   1&.2\p  1&.5&  7&  \p  5&  & -2.5&48.1&O9V  &C868& \\
B0457-6632    & 04576-6633 & 4 57 41.0& -66 30 36&   2&  \p   &.6&  2&.6\p   &.3& -0.4&47.7&B0V  &    & diffuse\\\hline
B0505-6807    & 05051-6807 & 5 ~5 ~5.7& -68 ~3 46&   1&.0\p   &.4&  1&.2\p   &.2& -0.3&47.4&BOV  &    & N23A\\
B0505-6807(N) & 05051-6807 & 5 ~5 ~9.7& -68 ~1 37&   1&.6\p   &.6&  1&.1\p   &.3& +0.5&47.6&BOV  &    & N23A\\\hline
B0510-6857(W) & 05101-6855A& 5 ~9 50.6& -68 53 ~5&  39&  \p  1&  & 26&  \p  1&  & +0.6&49.0&O6.5V&    & N105\\
B0510-6857(E) & 05101-6855A& 5 ~9 52.9& -68 53 ~0&  30&  \p  4&  & 31&  \p  2&  & -0.0&48.8&O7.5V&    & N105\\
B0510-6857(N) & 05101-6855A& 5 ~9 52.5& -68 52 47&   9&  \p  7&  &  9&.5\p  1&.5& -0.1&48.3&O9V  &    & diffuse, N105\\
B0510-6857(S) & 05101-6855A& 5 ~9 52.1& -68 53 25&  12&  \p  7&  & 14&  \p  2&  & -0.2&48.4&O8.5V&    & diffuse, N105\\\hline
B0519-6916(E) & 05196-6915 & 5 19 39.3& -69 13 43&   5&  \p  5&  &  2&.1\p   &.4& +1.2&47.6&B0V  &    & N119?\\
B0519-6916(N1)& 05195-6911 & 5 18 55.4& -69 ~9 ~0&   7&  \p  7&  &  4&  \p   &.5& +0.8&47.9&O9.5V&    & N119?\\
B0519-6916(N2)& 05195-6911 & 5 18 50.0& -69 ~9 32&   6&  \p  7&  &  5&.5\p   &.5& +0.1&48.1&O9.5V&    & N119?\\\hline
B0523-6806(NE)& 05233-6802 & 5 23 43.5& -68 ~0 34&  10&  \p 10&  & 10&  \p   &.5& -0.0&48.3&O9V  &    & N44M \\
B0522-5800    & {\parbox{1.7cm}{\renewcommand{\baselinestretch}{0.75}\scriptsize \rule{0ex}{2ex}05221-6800 +\rule[-1.ex]{0ex}{1.7ex}05223-6801}}
                          & 5 22 12.6& -67 58 32&    &  \n   &  & 34&  \p  1&.5&     &48.9&O7V  &    & N44B,C\\
B0523-6806(SE)& 05230-6807 & 5 23 24.7& -68 ~6 41&   4&.5\p  2&  &  4&.8\p  2&  & -0.1&48.0&O9.5V&    & N44?\\
B0523-6806(SW)& 05224-6807 & 5 22 19.6& -68 ~4 37&  11&  \p  1&.5& 10&  \p  1&  & +0.1&48.4&O8.5V&    & N44G,K\\
B0523-6806    & 05230-6807 & 5 22 55.2& -68 ~4 ~9&   3&  \p   &.3&  3&.3\p   &.3& -0.1&47.8&B0V  &    & compact source only, N44D\\\hline
B0525-6831    & 05253-6830 & 5 25 ~6~~& -68 28 15&  10&  \p  5&  & 10&  \p  5&  & +0.0&48.4&O8.5V&    & diffuse, N138A\\
B0525-6831(N) & 05253-6830 & 5 24 50.5& -68 26 55&    &  \m   &.5&   &.8\p   &.3&     &47.2&B0.5V&    & N138?\\
B0525-6831(W) & 05244-6832 & 5 24 10.5& -68 30 26&    &  \m  1&.5&   &.4\p   &.3&     &46.9&B0.5V&    & N138B,D\\\hline
B0531-7106    & 05320-7106 & 5 31 22.9& -71 ~4 ~9&   2&.1\p   &.5&  2&.1\p   &.3& +0.0&47.7&B0V  &    & compact core only, N206A\\
B0531-7106(SW1)&05313-7109 & 5 30 56.3& -71 ~6 ~2&   6&  \p  3&  &  6&  \p  3&  & +0.0&48.1&O9V  &    & diffuse, N206\\
B0531-7106(SW2)&05310-7110 & 5 30 20.6& -71 ~7 44&  20&  \p 20&  & 11&  \p  7&  & +0.9&48.4&O8.5V&    & diffuse, N206B\\
B0531-7106(SW3)&05310-7110 & 5 30 31.3& -71 ~8 56&   4&  \p  5&  &  2&.7\p   &.6& +0.6&47.8&O9.5V&    & N206B\\
B0531-7106(SE)& 05310-7106 & 5 32 18.6& -71 ~7 44&   9&  \p 10&  &  1&.7\p   &.8& +2.4&47.6&B0V  &    & N206?\\\hline
B0532-6629    & 05325-6629 & 5 32 32~~& -66 27 15&   1&  \p   &.2&   &.8\p   &.1& +0.3&47.4&B0V  &    & compact, N55A \\
B0532-6629(NE)& 05325-6629 & 5 32 33~~& -66 27 ~8&   5&  \p   &.8&  4&  \p   &.5& +0.3&48.1&O9.5V&    & diffuse, N55A \\\hline
B0538-7042(S) & 05389-7042 & 5 38 24.2& -70 44 26&   4&.5\p   &.5&  5&.5\p   &.6& -0.3&48.0&O9.5V&    & N213?\\
B0538-7042(E) & 05389-7042 & 5 38 26.9& -70 41 ~6&   1&.0\p   &.3&  1&.1\p   &.3& -0.1&47.4&B0V  &    & N213\\
B0538-7042    & 05389-7042 & 5 38 21.5& -70 41 ~5&  42&  \p  2&  & 26&  \p  2&  & +0.7&49.0&06.5V&    & N213, diffuse\\\hline
B0539-6931(1) & 05396-6931 & 5 39 15.7& -69 30 39&  12&  \p  2&  & 15&.5\p  2&  & -0.3&48.4&O8.5V&    & N158C\\
B0539-6931(2) & 05396-6931 & 5 39 17.4& -69 30 49&   3&.5\p  2&  &  3&.3\p  1&  & +0.1&47.9&O9.5V&    & N158C\\
B0539-6931(N) & 05391-6926 & 5 38 44.5& -69 24 38&  20&  \p 20&  &  5&.2\p  2&  & +2.0&48.1&O9V  &    & N158?\\
B0540-6935    & 05404-6933 & 5 39 40.0& -69 32 56&    &  \m 10&  &  2&.7\p  1&  &     &47.8&O9.5V&    & N158?\\\hline
B0540-6940(1) & 05401-6940 & 5 39 46.0& -69 38 39& 120&  \p  7&  &130&  \p  7&  & -0.1&49.4&O5V  &    & N160A2 \\
              &            &          &          & 111&  \p 10&  &127&  \p  9&  & -0.2&49.4&O5V  &C868& \\
B0540-6940(2) & 05401-6940 & 5 39 44.3& -69 38 48&  16&  \p  1&  & 16&  \p  1&  & +0.0&48.6&O8V  &    & N160A3? \\
              &            &          &          &  12&  \p  5&  & 15&  \p  5&  & -0.3&48.4&O8.5V&C868& \\
B0540-6940(3) & 05401-6940 & 5 39 43.4& -69 38 54&  50&  \p  5&  & 50&  \p  5&  & +0.0&49.1&O6.5V&    & N160A1 \\
              &            &          &          &  52&  \p 10&  & 47&  \p 10&  & +0.1&49.1&O6.5V&C868& \\
B0540-6940(4) & 05401-6940 & 5 39 39.0& -69 39 11&  12&  \p  1&  & 14&.5\p  1&  & -0.2&48.4&O8.5V&    & N160A, maser \\
              &            &          &          &  16&  \p  5&  & 11&  \p  5&  & +0.5&48.6&O8V  &C868& \\
B0540-6940(5) & 05401-6940 & 5 39 38.8& -69 39 04&   8&.0\p  1&  &  6&.0\p  1&  & +0.4&48.3&O9V  &    & N160A\\
              &            &          &          &    &  \m 10&  &  6&  \p 10&  &     &48.1&09V  &C868& \\
B0540-6940(E) & 05409-6942 & 5 40 25.2& -69 40 14&    &  \n   &  & 15&  \p  6&  &     &48.5&O8V  &    & N160C\\\hline
B0540-6946(1)\tablenotemark{a}	     						                 
              & 05405-6946 & 5 40 ~4.4& -69 44 37&  50&  \p 75&  & 70&  \p 15&  & -0.5&49.2&O6V  &    & N159D\\
              &            &          &          &  43&  \p 15&  & 64&  \p  8&  & -0.6&49.0&06.5V&C868& 3~cm calibration suspicious\\
              &            &          &          &  60&  \p 55&  & 55&  \p 10&  & +0.1&49.1&06.5V&C520& \\
              &            &          &          &  68&  \p 20&  & 70&  \p  7&  & -0.0&49.2&06V  &C462& \\
B0540-6946(4) & 05401-6947 & 5 39 37.5& -69 45 26&    &  \n   &  &120&  \p 15&  &     &49.4&O5.5V&    & N159A\\
              &            &          &          & 110&  \p 15&  &123&  \p  8&  & -0.2&49.4&O5.5V&C520& \\
              &            &          &          &  97&  \p 10&  &122&  \p  8&  & -0.3&49.4&O5.5V&C868& \\
B0540-6946(5) & 05401-6947 & 5 39 37.5& -69 46 10&    &  \n   &  & 30&  \p  5&  &     &48.8&O7.5V&    & N159A\\
              &            &          &          &  35&  \p 20&  & 32&  \p 10&  & +0.1&48.9&O7V  &C520& \\\hline
B0542-7121    & 05423-7120 & 5 41 37.5& -71 19 02&  19&  \p  2&  & 18&  \p  1&.5& +0.1&48.6&O8V  &    & S169, N214C, knots w/halo\\
B0542-7121(N) & 05423-7120 & 5 41 17.8& -71 16 39&   1&.8\p   &.6&   &.9\p   &.3& +1.0&47.6&B0V  &    & N214H\\\hline
B0545-6947(N) & 05458-6947 & 5 45 ~4.3& -69 39 28&    &  \n   &  &  7&.3\p  1&  &     &48.2&O9.5V&    & \\
B0545-6947(NE)& 05458-6947 & 5 45 57.3& -69 43 56&   1&.3\p  1&.5&  2&.1\p   &.5& -0.7&47.7&B0V  &    & N168?\\
B0545-6947(1) & 05458-6947 & 5 45 27.8& -69 46 23&  12&.3\p  1&.5& 13&.1\p  1&.5& -0.1&48.5&O8V  &    & N168A\\
B0545-6947(2) & 05458-6947 & 5 45 20.0& -69 46 45&   2&.7\p   &.5&  2&.7\p   &.5& +0.0&47.8&O9.5V&    & N168B\\
\enddata
\tablenotetext{a}{numbering scheme follows \citet{huntwhit94}.}
\tablecomments{All units and columns as Table~\ref{smcflux}. Adopted distance to the LMC is 50~kpc \citep{alcock}.}
\end{deluxetable}
\clearpage

\label{population}

The properties of the population of detected compact \hii regions
should be sensitive to the physics of star formation in the Magellanic
Clouds.  As our sample was selected to probe the most compact and
densest \hii regions, we probe both the individual star formation
mechanism and the formation mechanism of clusters, if indeed these are
different.  In the following sections we first compare the IRAS and
radio properties of the sample.  We also model the total luminosity
and Lyman-continuum flux distributions, which are sensitive to the
stellar and cluster initial mass functions.  Finally we look for any
trends involving the radio spectral index.  Analysis of the
high-resolution IR properties of the sample will be presented in a
future paper.

\subsection{Spectral Indexes and FIR Colors}

\begin{figure}
\plotone{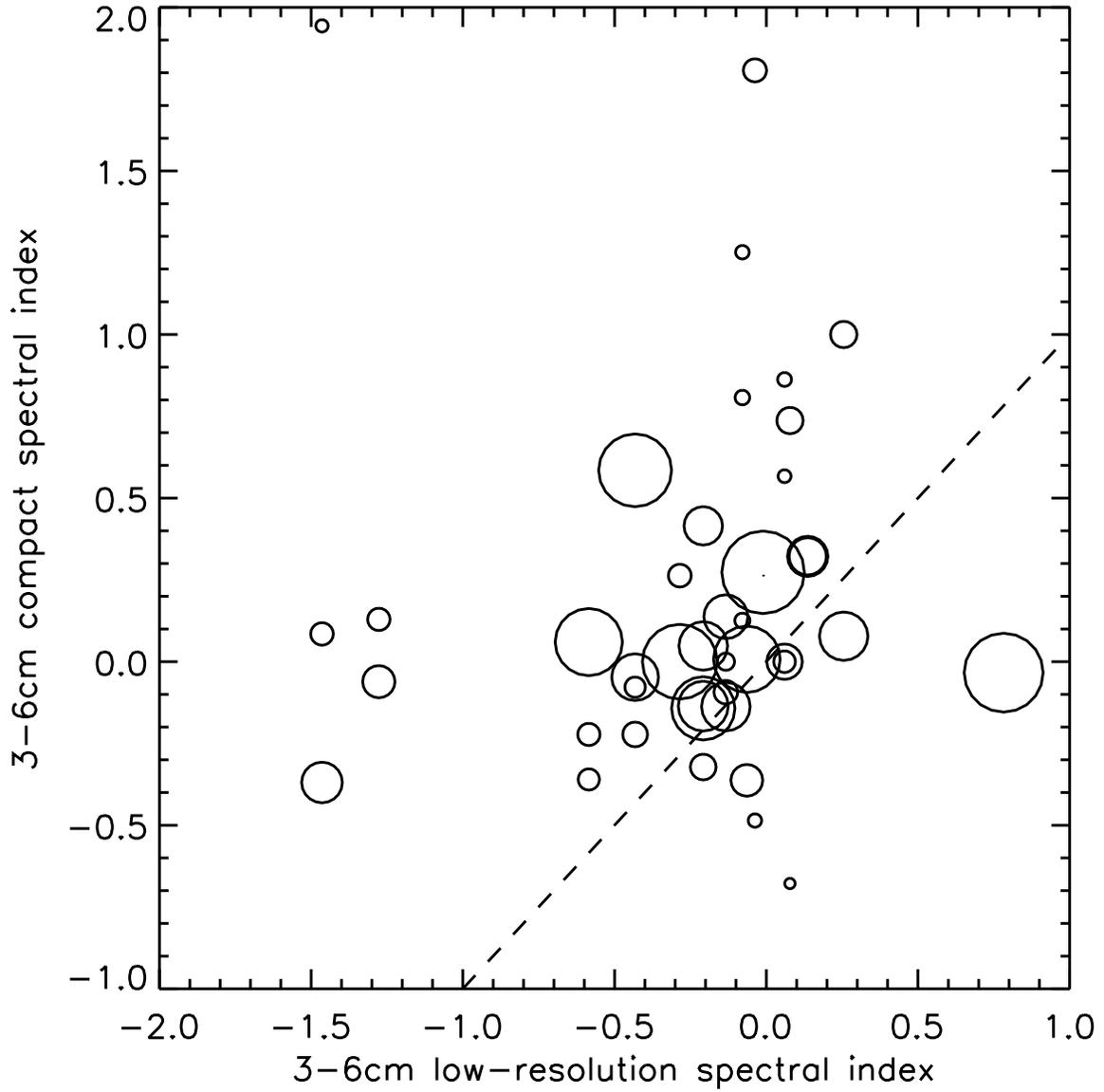}
\caption{\label{spix}
3-6~cm spectral indexes of detected sources, assuming that the flux density 
is a power-law function of frequency $F_\nu\propto\nu^\alpha$.  
The size of the data point scales with the signal-to-noise of the 
detected source.  The spectral indexes tend to be more positive
(more optically thick thermal) at higher resolution.}
\end{figure}

One can describe a source's spectral shape near $\lambda$=3--6~cm with
a single number $\alpha$ if one assumes that the flux density is a
power-law function of frequency $F_\nu\propto\nu^\alpha$.  Optically
thin and optically thick thermal radiation have spectral indexes of
$\alpha$=-0.1 and $\alpha$=2 (blackbody limit), respectively.
Figure~\ref{spix} shows the spectral indexes of detected sources,
showing many that are consistent with thermal radiation, as well as
some which are probably background nonthermal sources.  The spectral
index of the compact sources detected is often larger (more consistent
with thermal radiation or a higher optical depth) than the spectral
index calculated for the same source from the relatively low
resolution Parkes data.  This is expected physically, because the more
diffuse emission around a compact embedded source will be optically
thin, driving the spectral index from a positive value toward zero.

\begin{figure} 
\plottwo{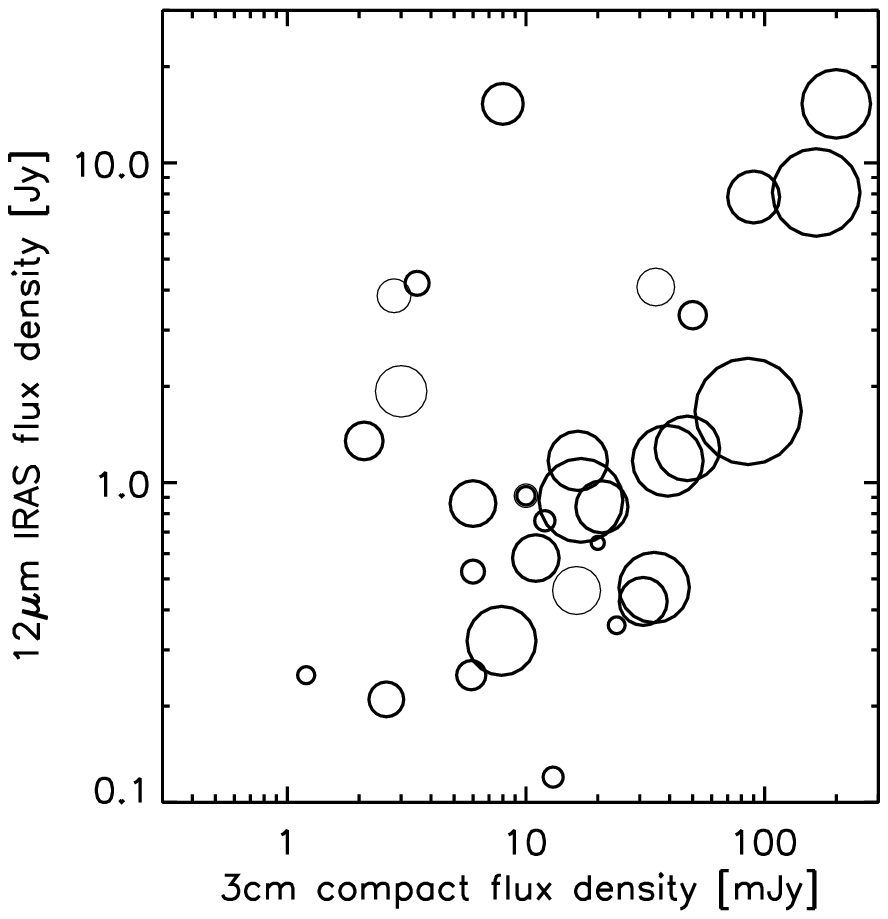}{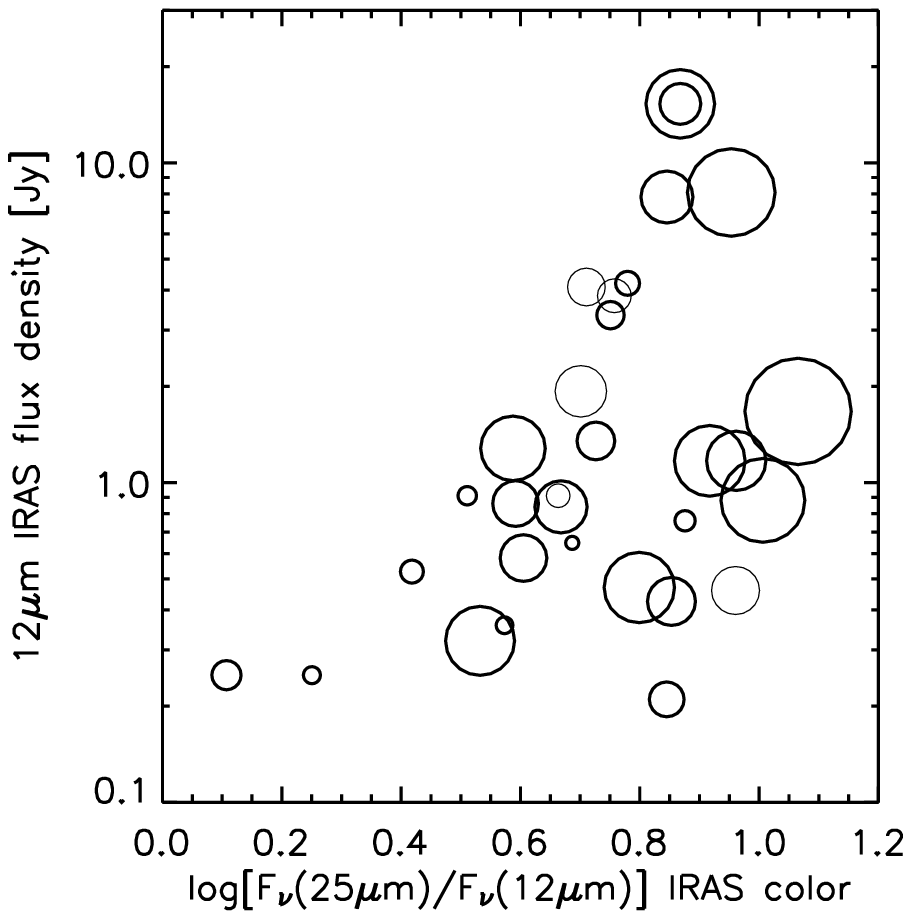}
\plottwo{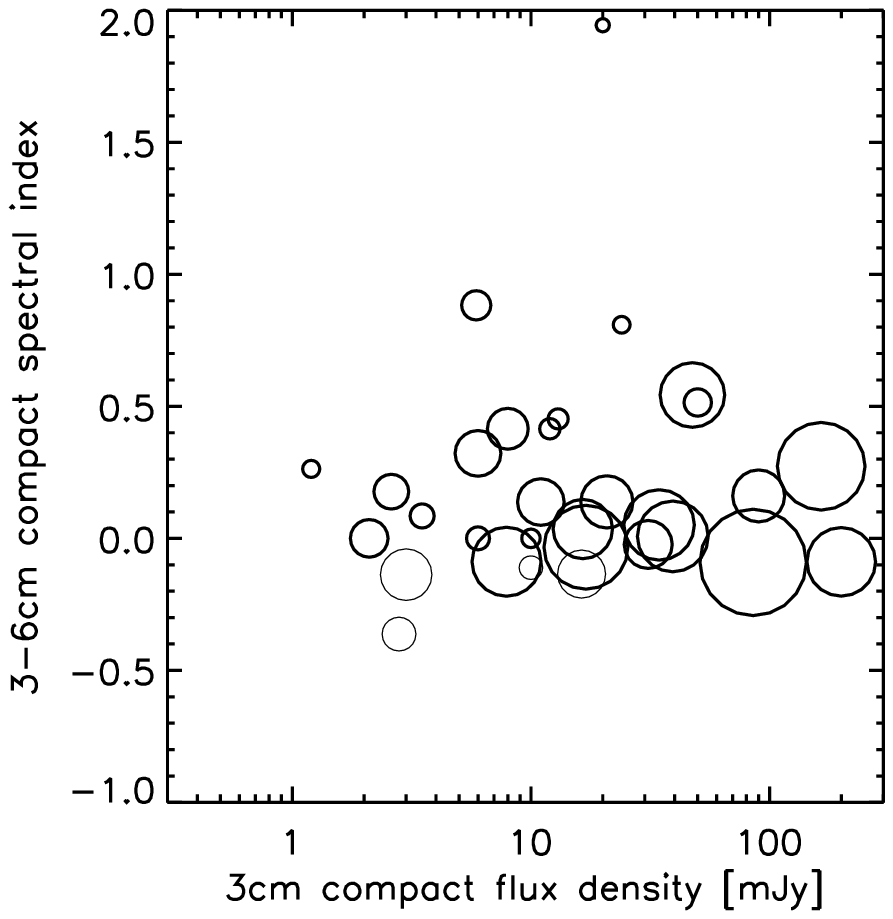}{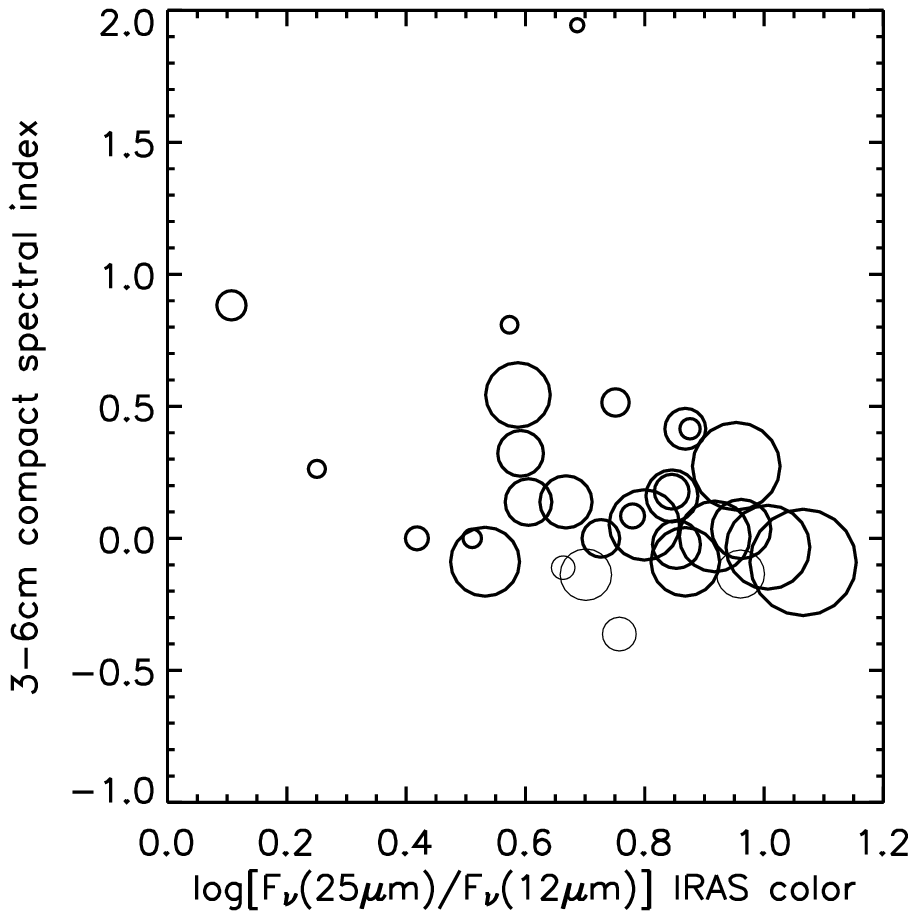}
\caption{\label{colcol} The various relationships between IRAS color,
IRAS flux density, 3~cm compact source flux density, and 3/6~cm
compact source spectral index are explored.  The size of the data
point scales with the signal-to-noise of the detected source.  The
flux densities of several compact radio sources are summed in cases
where more than one compact source is associated with a single IRAS
source.  Nonthermal sources are plotted with a thin line and excluded
from analysis of correlations (see text).}
\end{figure}

We used the WC89 FIR color criteria to select our sources, in order
the maximize the chances of finding compact thermal sources, or
ultracompact \hii regions.  While the selection criterion did yield a
very high detection rate of compact sources, many of which have flat
or inverted (optically thick thermal) spectral indexes, there do not
appear to be any clear correlations between the FIR color and the
optical depth of associated compact sources.  Figure~\ref{colcol}
explores various possible correlations between FIR (low-resolution)
color and flux density, and high-resolution cm spectral index and flux
density.  Nonthermal sources are plotted with thin lines and excluded
from analysis of correlations. There is a weak trend for sources with
higher 12$\mu$m flux densities to have higher 3~cm flux densities in
compact sources.  The Pearson correlation coefficient is 0.51 and the
slope between log($F_{12\mu m}$) and log($F_{3cm}$) is 0.50$\pm$0.17.
Including nonthermal sources decreases the correlation coefficient to
0.41, and the slope to 0.40$\pm$0.16.  There is also a weak trend for
brighter 12$\mu$m IRAS sources to have and envelope of redder
24/12$\mu$m colors.  The Pearson correlation coefficient between IRAS
color and IRAS 12$\mu$m flux density is only 0.14, consistent with an
envelope effect.

Our sample could address whether there is any correlation between the
luminosity of an \hii region and how embedded it is (as indicated by
the radio spectral index).  It has been suggested
\citep[e.g.][]{beckman00} that there is a change from
ionization-bounded to density-bounded as the luminosity of an \hii
region increases.  These investigators also find that more
H$\alpha$-luminous \hii regions leak more ionizing radiation than less
H$\alpha$-luminous ones.  The lower left panel of Figure~\ref{colcol}
shows the spectral index of our \hii regions as a function of 3~cm
flux density.  There is no discernible trend -- \hii regions powered
by more numerous or more massive ionizing stars show the same range in
\hii region density as less luminous regions.  There is also no
correlation between IRAS color and radio spectral index -- a redder
IRAS color does not predict a more optically thick compact \ion H2
region.  The correlation coefficients for radio spectral index with
radio flux density and IRAS color are -0.02 and -0.16, respectively.

\subsection{Modeled Size and Density}

\begin{figure}
\plotone{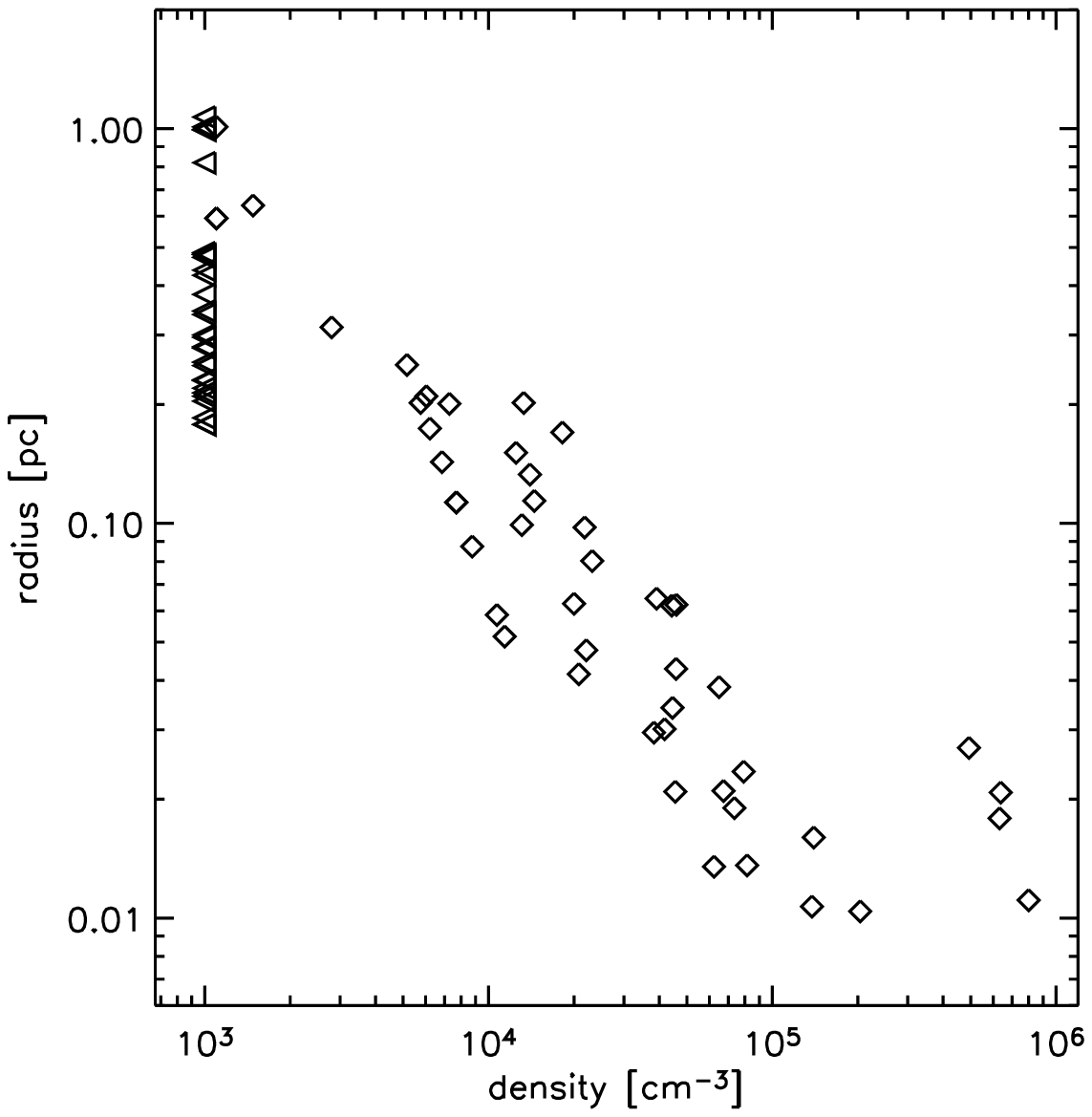}
\caption{\label{model}
Electron densities and radii of constant-density spheres that fit the radio 
flux densities of sources consistent with having thermal emission.}
\end{figure}

We can model the radio spectral energy distribution of each source as
a compact spherical \hii region of constant density and electron
temperature.  These simple models allow for only two free parameters,
radius and electron density.  It is assumed that the radio emission is
purely thermal in nature, but the free-free emission may be
self-absorbed at in order to reproduce spectral indices of $\alpha
>-0.1$ (sufficiently large electron densities cause the radio spectral
energy distribution to become ``inverted'' due to self-absorption at
lower frequencies).  The radio emission is calculated for each chord
through the sphere parallel to the line of sight, and this is
integrated over the projected circle perpendicular to the line of
sight (simple radiative transfer without scattering).  More complex
models are possible, but not warranted since we only have two data
points for each region.  Figure~\ref{model} shows the modeled radii
and densities of the sources that are consistent with thermal
emission.  The spectral shape of thermal emission at frequencies above
that at which the emitting source is optically thin is fairly
insensitive to the source density. Therefore we only model source
densities above 10\ts{3}~cm\ts{3}.  For sources consistent with lower
densities, we give the radius of a 10\ts{3}~cm\ts{3} source as a lower
limit to the radius.

\subsection{Exciting Star(s) Ionizing Flux}

If one assumes that each thermal compact radio source is an \hii
region and that the radio continuum emission comes exclusively from
optically thin thermal bremsstrahlung radiation, the number of
ionizing photons required to ionize the source is derived by
\citep{condon},
\[ Q \ge 6.3 \times 10^{52} s^{-1} 
\left({T_e}\over{10^4 K}\right)^{-0.45}
\left({\nu}\over{GHz}\right)^{0.1} \left({L_{thermal}}\over{10^27 erg
s^{-1} Hz^{-1}}\right),\]
where $T_e$ is the electron temperature of the nebula, $\nu$ the
frequency of observation, and $L_{thermal}$ the observed luminosity.
The Lyman continuum flux $Q$ calculated from the 3~cm flux density (or
the 6~cm flux density for sources with no 3~cm detection) is included
in Tables \ref{smcflux} and \ref{lmcflux}.  Note that these values are
lower limits due to the assumption that the emission is optically
thin.  The spectral type of a single star required to produce the
given ionizing flux is also tabulated, using the conversions from
\citet{smith02} and Crowther (private communication, 2003).

\begin{figure}
\plotone{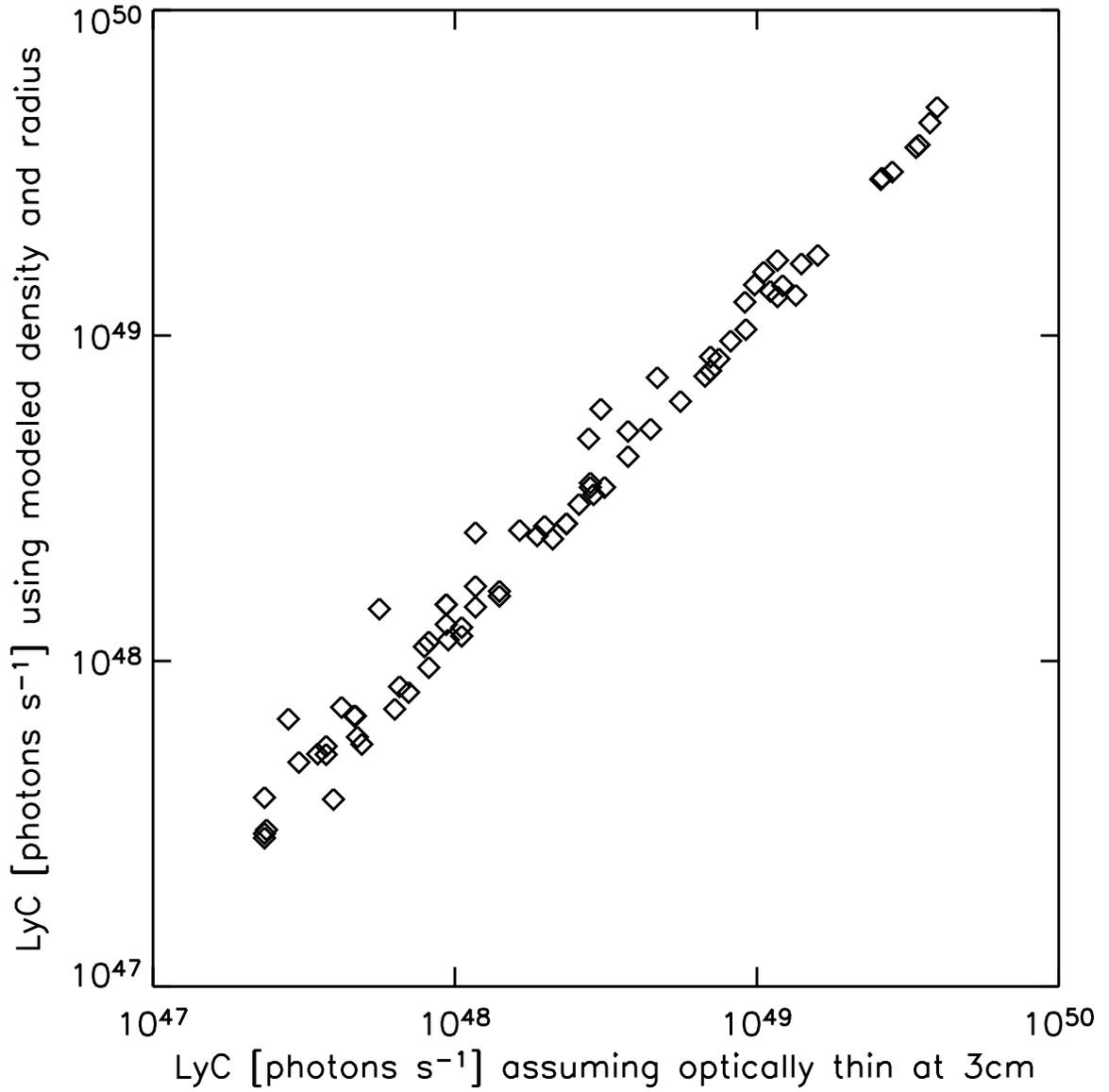}
\caption{\label{lyclyc} Lyman continuum fluxes calculated from the
\citet{condon} formula and the 3~cm flux density, and calculated from the
density and radius modeled to fit both flux densities.  The methods agree
well, with the model flux sometimes larger than the 3~cm-derived flux ,
probably for sources that are not completely optically thin at 3~cm.
}
\end{figure}

As a self-consistency check on the ionizing flux calculated using the
simple \citet{condon} formula and on the simple constant-density
models, we calculate the number of ionizing photons required to ionize
each modeled region, assuming photoionization equilibrium.
Figure~\ref{lyclyc} shows that the $Q$ values calculated using the two
methods agrees favorably.  The main source of the small discrepancies
in the assumption of the emission being optically thin in the Condon
relation above; if the emission is self-absorbed, $L_{thermal}$ will
be underestimated.
%

\subsection{Cluster Luminosity Function}

The luminosity function of extragalactic \hii regions can be fit with
a power law of index $a$=-2.0$\pm$0.5, i.e. $dN(L) \propto L^a dL$.
This result is similar for H$\alpha$ luminosities
\citep[e.g.][]{keh89}, radio luminosities \citep[e.g.][and references
therein]{mckee97}, and optical luminosities \citep[e.g.][in particular
$a$=-2.01$\pm$0.08 for the LMC]{larsen02}.  \citet{ee97} propose that
this is due to a universal formation mechanism: all types of clusters
form with constant efficiency in molecular clouds, so the cluster mass
distribution reflects the interstellar cloud mass distribution, which
has an index $\sim -2$.

The \hii region or cluster luminosity function is related to the
distribution of the number of cluster member stars $N_\star$ (or
cluster mass), and to the stellar initial mass function (IMF).
\citet{oey98} provide a detailed explanation of how these
distributions are related.  They draw attention to the particularly
important effect of small-number statistics.  In rich, luminous
clusters with many stars, the stellar IMF is statistically populated
out to a fairly high mass, and the cluster luminosity simply reflects
the total cluster mass.  The cluster luminosity function in this
regime, which \citet{oey98} call ``saturated'' then simply reflects
the cluster mass distribution, a power-law $dN(M_c) \propto
M_c^\beta dM_c$, with index $\beta$=-2.  Sparser clusters are
dominated by one or a few high-mass members -- the statistical
variation in luminosity of clusters of a given mass can become large
due to poor sampling of the stellar IMF.  This flattens the cluster
luminosity function at the faint end, and in fact can transform
power-law behavior into a more rounded distribution \citep[see
e.g. Fig.~4 in ][]{oey98}.

\begin{figure}
\plotone{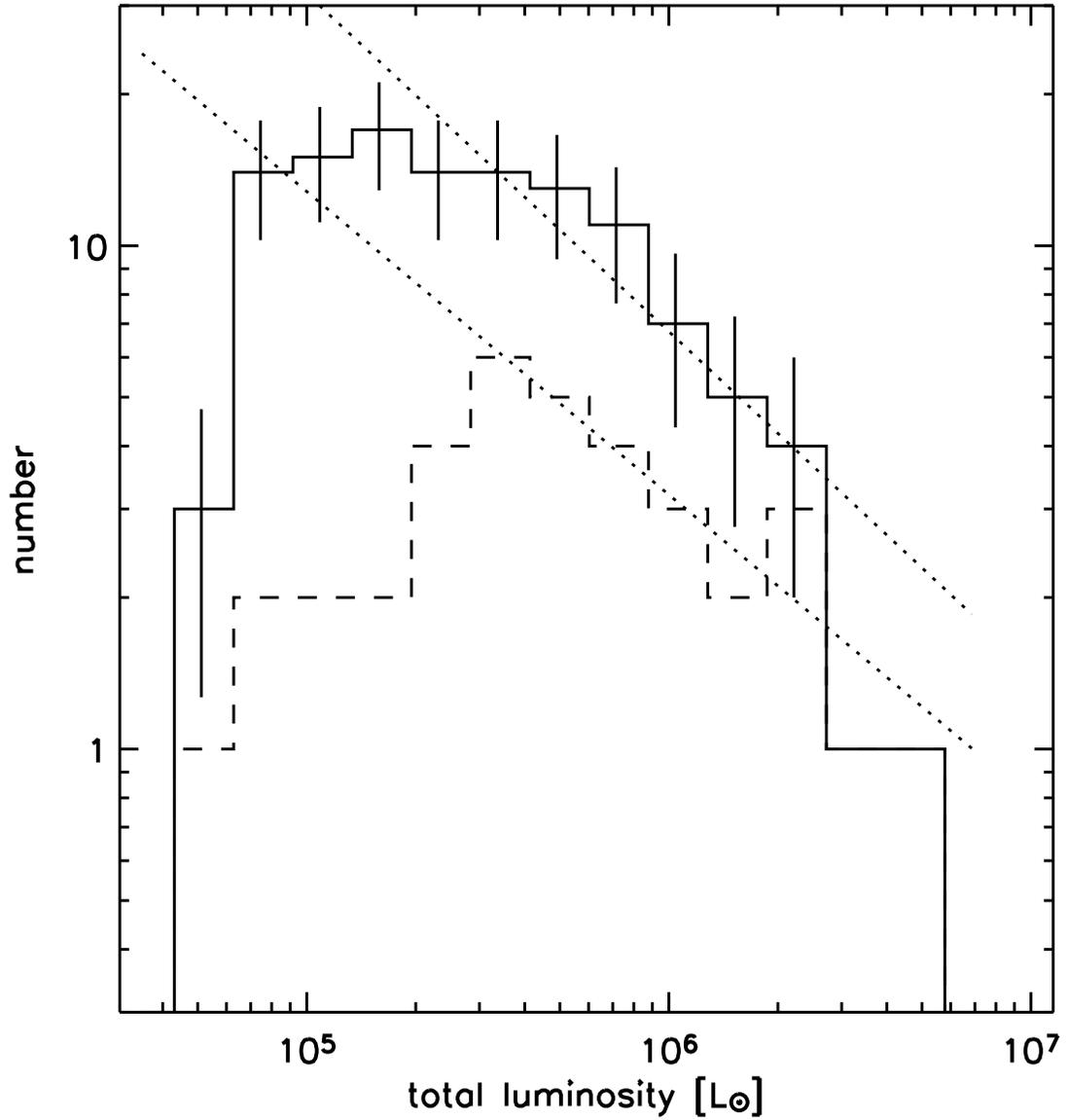}
\caption{\label{iras_clf} Total luminosity function of all IRAS
sources in the Magellanic Clouds that satisfy the WC89 color criteria
for \uchii regions (solid), and of our candidate sources (selection
criteria described in the text). Power laws are fitted to the
unconfused bright end of the distribution, and have indices of
-1.67$\pm$0.3 and -1.60$\pm$0.5 for the full and subselected
populations ($dN(L) \propto L^{-1.67} dL$).}
\end{figure}

Figure~\ref{iras_clf} shows the luminosity distributions of all IRAS
sources satisfying the WC89 color criteria, and all of our selected
candidate \uchii sources.  The bolometric luminosity has been
estimated for each \hii region by the sum over the 4 IRAS bands of
$\sum_{i=1}^{4} \nu(i) F_{\nu(i)}$ \citep[e.g.][]{casassus00}.  The
distributions flatten, probably due to the statistical effect
described above.  Single massive stars have luminosities of
$\sim$10$^5$L$_\sun$, so most of our sample is probably in the regime
of ``unsaturated'' stellar IMF statistics.
The turnover of the luminosity function at the faint end ($L \lesssim
10^5 L_{\sun}$) is due to confusion-induced incompleteness resulting
from the quite large IRAS beam (e.g. $\sim$50pc at 60$\mu$m at the
distance of the Magellanic Clouds).  The confusion limit for sources
\citep{condon74} with a uniform spatial distribution and following a
power law luminosity distribution with is about 10 beams/source.  The
effective resolution of the IRAS point source catalog is a nontrivial
function of the spatial distribution of emission, but the expected
confusion limited source densities are about 40, 40, 15, and 5
sources~deg\ts{-2} at 12, 25, 60, and 100~$\mu$m, respectively
(compare to the source densities at which special ``high density''
algorithms were invoked in constructing the IRAS catalog: 45, 45, 16,
and 6 sources~deg\ts{-2}).  This translates into confusion limited
flux densities of 0.3, 0.3, 2, and 20~Jy at 12, 25, 60, and 100~$\mu$m
(verified by inspection of the point source catalog in the LMC), and
an estimated confusion limited total luminosity in the Clouds of
2.5\up{38}~erg/s (7\up{4}~L$_\sun$).  The luminosity function of all
IRAS sources with \hii region colors (solid curve in
Fig.~\ref{iras_clf}) indeed turns over about that confusion limited
luminosity.

It is reassuring to note that the distributions above their peaks have
the same power-law indexes within errors (-1.67$\pm$0.3 for the full
color-selected sample and -1.60$\pm$0.5 for our ``radio-aware''
selection), and despite a slight bias toward observing brighter
sources, we have not apparently affected our sample with selection
effects beyond those unavoidable results of low spatial resolution
previous observations.

It is possible to quantify the constraints that our sample, selected
to probe the most compact and densest \hii regions, places on the
stellar and cluster mass functions.  We use Monte-Carlo methods to
model the cluster luminosity function, similar to the process followed
by \citet{oey98} and \citet{casassus00}.  We assume that the cluster
masses $M_c$ follow a power-law distribution \[ p(M_c) \propto
M_c^{-\beta} \] and that the stars in the cluster follow a standard
stellar initial mass function (IMF) \[ p(M_\star) \propto
M_\star^{-(1+\gamma)}.\] \citet{casassus00} imposed a power-law on the
number of stars per cluster \[ p(N_\star) \propto N_\star^{-\beta}, \]
rather than on the cluster mass.  We prefer to use the cluster mass
because it has physical meaning and may be related to the interstellar
cloud mass distribution.  The synthetic luminosity functions
constructed either way are quantitatively consistent.
We assume that because we selected for young star-forming \hii
regions, that our cluster population can be adequately modeled as
unevolved.  \citet{oey98} showed that evolution can be an important
effect for older \hii regions (e.g. in interarm regions of spiral
galaxies), as the most massive stars in the cluster die off, and the
more numerous less massive stars become more important to ionizing the
cluster.

\begin{figure}
\plotone{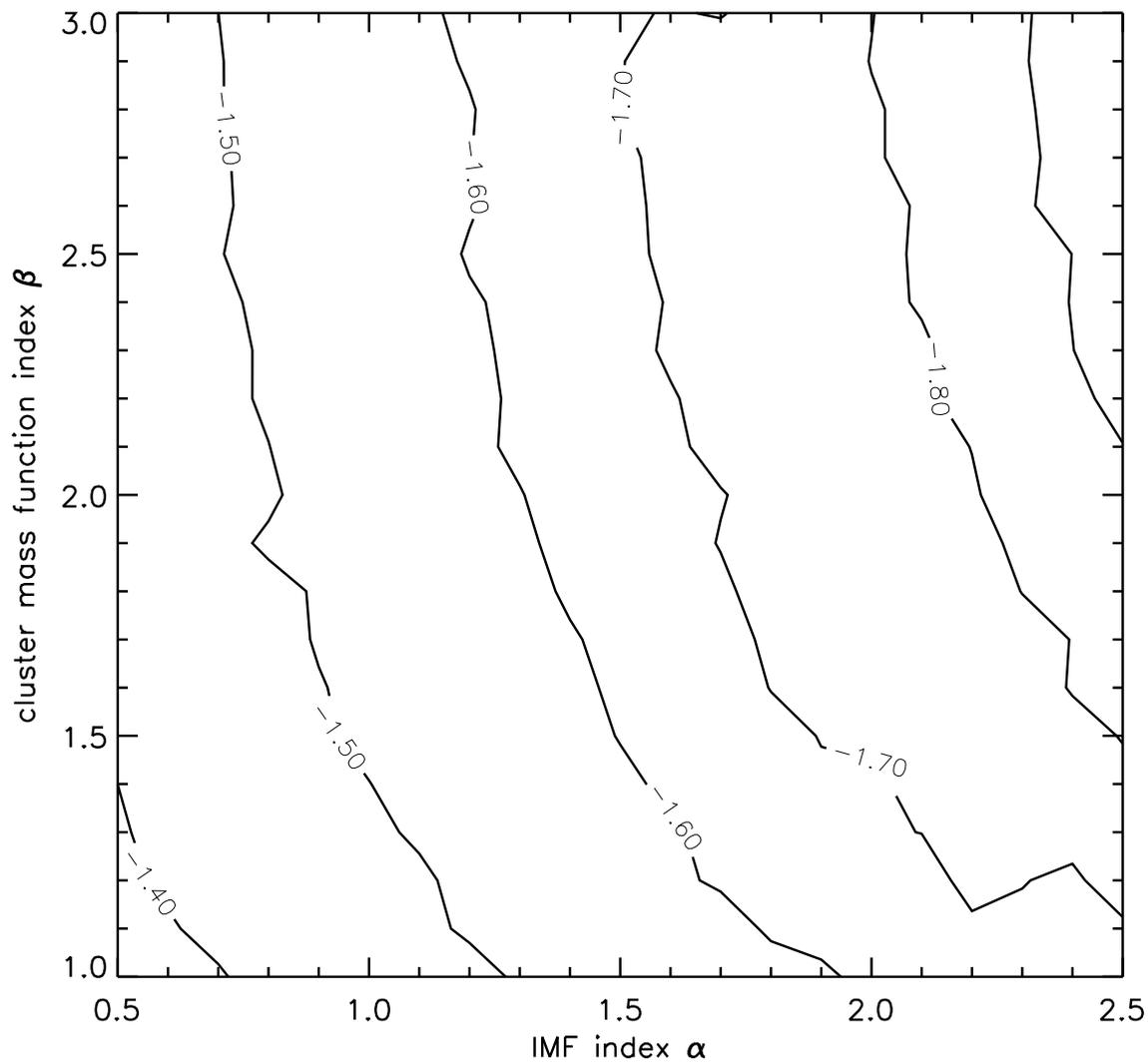}
\caption{\label{lmodel} The index of a power-law
fit to the bright end of a synthetic cluster luminosity function, as a
function of stellar IMF index $\gamma$ and cluster mass function index
$\beta$. For this sample, the cluster luminosity function index is $a \approx -1.65$,
which corresponds to a standard stellar initial mass function of 
$\gamma \approx 1.5$, but is fairly insensitive to the cluster mass function. }
\end{figure}

Figure~\ref{lmodel} shows the index of a power-law fit to the bright
end of the cluster luminosity function as a function of IMF power-law
index $\gamma$ and of the cluster mass function power-law index
$\beta$.  The results are insensitive to the stellar and cluster mass
upper and lower cutoffs (for this particular run (0.5,120)~M$_\sun$
and (4,12000)~M$_\sun$, respectively).  Such a fitted power-law is
sensitive to the range over which one fits, since the luminosity
function rolls over, so Figure~\ref{model} should be used to
understand qualitative trends, with possible systematic offsets in the
index (by 0.1--0.2) if a different luminosity range is used.
The effects of poor statistics on the form of the luminosity function
have been well-described by \citet{oey98}: The high-L end of the
distribution is defined by a chance collection of particularly massive
stars and can fall off very quickly.  At the faint end, the
distribution is flattened by the spread in L for a given cluster mass.
In this statistically unsaturated regime, the {\it cluster} luminosity
function index is much more strongly dependent on the {\it stellar}
mass function index than the cluster mass function index.  Due to the
steepness of the mass-luminosity relation for main-sequence stars, it
is not hard to make a fairly bright cluster with a small number of
massive stars.  Instead of enforcing a power-law on either data or
model, we fit the bright (unconfused) part of the observed luminosity
function to theoretical models.  The best fit is for
$\gamma$=1.5$\pm$0.25 and $\beta$=2.0$\pm$0.5.

\begin{figure}
\plotone{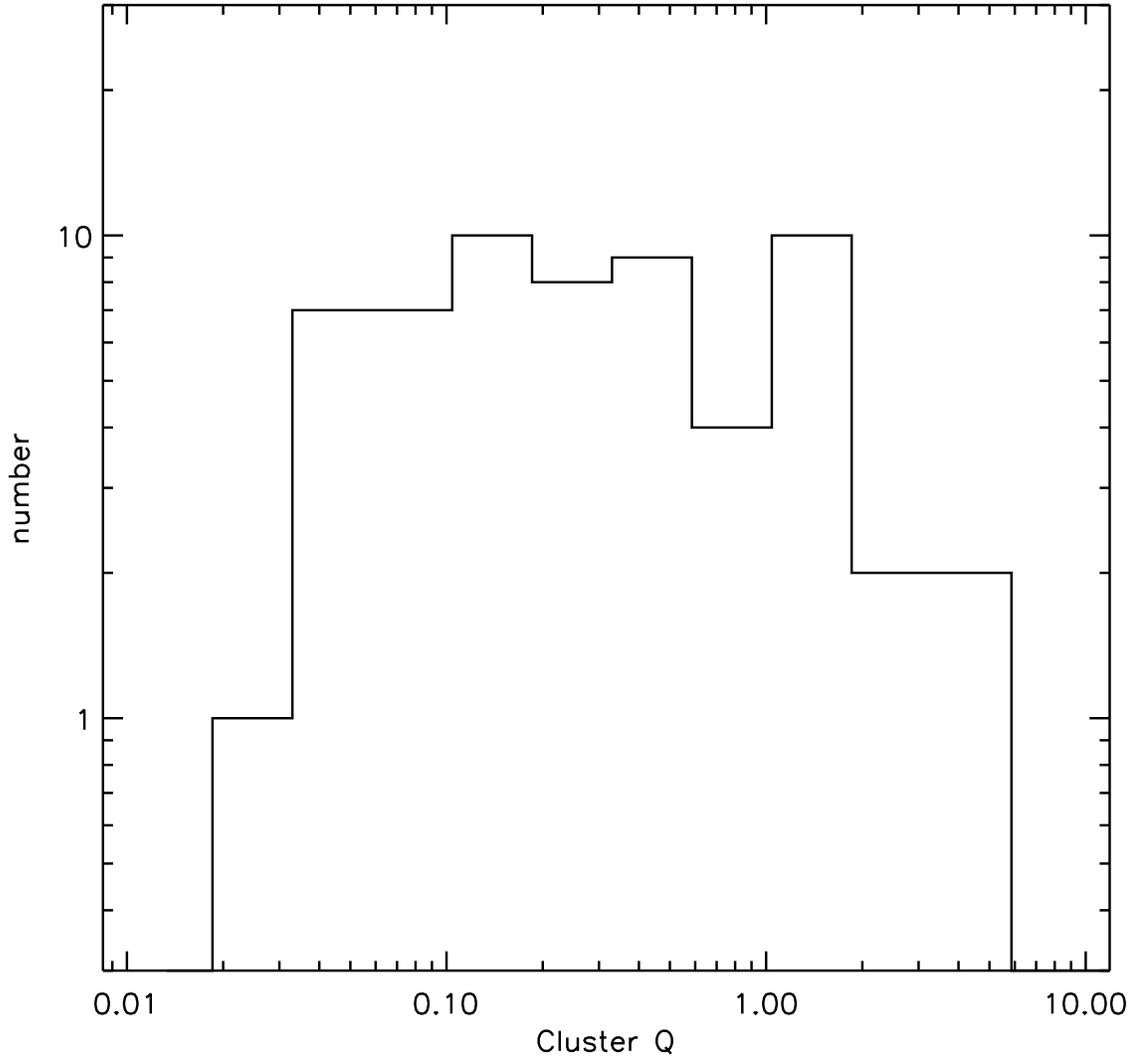}
\caption{\label{qdist} The distribution of Lyman continuum fluxes $Q$
implied for the observed compact radio sources.  $Q$ is quoted in
units of 1\up{49}~photons~s\ts{-1}.  
}
\end{figure}

\begin{figure}
\plottwo{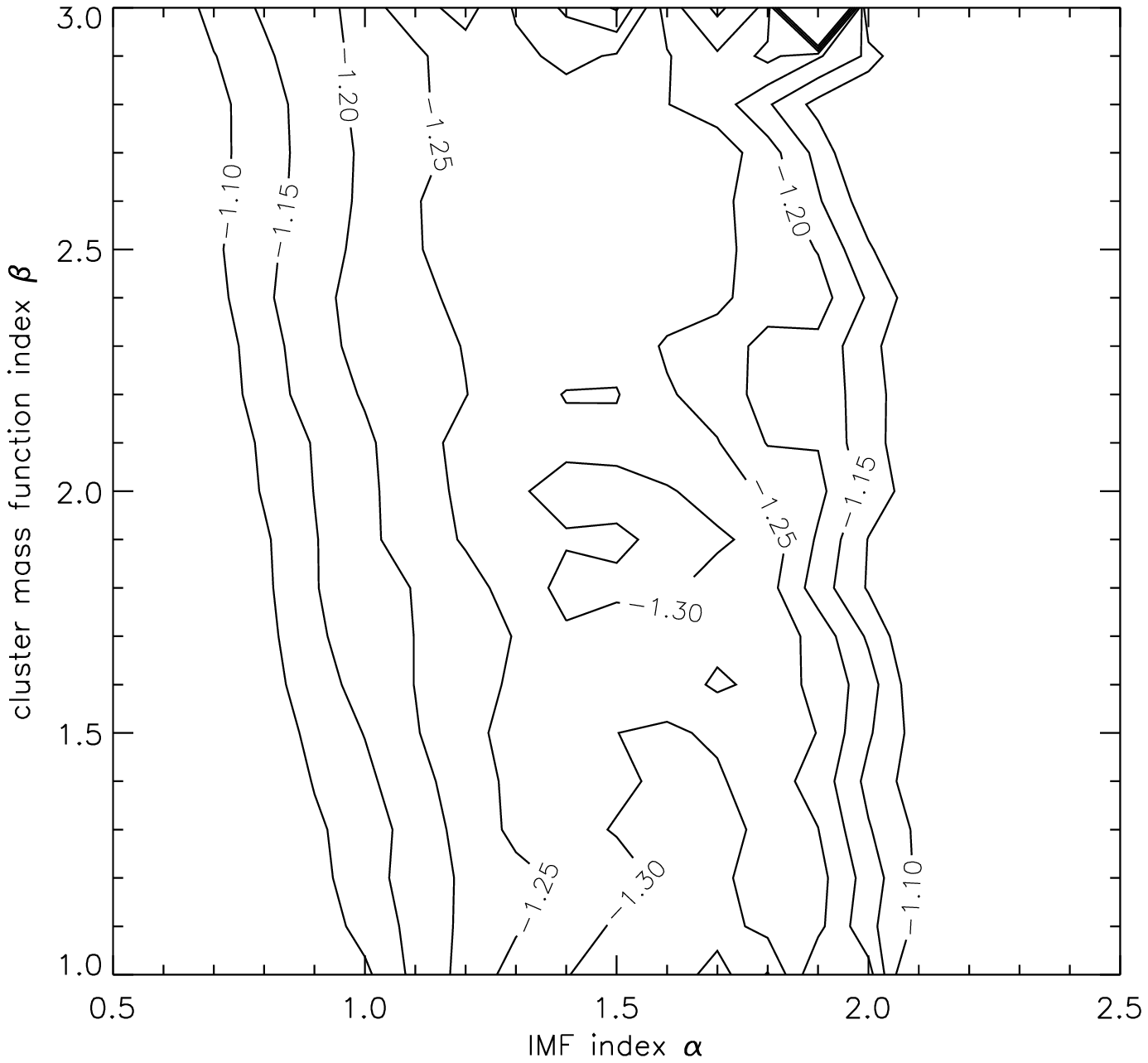}{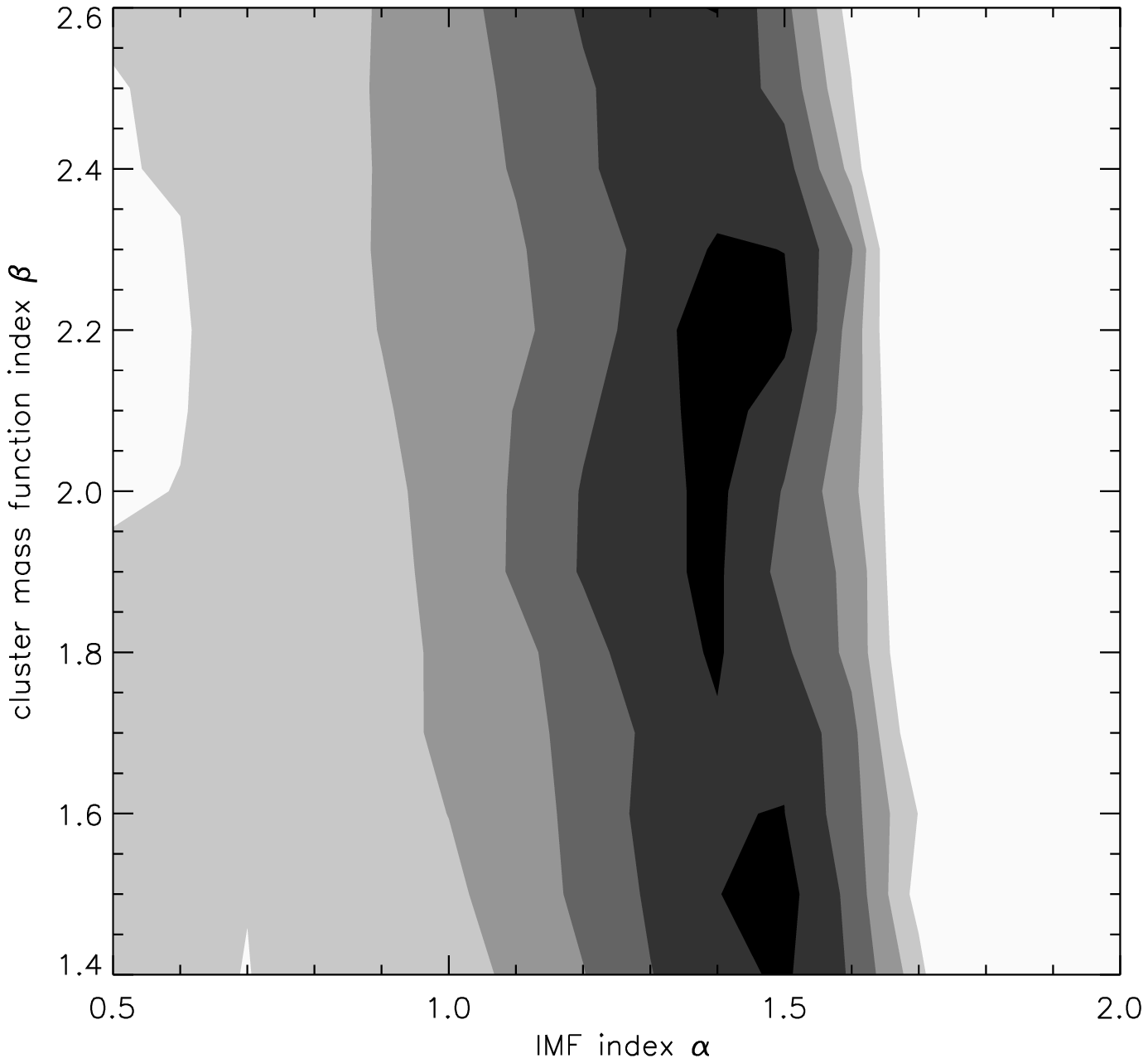}
\caption{\label{qfit} The first panel shows the index of a power-law
fit to the upper end of a synthetic $Q$ distribution, as a function of
stellar IMF index $\gamma$ and cluster mass function index $\beta$.
It is more robust not to impose a power-law on either the observed or
model $Q$ distribution, so confidence contours for the fitting the
data with this type of model are shown in the second panel.  The data
is best fit by the white regions, and the goodness-of-fit decreases
with darker shades. }
\end{figure}

It is even more interesting to fit the distribution of Lyman continuum
photon fluxes $Q$.  $Q$ is determined here from our radio
observations, so this constitutes analysis of our observations, rather
than analysis of the input sample (the total luminosities were
calculated from IRAS flux densities). The $Q$ distribution of our
observed compact radio sources is shown in Figure~\ref{qdist}.  The
observed (and modeled) $Q$ functions are even less easily understood
as power laws; the $Q$-distribution is even further into the
statistically unsaturated regime, because of the extremely steep
mass-$Q$ relation for main sequence stars (even steeper than the
mass-luminosity relation).  For illustration only, we proceed to fit a
power-law to the portion of the distribution which is relatively
straight, and show the index of this power-law as a function of
$\gamma$ and $\beta$ in Figure~\ref{qfit}.  As with the total cluster
luminosity, the $Q$ distribution is more sensitive to the stellar IMF
index $\gamma$ than the cluster mass function index $\beta$.  Again,
we search for the model distribution which best fits the data without
imposing a power-law, and find $\gamma$=1.4$\pm$0.2 and
$\beta$=2.1$\pm$0.3.  These indexes agree with those determined from
the total IRAS luminosity (at much lower resolution), and with
observations and theories of star formation in the Milky Way.

\section{Conclusions}
\label{conclude}

We conducted a radio continuum survey of \uchii candidates in the
Magellanic Clouds using the highest resolution configuration of the
Australia Telescope Compact Array.  Candidates were selected from IRAS
and lower resolution radio continuum measurements, using the far-IR
colors determined by \citet{wc89b} to be characteristic of \uchii
regions.  We find a very high success rate for detecting compact radio
sources with spectral indexes consistent with thermal emission based
on these selection criteria, even with the relatively poor spatial
resolution of the IRAS survey.  This detection rate is at least in
part due to the thermal infrared luminosities of \uchii regions;
\uchii regions are among the most luminous Galactic objects in the
IRAS catalog \citep[$>60$\% of the IRAS sources with flux densities
$>10^4$~Jy at 100$\mu$m are \uchii regions][]{wc89a}.  Consequently,
\uchii regions will tend to be dominant sources at thermal infrared
wavelengths, making them relatively easy to detect based on their
infrared colors despite contaminating sources within the resolution
element.  This is good news for inferences drawn from infrared
observations with the {\it Spitzer} Space Telescope and ASTRO-E
satellites, which probe similar spatial scales in nearby galaxies as
IRAS did in the Magellanic Clouds.

In many of the cases where such information is available, we find that
the compact radio sources are found in the densest, highest excitation
parts of \hii region complexes.  Simple models of the radio sources as
constant-density spheres have electron densities of a few times
10\ts{3} to 10\ts{6}~cm\ts{-3}, and sizes ranging from 0.01 to 0.1
parsecs, reasonable physical characteristics for compact and
ultra-compact \hii regions.  We are clearly sampling the population of
youngest embedded \hii regions in the Clouds.

We compare the radio flux densities of our detected compact sources
with the infrared flux densities of the candidate regions, and with
radio flux densities at lower spatial resolution. Many of the radio
sources have spectral indexes that are consistent with thermal
radiation, and there are some that are probably background nonthermal
sources.  The spectral index of the compact sources detected is often
larger (more consistent with thermal radiation or a higher optical
depth) than the spectral index calculated for the same source from the
relatively low resolution Parkes data.  This is expected physically,
because the more diffuse emission around a compact embedded source
will be optically thin, driving the spectral index from a positive
value toward zero.  Although most candidates with WC89 infrared colors
apparently contain compact \hii regions, no obvious trends that more
extreme infrared colors indicate more embedded radio sources were seen
{\it at these resolutions}.

We model the total luminosity and Lyman flux ($Q$) distributions of
the regions, using a Monte-Carlo method to model those {\it cluster}
number distributions as a function of the {\it stellar} mass function
within a cluster and the cluster number or mass distribution.  We find
that the population of young clusters in the Magellanic Clouds, as
observed through their compact \hii regions, is consistent with a
stellar IMF that one would find reasonable in the Milky Way: slightly
steeper than Salpeter, with a power-law index $\gamma\simeq$1.5.  We
find that the Magellanic Cloud cluster distribution is consistent with
a fairly broad range of cluster mass functions, but the best-fit has a
power-law index $\beta\simeq$-2.0, which is expected from the mass
distribution of interstellar molecular clouds on many scales.

\section{Acknowledgments}

We thank Ed Churchwell for useful discussion.  This work would not
have been possible without the dedicated work of Robin Wark, Jim
Caswell, and the other staff at the Australia Telescope National
Facility.  R.I. was supported at the start of this investigation by an
NSF Graduate Student Fellowship to the University of Colorado, and
currently supported by NASA (GLIMPSE {\it Spitzer} Legacy Grant to
E. Churchwell, U. Wisconsin). K.J. is currently supported by an NSF
A\&AP postdoctoral fellowship.  P.C. acknowledges support from NSF
(AST 9731570).

The Second Palomar Observatory Sky Survey (POSS-II) was made by the
California Institute of Technology with funds from the NSF, NASA,
the National Geographic Society, the Sloan Foundation, the Samuel
Oschin Foundation, and the Eastman Kodak Corporation.  
%
This research made use of the NASA/ IPAC Infrared Science Archive,
which is operated by the Jet Propulsion Laboratory, California
Institute of Technology, under contract with the National Aeronautics
and Space Administration.
This research also made extensive use of NASA's Astrophysics Data
System Bibliographic Services, and of the SIMBAD database operated at
CDS, Strasbourg, France.

\clearpage

\appendix
\section{Detections in Individual Regions}
\label{individuals}

\subsection{Small Magellanic Cloud}

\subsubsection{N66}

\begin{figure}
\plottwo{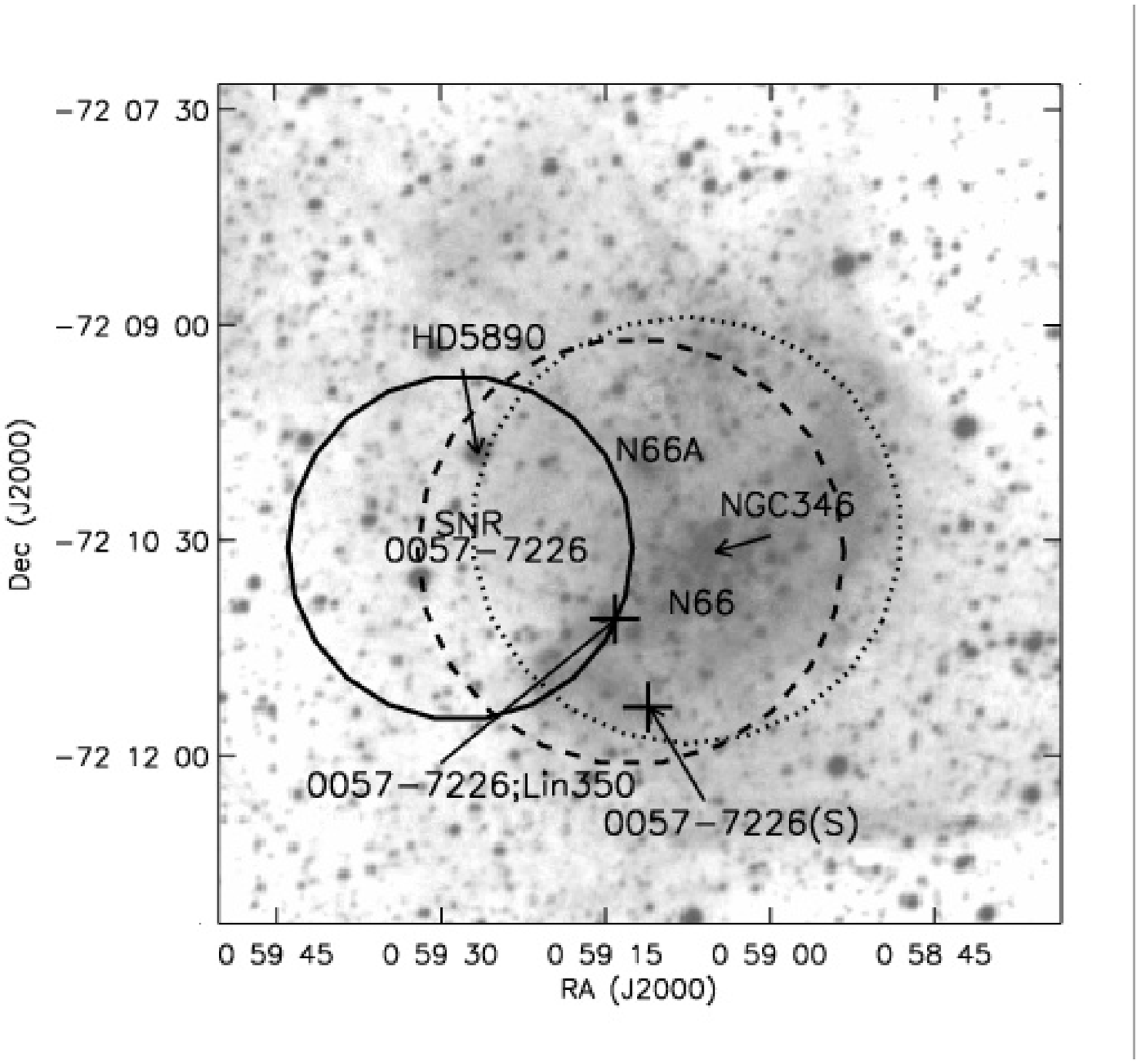}{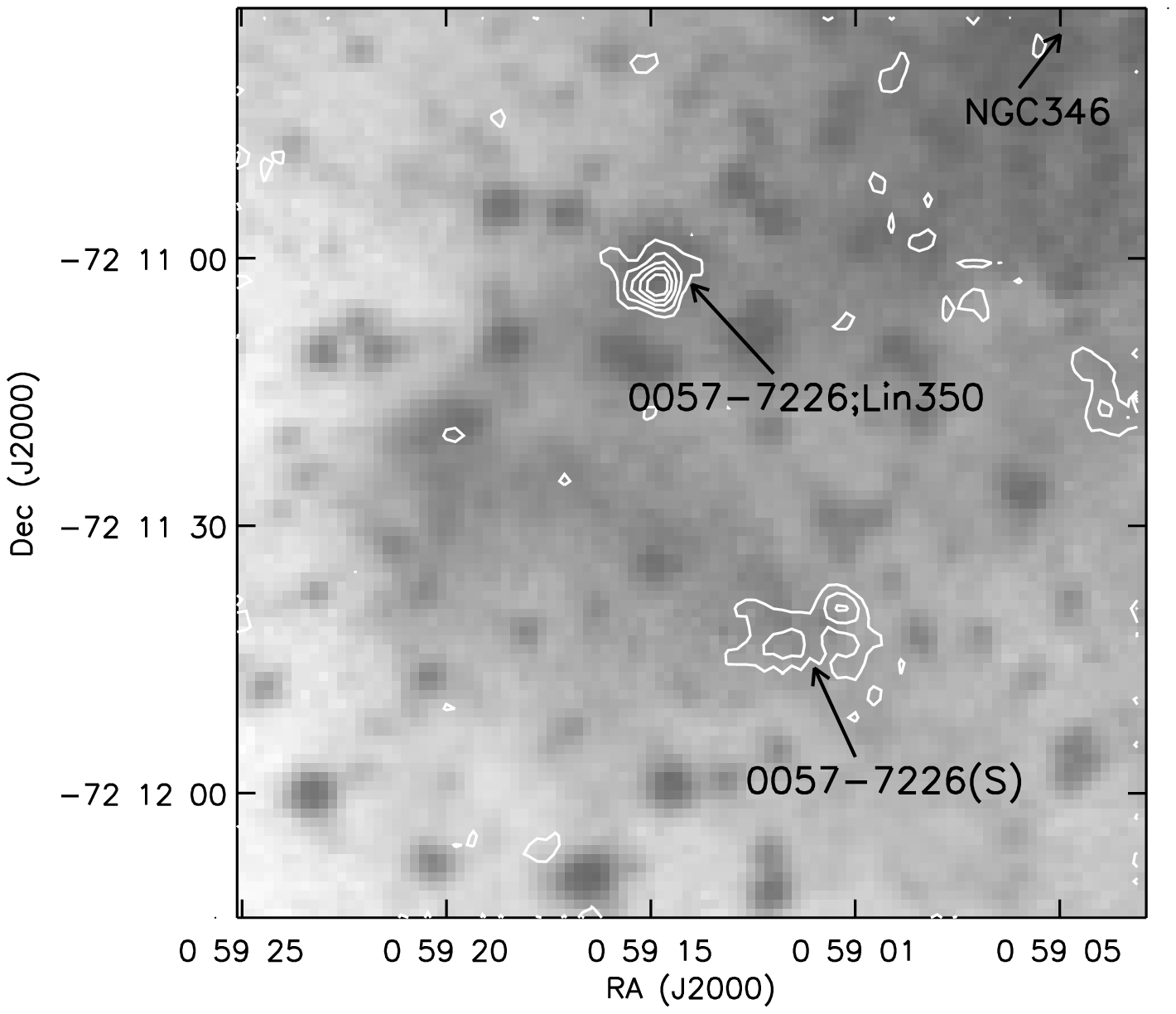}
\epsscale{0.5}
\plotone{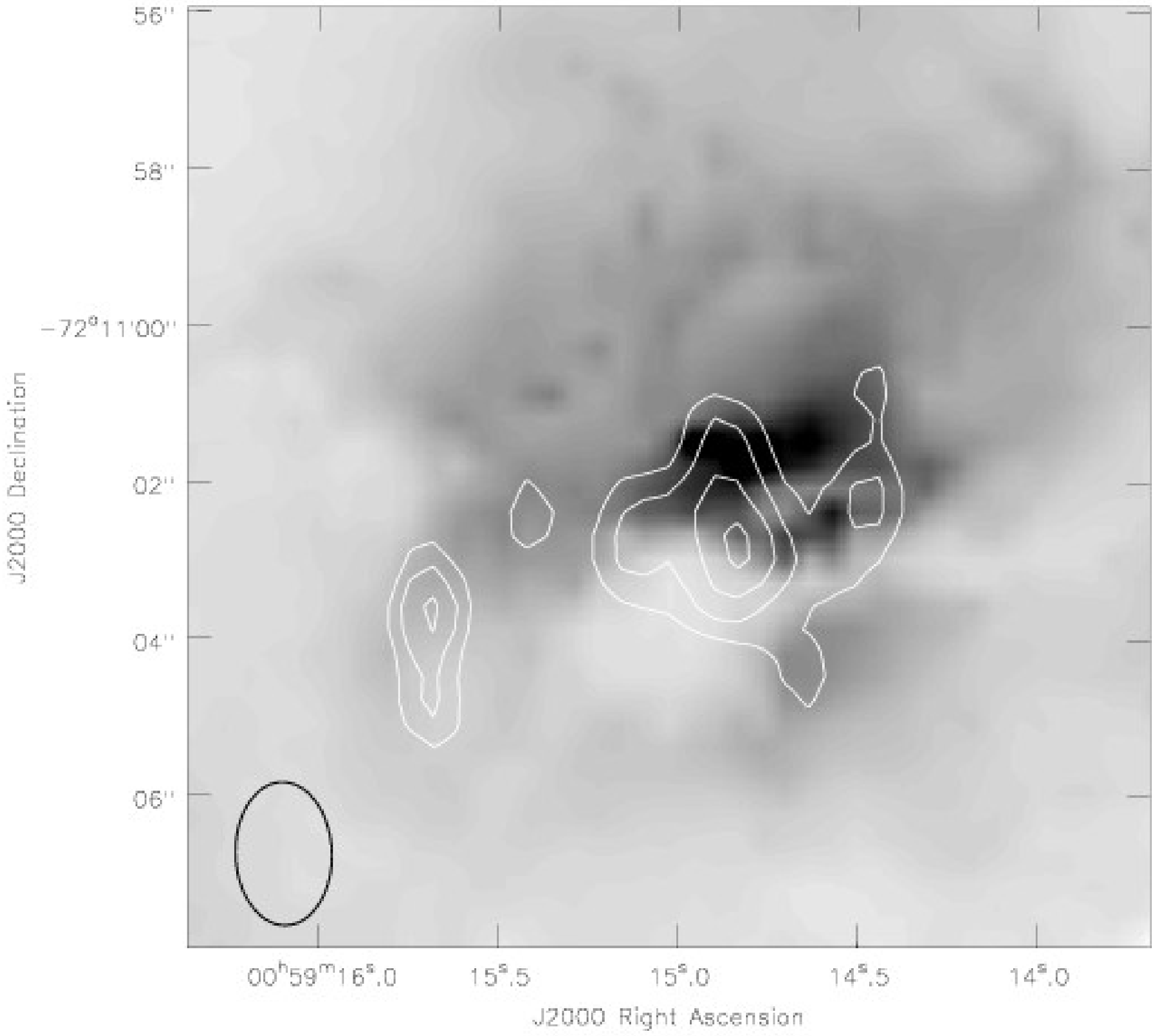}
\epsscale{1.0}
\caption{\label{n66fig} Panel 1 is a POSS-II R image of the N66 \hii
complex, with optically identified subregions labeled.  The Parkes 3
and 6~cm source \citep{filip98a} is indicated by the dashed line, and
the IRAS point source by the dotted line (where the size of the circle
is the spatial resolution of those surveys).  The compact radio
sources detected in this survey are marked with crosses. The supernova
remnant 0057-7226 is marked with a solid circle.  The second panel
shows 6~cm contours from this dataset on a zoomed-in optical image.
The third panel shows contours of 3~cm compact emission superimposed
on an {\it HST} H$\alpha$ image.}
\end{figure}

N66 is the brightest \hii region in the SMC, and should not be
confused with a famous planetary nebula of the same name in the LMC.
The region does not appear to have significant atomic
\citep{stavely97} or molecular gas \citep[except for a cloud to the
NE,][]{contursi00}.  Figure~\ref{n66fig} shows the POSS-II R image of
N66 with important objects marked, including the location of SNR
0057-7226 \citep{ye91}.  Interpretation of the region is complicated
by the presence of that SNR and the spatially coincident wind-blown
bubble from HD5980, which may be behind the SNR \citep[see cartoon of
the region in][Figure~6]{danforth03}.  The main \hii region (elongated
to the NW-SE) is likely powered by the cluster NGC~346 visible in
Figure~\ref{n66fig}.  \citet{massey89} performed an extensive study of
the stellar content of the region, and find several very massive
(40--85~M$_\sun$) stars and many B-type stars.  Our radio contours
show two compact sources to the southeast of NGC~346.

High-resolution 3~cm radio contours if the northern source (0057-7226)
are shown superimposed on H$\alpha$ emission in the third panel of
Figure~\ref{n66fig}.  Filamentary structure seen in H$\alpha$ emission
is reflected by complex morphology of the radio continuum emission in
N66.  This source is coincident with the mid-IR peak ``H'' in
\citet{contursi00}, which they find to have an unremarkable
mid-infrared spectrum.  The source has complex morphology in both
radio continuum and H$\alpha$.  While it does not appear to be
associated with any of the stars \citet{massey89} identified as
massive (and relatively unembedded) based on U-B color, it is
spatially coincident with the emission-line object Lin~350
\citep[][who identify it as a compact \hii region based on H$\alpha$
spectral line morphology]{me93}. This radio source also lies within
the CO(2-1) contours in the observations of \citet{rubio00}.

The other source, 0057-7226(S) is also associated with the \hii region
N66 (as above).  However, it does not appear to have a 7$\mu$m
counterpart in the \citet{contursi00} ISOCAM observations, and it does
not fall within the CO(2-1) contours of \citet{rubio00}.  In
combination with its spectral index ($\alpha\simeq$-0.5), this
suggests that it may not be an \uchii region, but perhaps a background
object (e.g. AGN), or supernova remnant.

\subsubsection{N81 and N88}

\begin{figure}
\plottwo{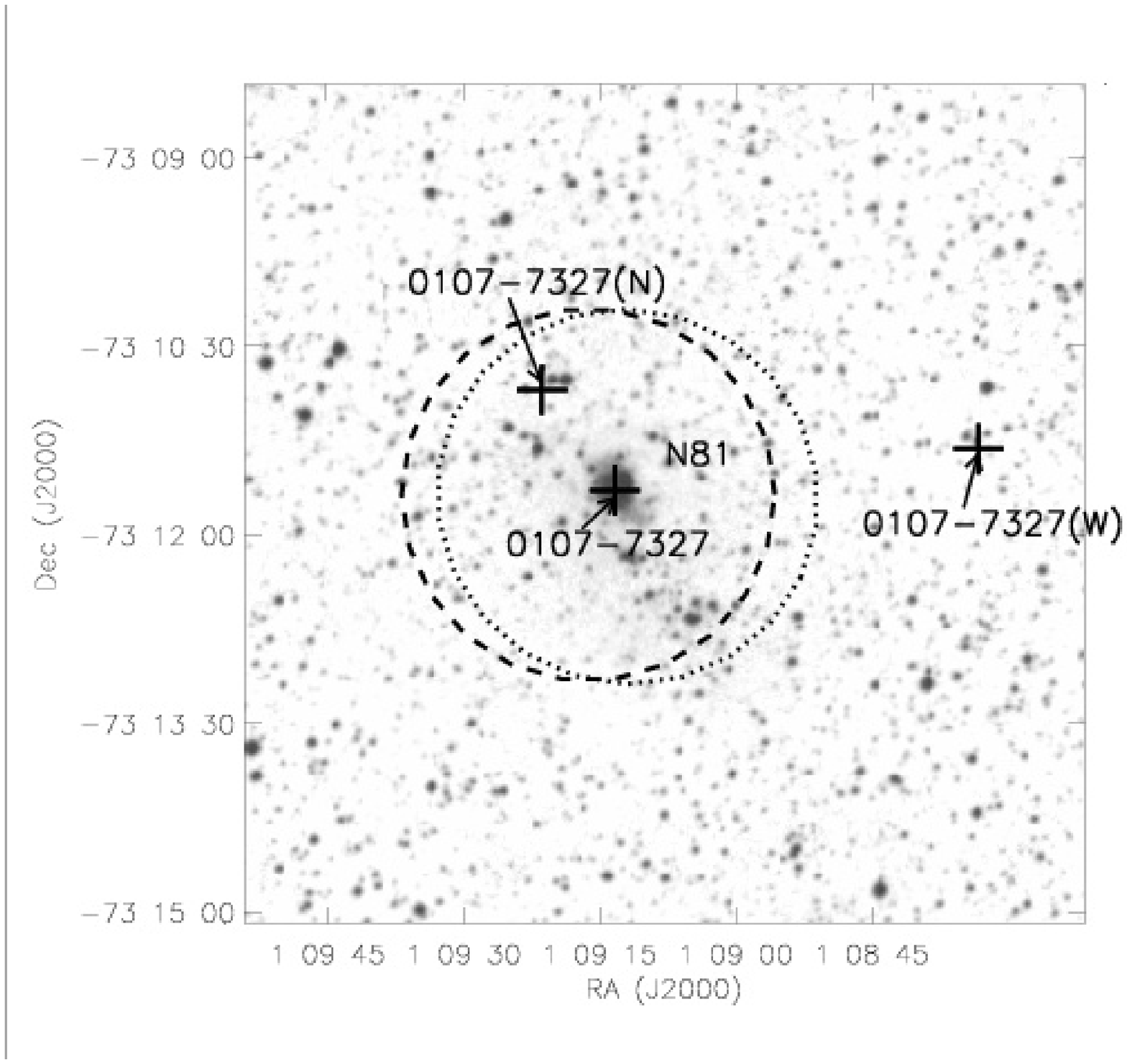}{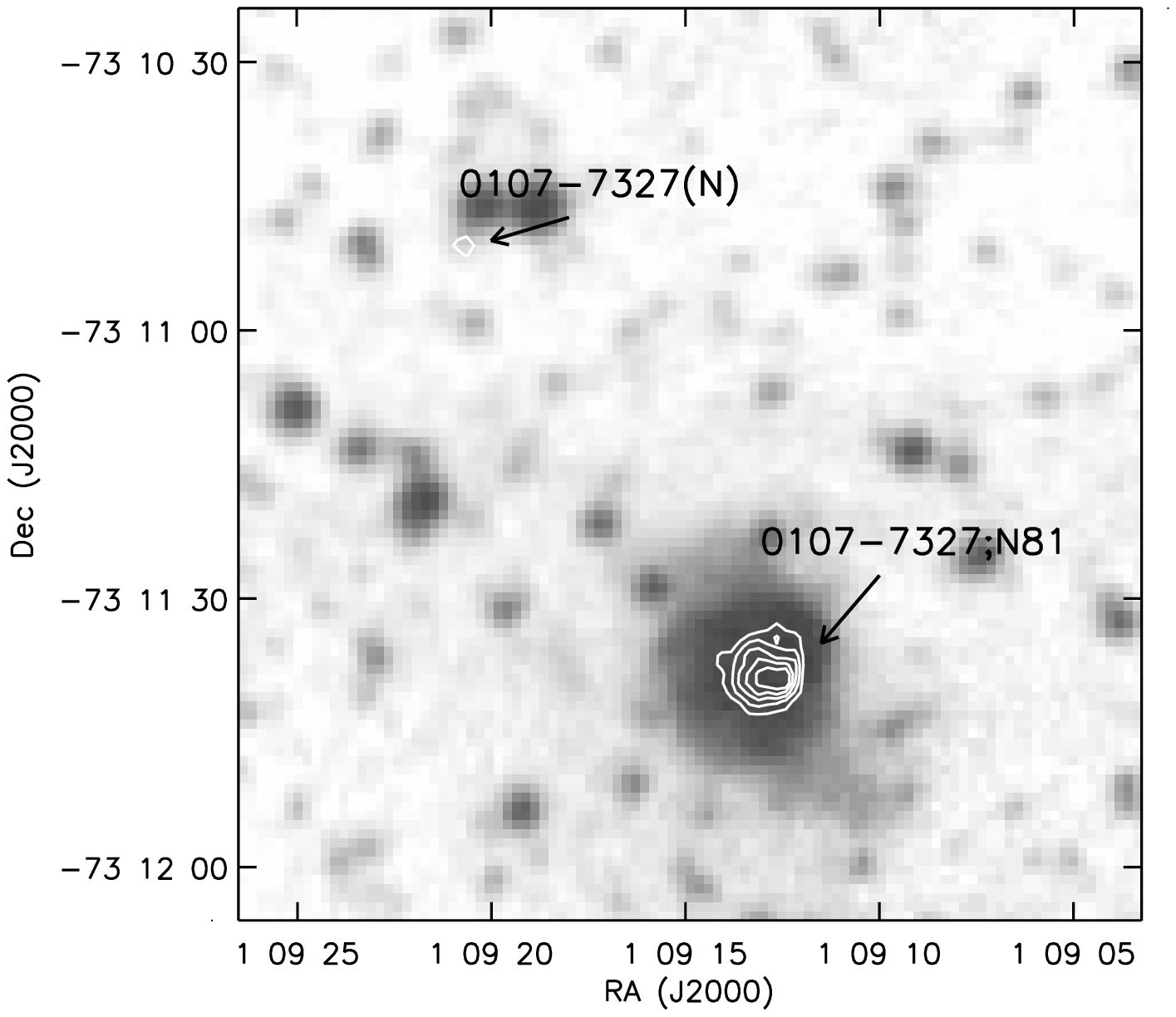}
\epsscale{0.5}
\plotone{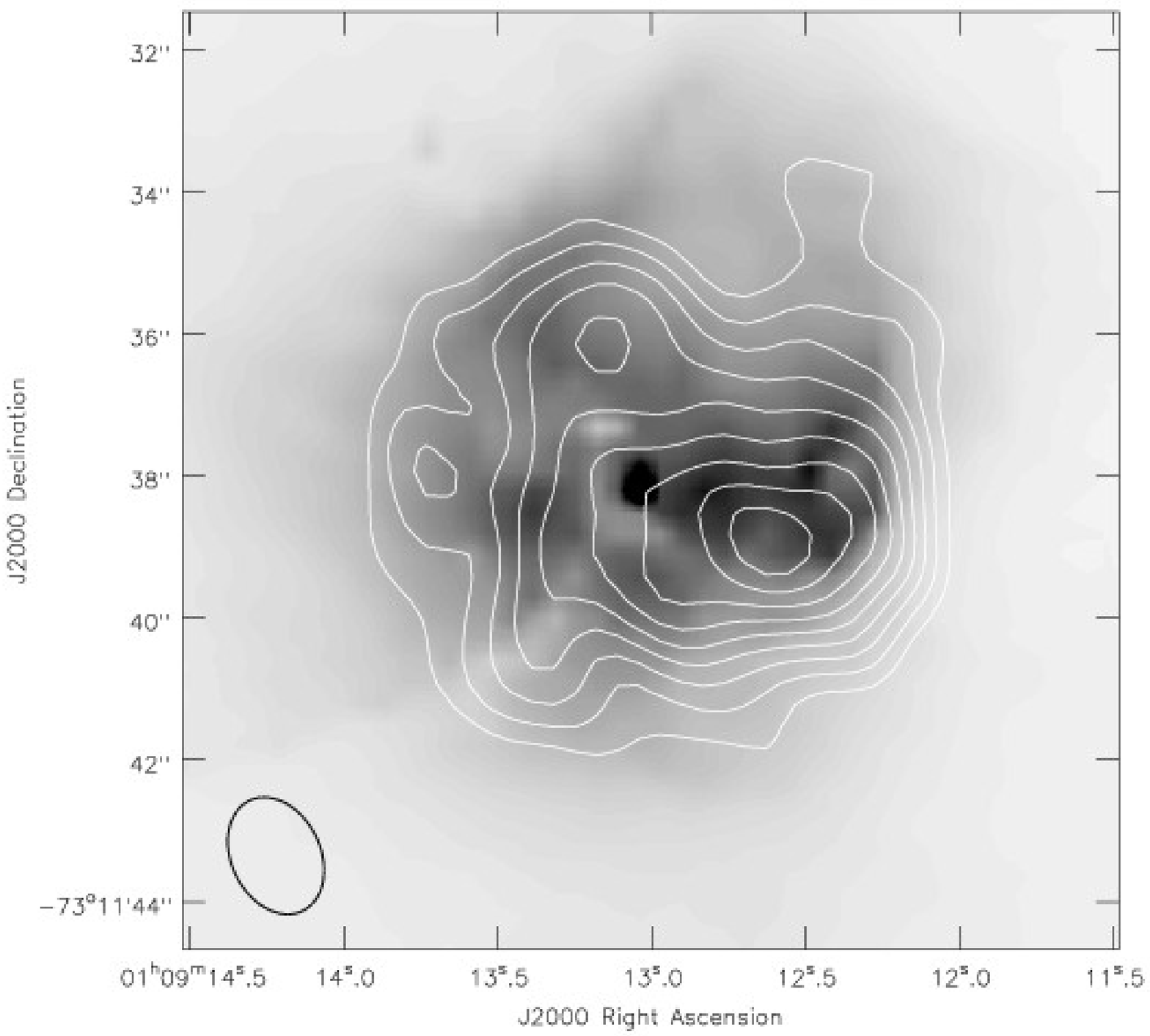}
\epsscale{1.0}
\caption{\label{n81fig} The first two panels show POSS-II R images of the
N81 \hii complex, with optically identified subregions labeled in
panel (a).  Parkes 3 and 6~cm sources \citep{filip98a} are indicated
by dashed lines, and IRAS point sources by dotted lines (with the size
of the circle indicating the spatial resolution of those surveys).
The compact radio sources detected in this survey are marked with
crosses.  In the second panel, 6~cm contours are shown on a zoomed-in
optical image.  The third panel shows contours of high-resolution
3~cm radio emission superimposed on an {\it HST} H$\alpha$ image.}
\end{figure}

\begin{figure}
\plottwo{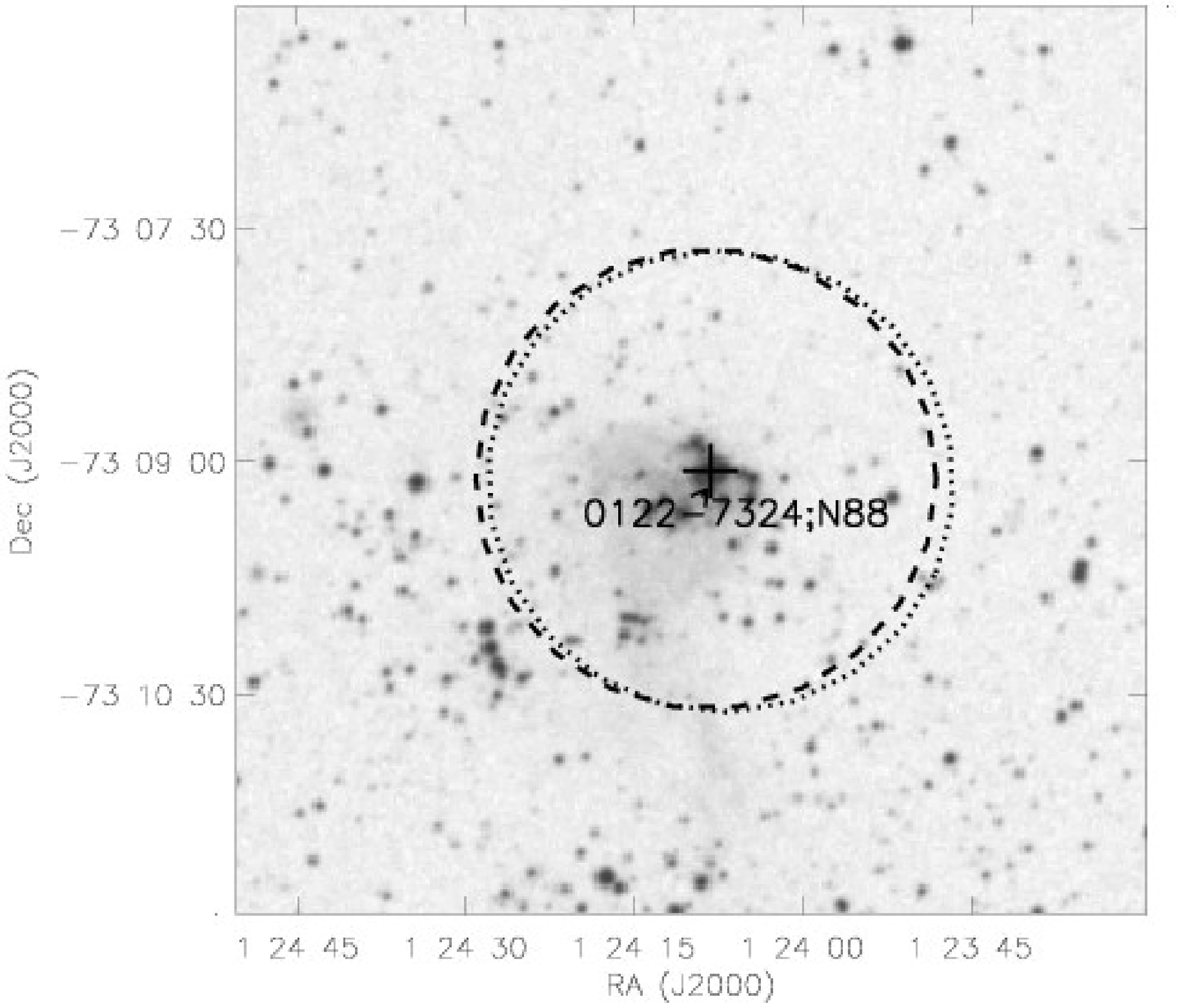}{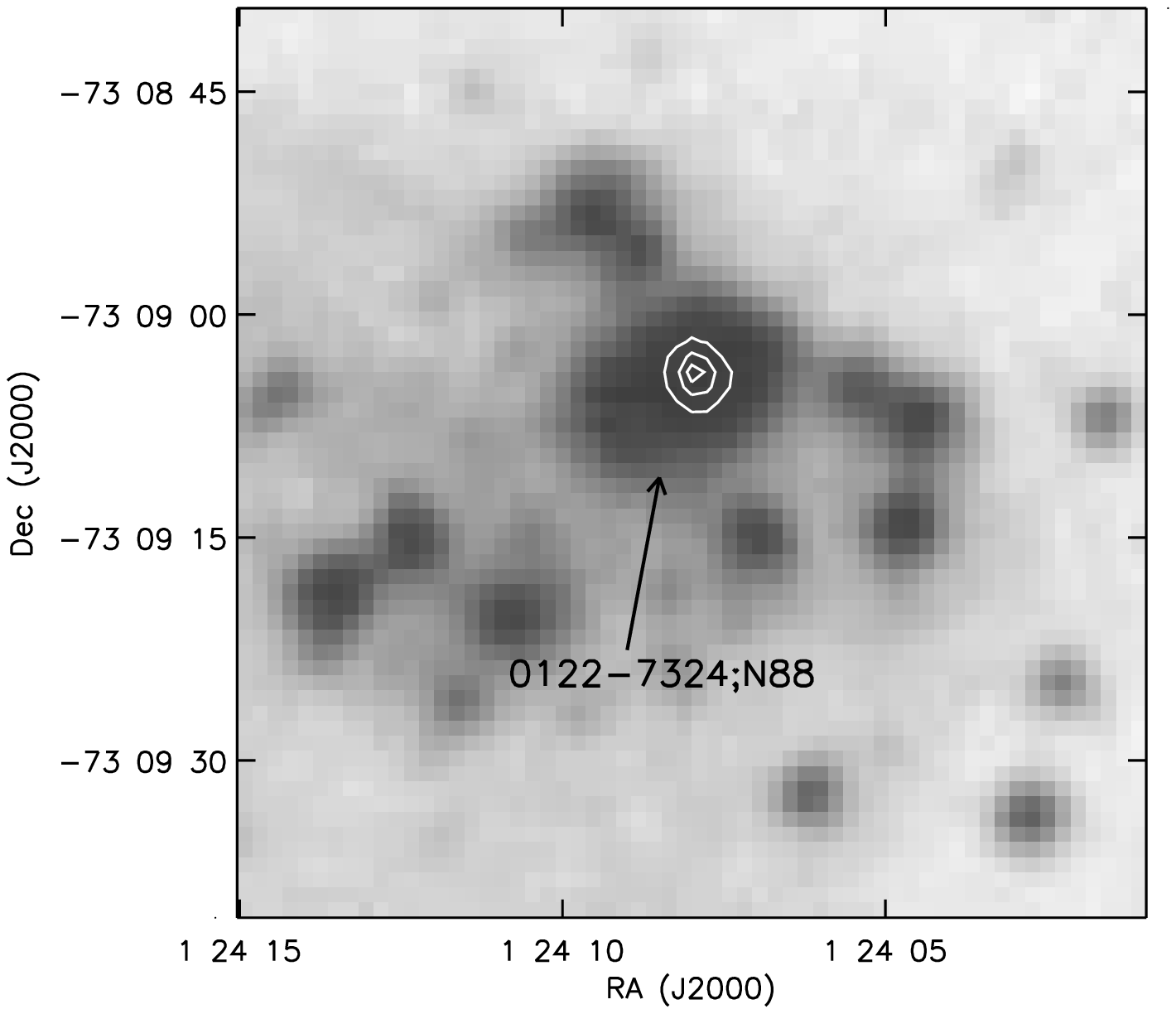}
\epsscale{0.5}
\plotone{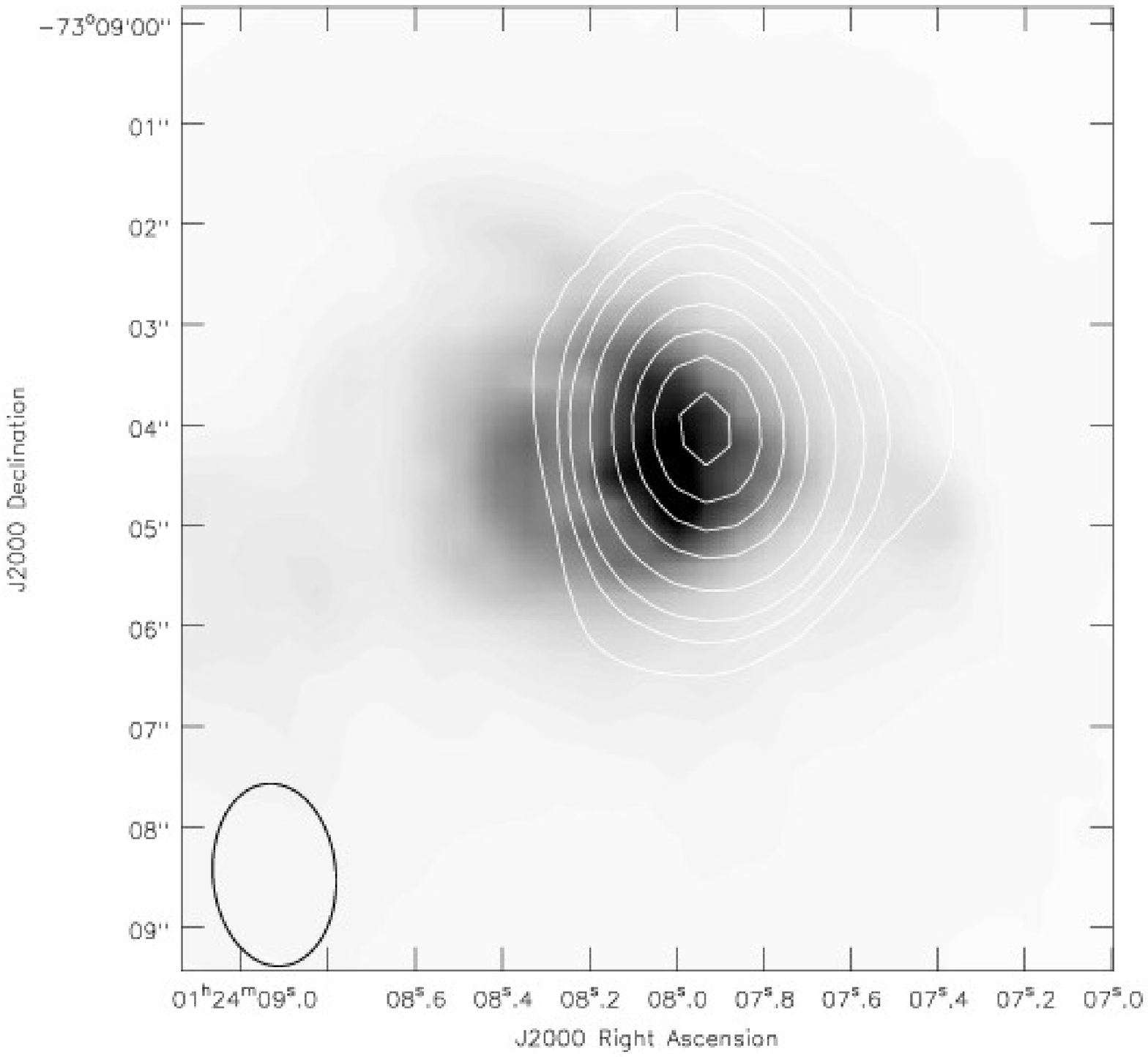}
\epsscale{1.0}
\caption{\label{n88fig} The first two panels show POSS-II R images of
the N88 \hii complex, with subregions labeled as in
Figures~\ref{n66fig} and \ref{n81fig}.  The third panel shows contours
of high-resolution 3~cm radio emission superimposed on an {\it HST}
H$\alpha$ image.}
\end{figure}

N81 (DEM~138, IC~1644) and N88 are bright \hii regions in the Shapley
Wing 1$\deg$.2 SE of the main bar of the SMC.  They have two of the
highest optical surface brightness and magnitudes among SMC \hii
regions, and may be located in neighboring and interacting \ion H1
clouds \citep[][and references therein]{heydari88-n81}.
Figures~\ref{n81fig} and \ref{n88fig} show the optical appearance of
the two regions, with compact radio sources, IRAS, and Parkes radio
sources marked.  Radio contours are also shown on zoomed-in optical
images.  The compactness and apparent simplicity of these two regions
make them good candidates for single-generation simple clusters.

The source 0107-7327 is clearly associated with N81.  The {\it HST}
observations of \citet{heydari-n81} indicate that N81 is powered by a
small group of newly born massive stars.  This radio source has also
been detected in the CO(1-0) observations of \citet{israel93}.  The
sources 0107-7327(N) and 0107-7327(W) are weak detections, but the
northern source could also be associated with N81.  Similarly,
0122-7324 is clearly associated with N88.

The third panels of Figures~\ref{n81fig} and \ref{n88fig} show
contours of 3~cm radio emission superimposed on H$\alpha$ images of
N81 and N88, respectively.  In both cases the radio and optical
morphologies agree fairly well.  The radio emission from N88A appears
to be offset somewhat to the west of the H$\alpha$ emission, although
the $\sim$1\arcsec\ pointing uncertainty of {\it HST} should be kept in mind
in interpreting this offset.  \citet{heydari-n88} describe the sharp
north-western boundary of H$\alpha$ emission as an ionization front,
but if the radio emission extends further in that direction, the lack
of H$\alpha$ could be an obscuration effect.

\citet{heydari-n81} require 1.4\up{49}~photons~s\ts{-1} of ionizing
radiation to account for the H$\beta$ flux of N81. They estimate the
brightest star which they observe in Str\"omgren $y$ is consistent
with an O6.5V, which would produce sufficient ionizing radiation, but
argue convincingly based on their images that the region is powered by
a small cluster and not a single star.
In N88A, \citet{heydari-n88} find that the average extinction is
$A_v=1.5$ and rises to values at least as high as 3.5, which is
unusual for low metallicity SMC \hii regions \citep[see
also][]{kurt99}.  Due to the heavy extinction, the exciting star (or
stars) is not detected.  Based on the H$\beta$ flux, they estimate a
Lyman continuum flux for this object of 2\up{49}~photons~s\ts{-1}
The Lyman continuum fluxes implied by our radio flux densities for the
central sources of N81 and the source in N88A are 1.3 and
3.3\up{49}~photons~s\ts{-1}, respectively.  The somewhat higher flux
required by the radio observations than the H$\beta$ in N88A is
consistent with the high extinction in the area, although the flat
radio spectral index indicates that the ionized gas it not itself at
remarkably high density.


\subsubsection{N12, N26, N33, \& N78}
Two sources were detected near \ion H2 region N12, but it is not clear
that either is directly related to the optical or infrared emission.
The source 0043-7321 may be associated with the \ion H2 region NGC~249
(N12B).  \citet{copetti89} estimates that NGC~249 has a Lyman
continuum flux of $\sim 3.2 \times 10^{49}$~s$^{-1}$, and the age of
the \ion H2 region is $< 2$~Myr.  The source 0043-7321(N) may be
associated with the \ion H1 shell [SSH97]106.  \citet{stavely97}
estimate that this \ion H1 shell has a radius of $\sim 4.'7$, and
expansion velocity of 9.2~km~s${-1}$, and an age of 5.2~Myr.  For each
of the regions N12, N26, N33, and N78, Figure~\ref{many_smc} shows a
POSS-II R image with important objects labeled following the
convention of Figures~\ref{n66fig}--\ref{n88fig}, and a zoomed-in
image showing our 6~cm radio contours.

The source 0046-7333 is nearest to the \ion H2 region N26, but amid a
high density of \ion H2 regions, including N20, N21, N23, N25, N26,
N28, and N12 = DEM~15.  This radio source is also near the dark
nebulae [H74]11--14 \citep{hodge74} and co-spatial with the molecular
cloud SMC~B2 \citep{rubio93}.  \citet{testor01} estimates the Lyman
continuum flux powering N26 to be $\sim 2.8 \times 10^{48}$.

The source 0047-7343 appears to be associated with N33 (Lin 138).
0047-7343(N) may be associated with N31 (DEM~44, Lin 120), which has
an optical extent of 4$\times$7~arcmin, so is broadly associated with
everything in this field.  Also nearby is the planetary nebula
candidate [JD2002]~7 \citep{jd2002}, but as that has a diameter of
only 7\arcsec, it is probably unrelated.
 
The sources 0103-7216 and 0103-7216(N) may be associated with the \hii
region N78 (DEM 126).  \citet{copetti89} estimates that DEM~126 has a
Lyman continuum flux of $\sim 5 \times 10^{49}$~s$^{-1}$, and the age
of the \ion H2 region is $3-4$~Myr.  The star cluster IC1624 is nearby
to the south but apparently unrelated.

\begin{figure}
\plottwo{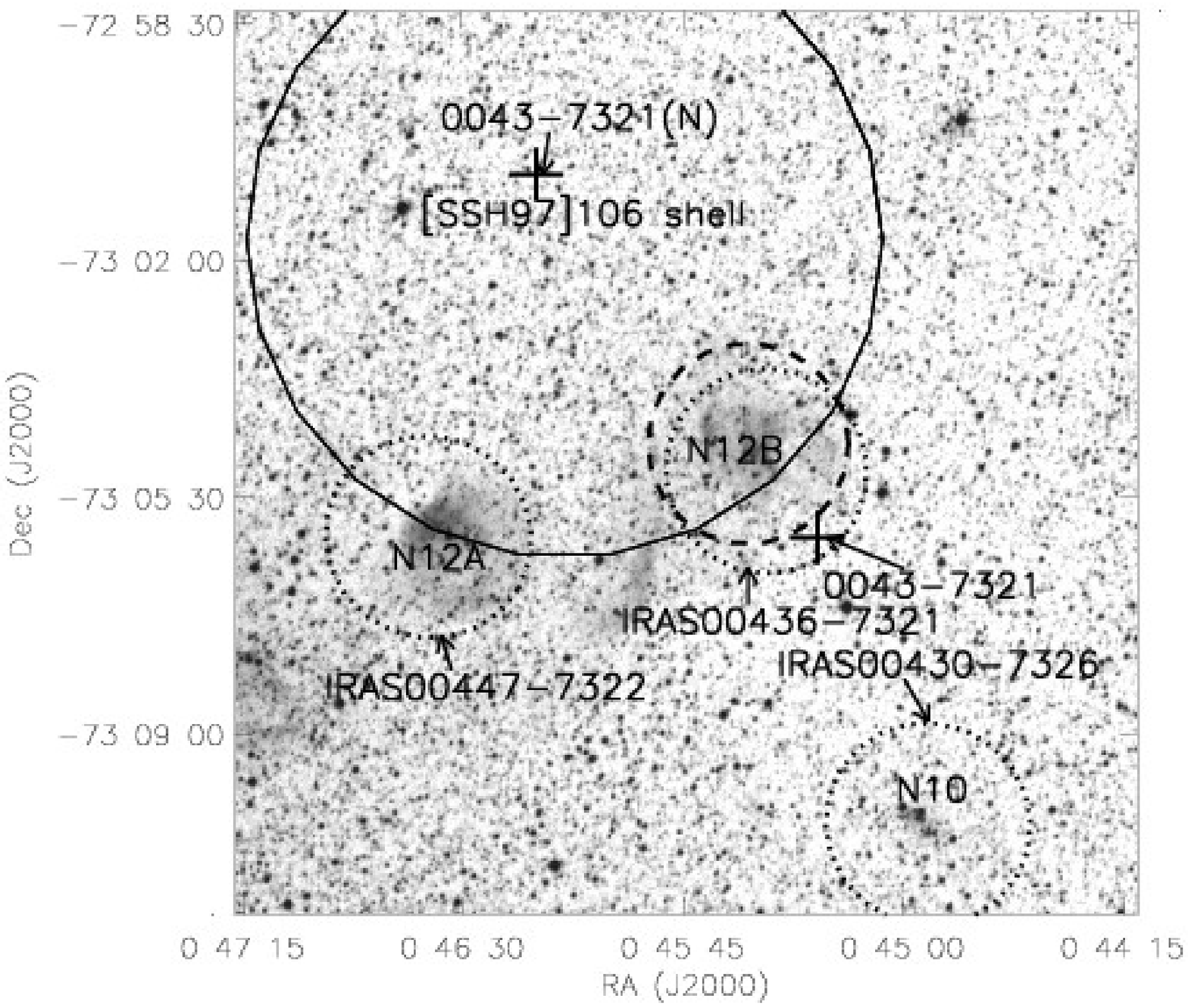}{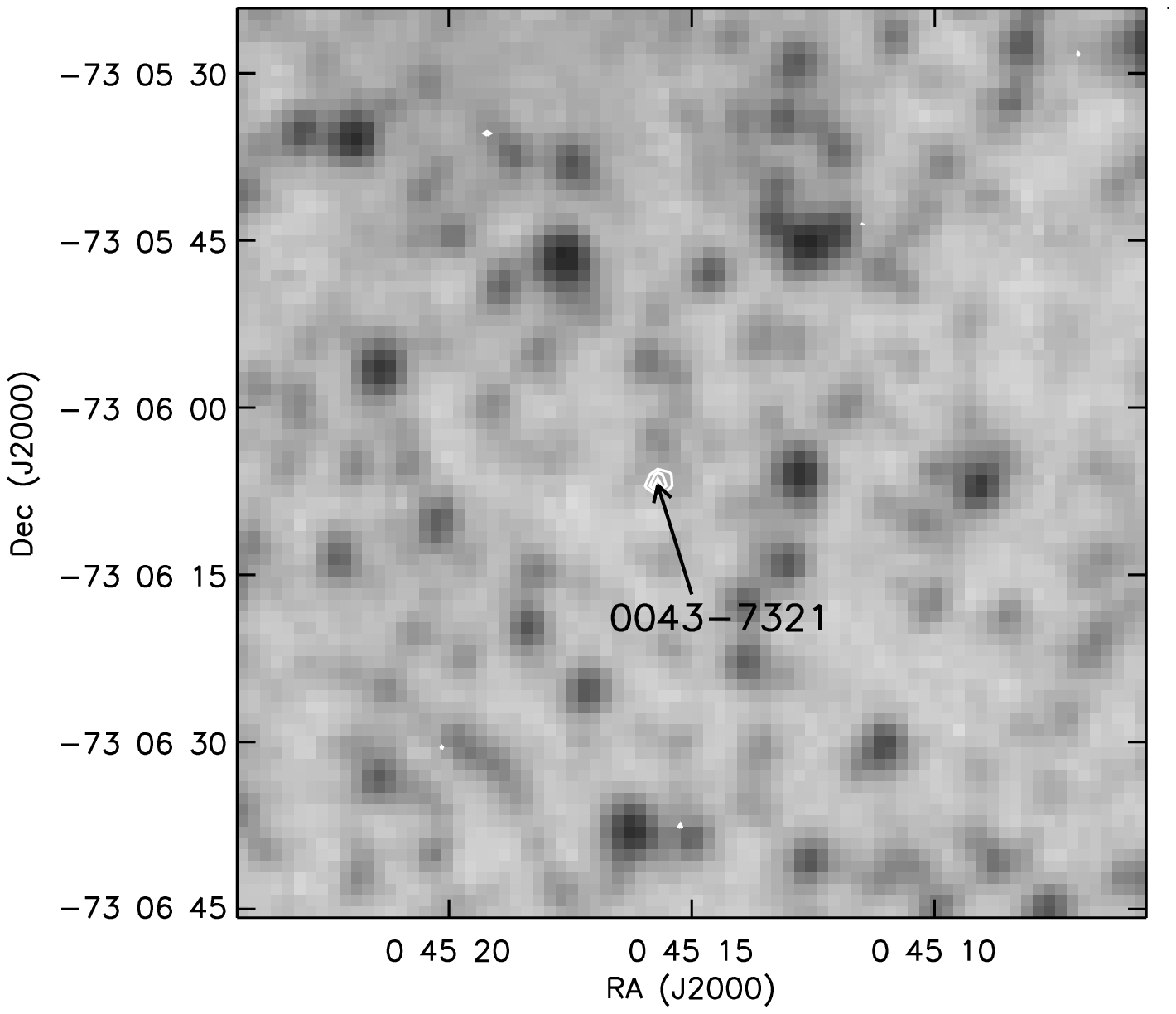}
\plottwo{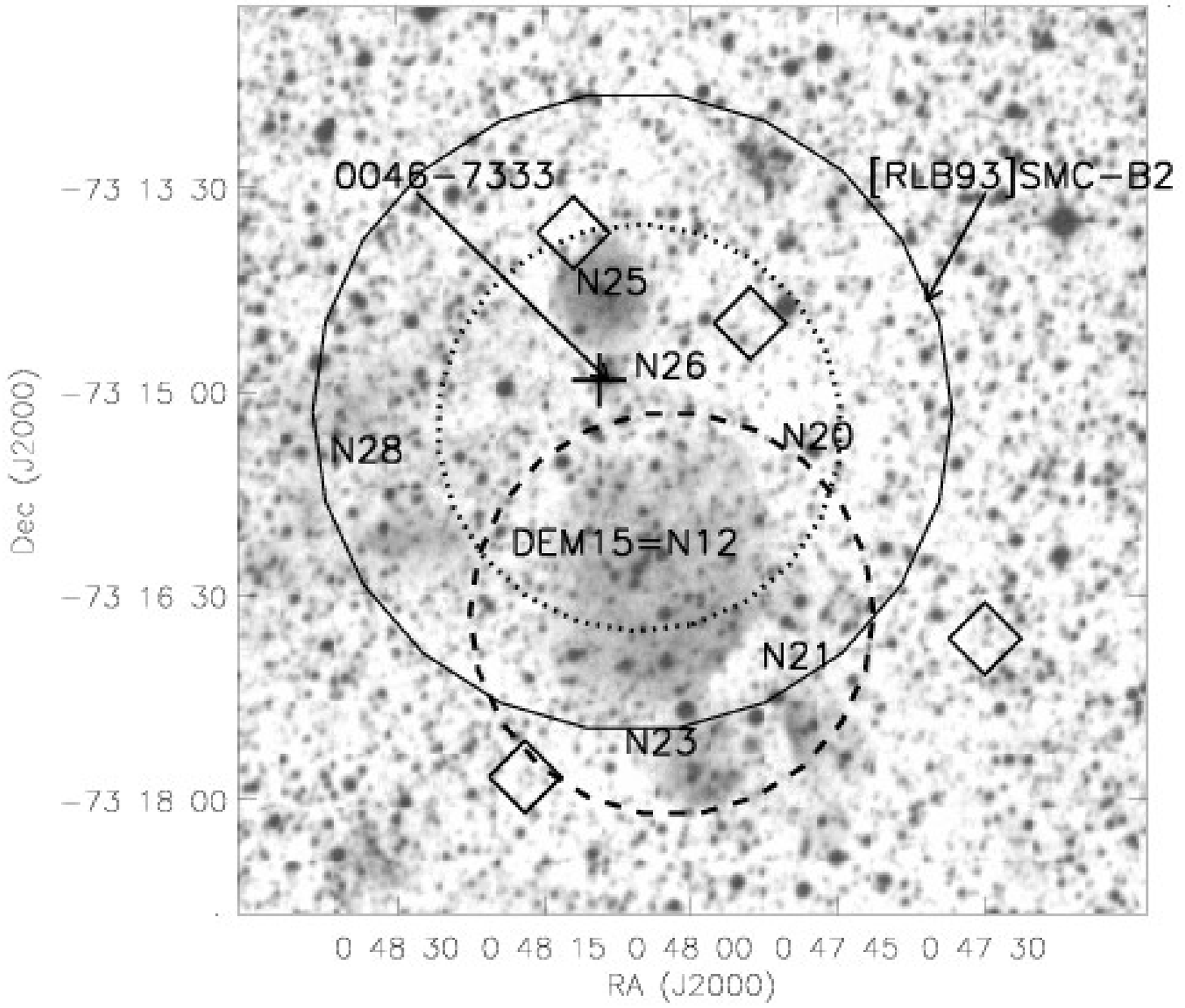}{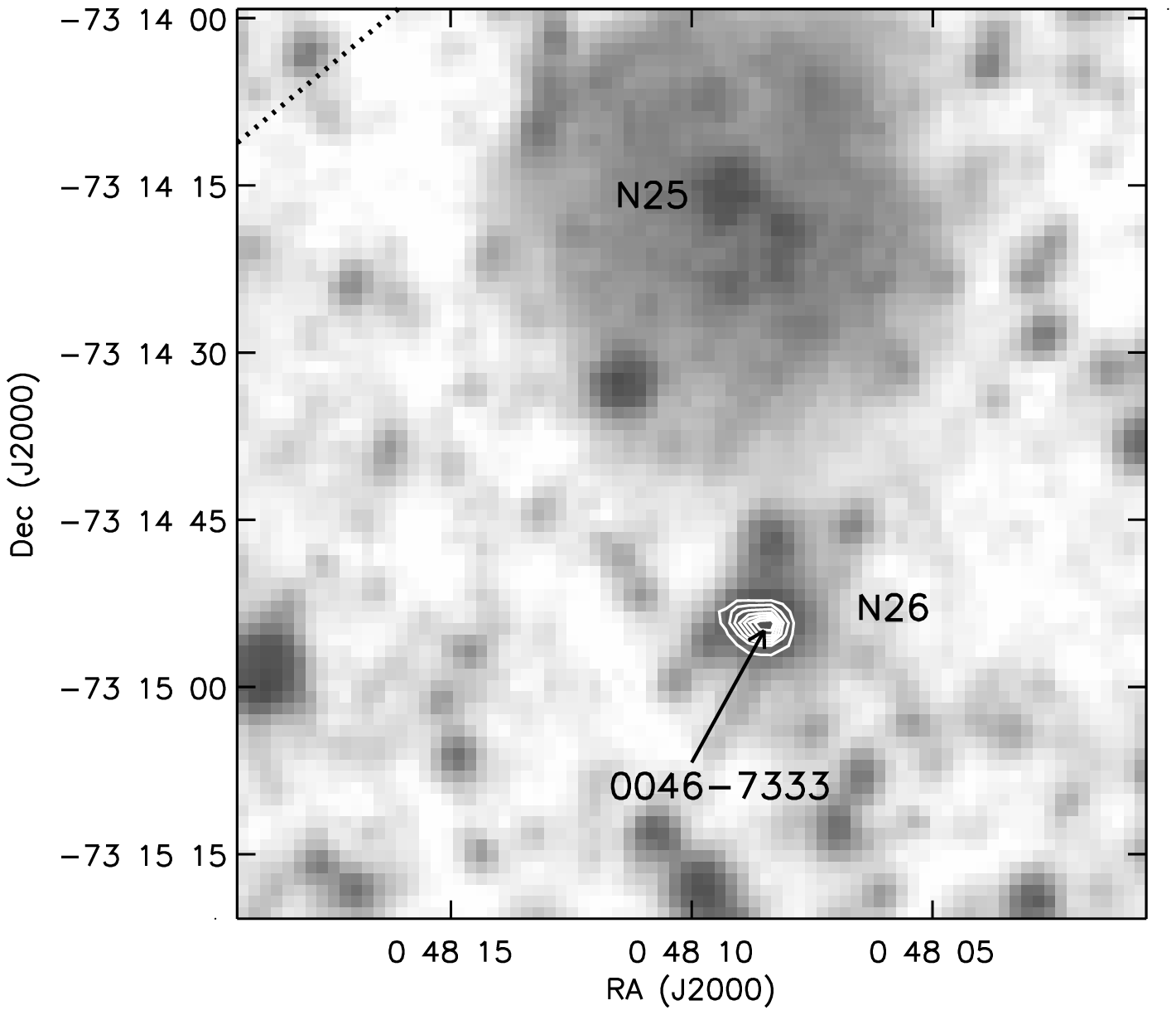}
\caption{\label{many_smc} Optical images of the N12, N26, N33, and N78
regions with compact sources from this dataset, previously known
objects, IRAS and Parkes radio sources marked.  Notation as
Figure~\ref{n81fig}.  In the first of the two panels showing N26, the
large diamonds denote the positions of the dark nebulae [H74]11--14
(numbers are in order of increasing R.A.), presumably associated with
dense gas.
}
\end{figure}

\addtocounter{figure}{-1}
\begin{figure}
\plottwo{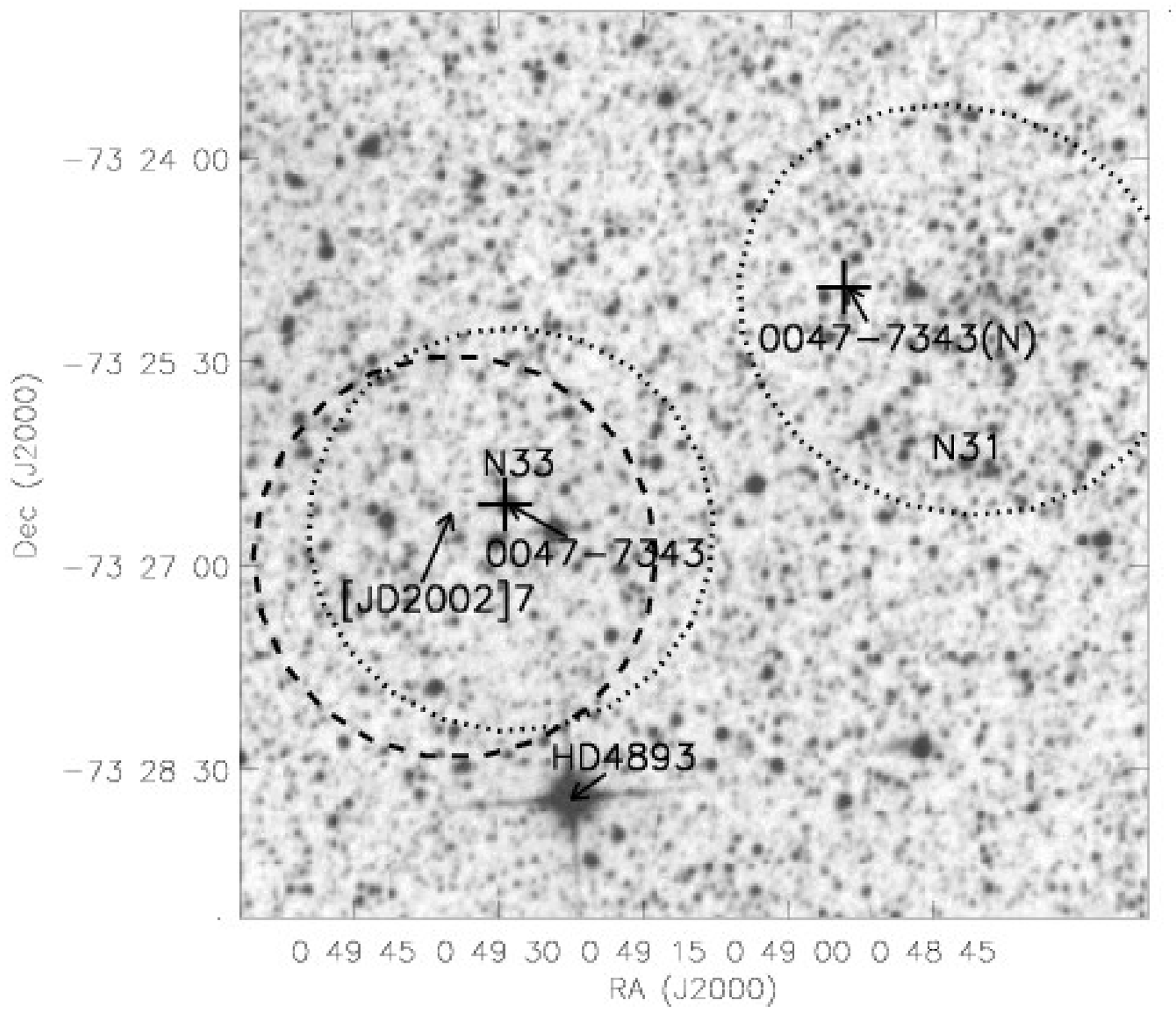}{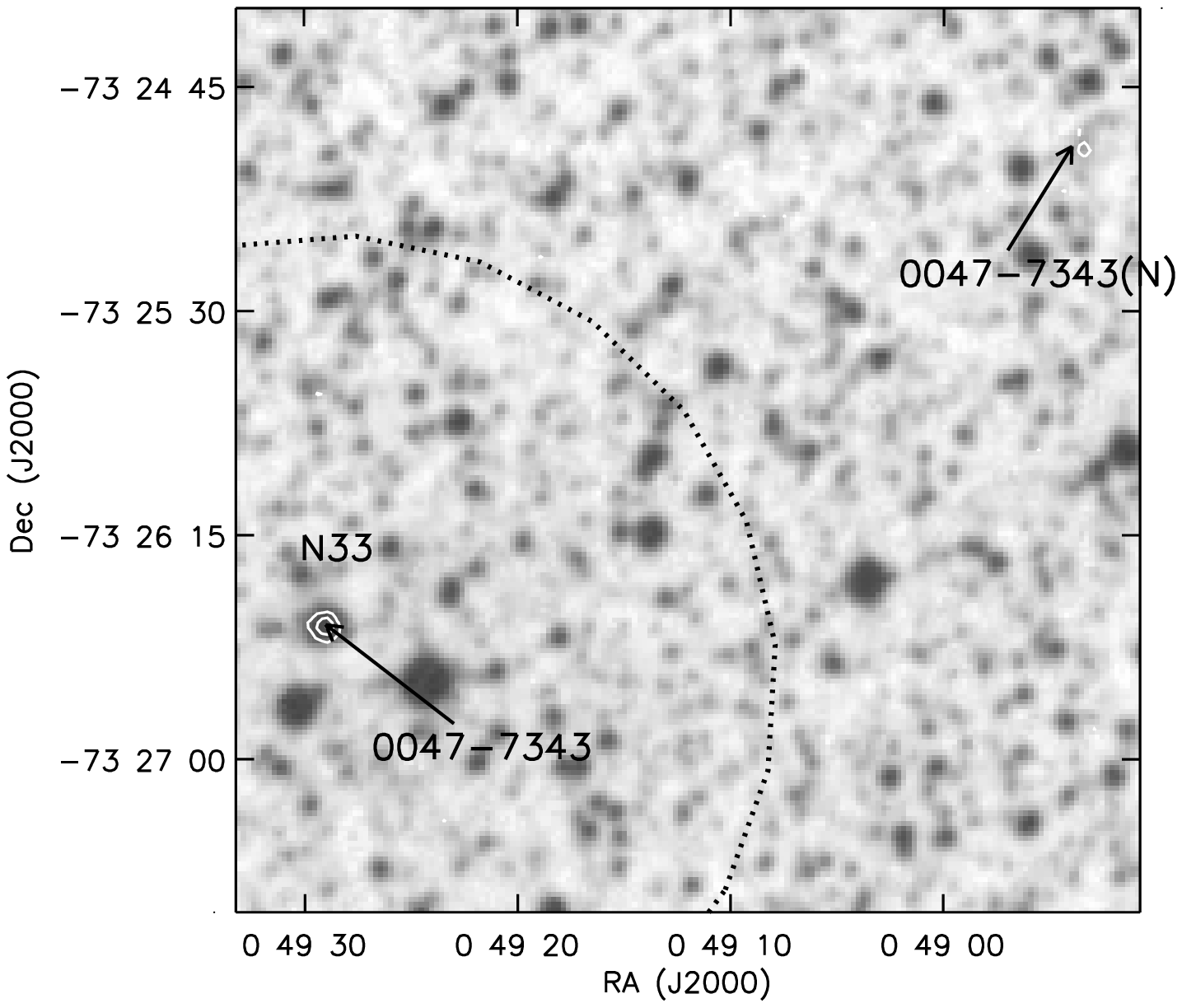}
\plottwo{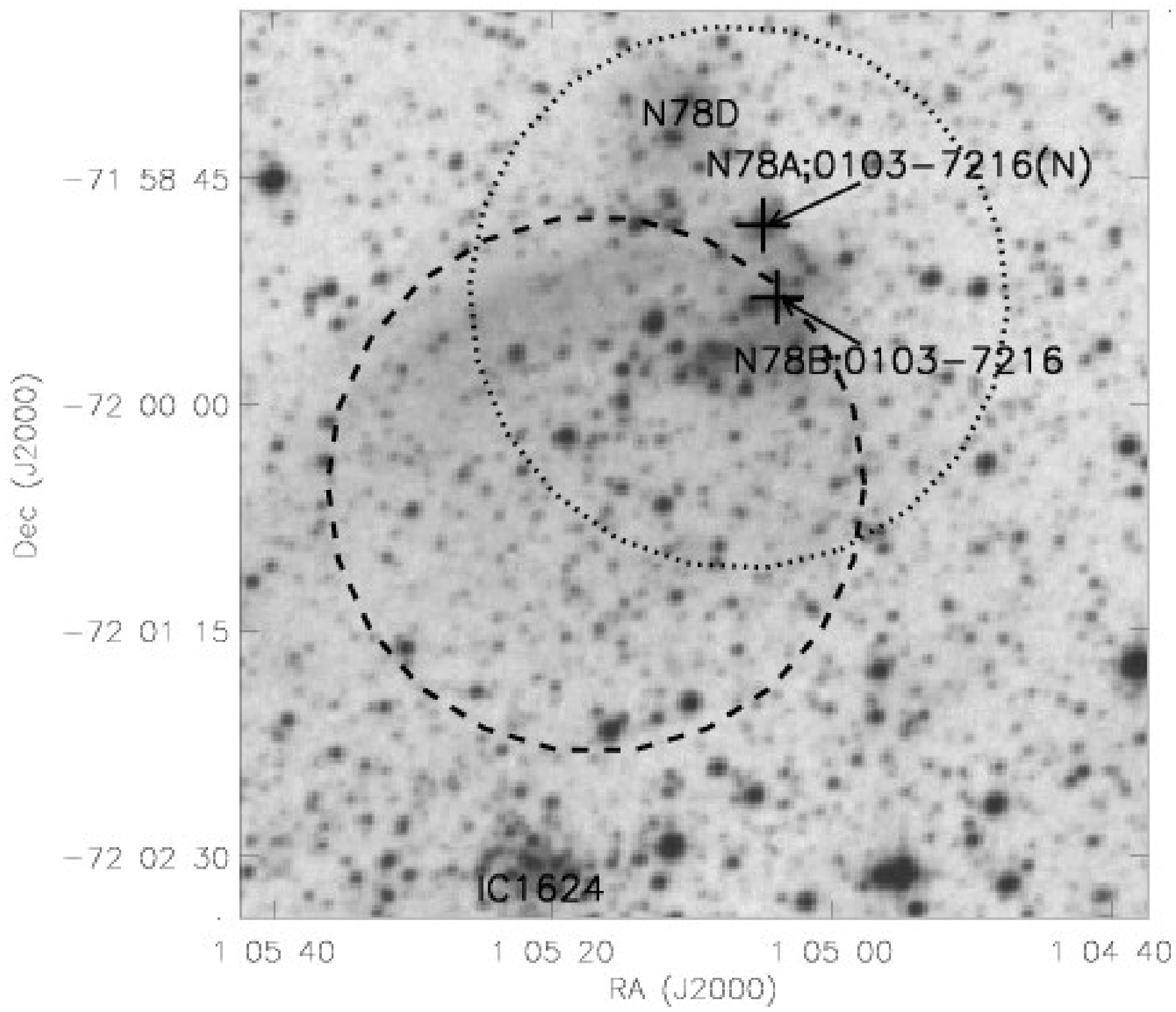}{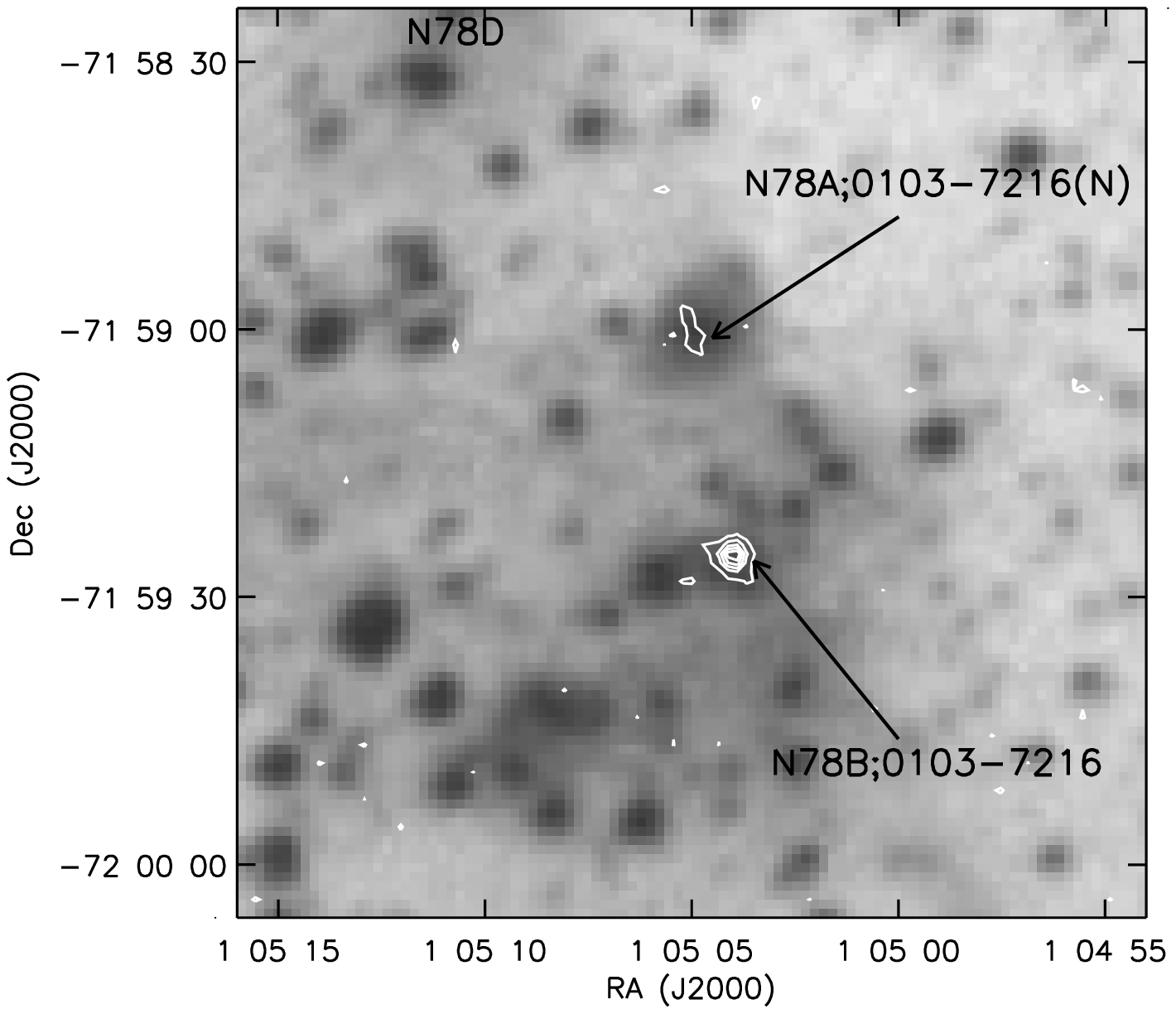}
\caption{continued.}
\end{figure}

\subsection{Large Magellanic Cloud}

\subsubsection{N159/N160}

\begin{figure}
\plotone{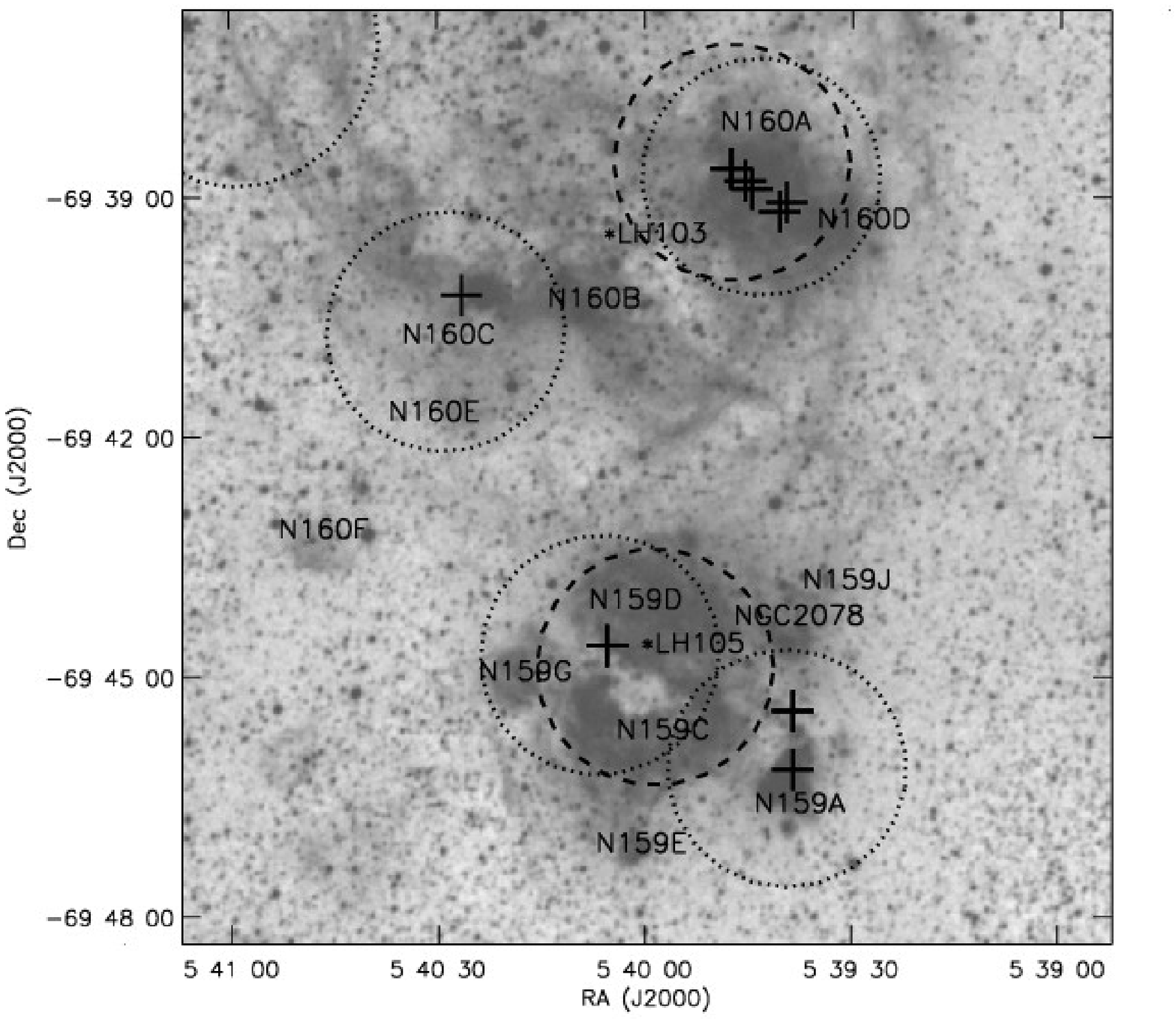}
\caption{\label{n160msx} 
POSS-II NIR image of the N159/N160 region, in which the main optical
\hii regions are labeled.  Parkes 3 and 6~cm sources \citep{filip98a}
are indicated by dashed lines, and IRAS point sources by dotted lines
(with the size of the circle indicating the spatial resolution of
those surveys).  The compact radio sources detected in this survey are
marked with crosses.}
\end{figure}

The N159/N160 \hii region complexes are located $\sim$40\arcmin\ to
the south of 30~Doradus, in what appears to be a region in which star
formation is propagating from the relatively evolved 30~Dor region out
into the more quiescent southern CO arm \citep[][and references
therein]{bolatto00}.  Figure~\ref{n160msx} shows a POSS-II NIR image
of the region, in which the main components are visible, and optical
\hii regions are labeled.  The star clusters LH103 and LH105 are
$\gtrsim$5\arcmin in diameter, and contain the entire N160 and N159
regions, respectively.  See \citep{bicacat} for a complete LMC object
catalog with sizes and positions.  Parkes 3 and 6~cm sources
\citep{filip98a} are indicated by dashed lines, and IRAS point sources
by dotted lines (with the size of the circle indicating the spatial
resolution of those surveys).  The compact radio sources detected in
this survey are marked with crosses.

\begin{figure}
\plotone{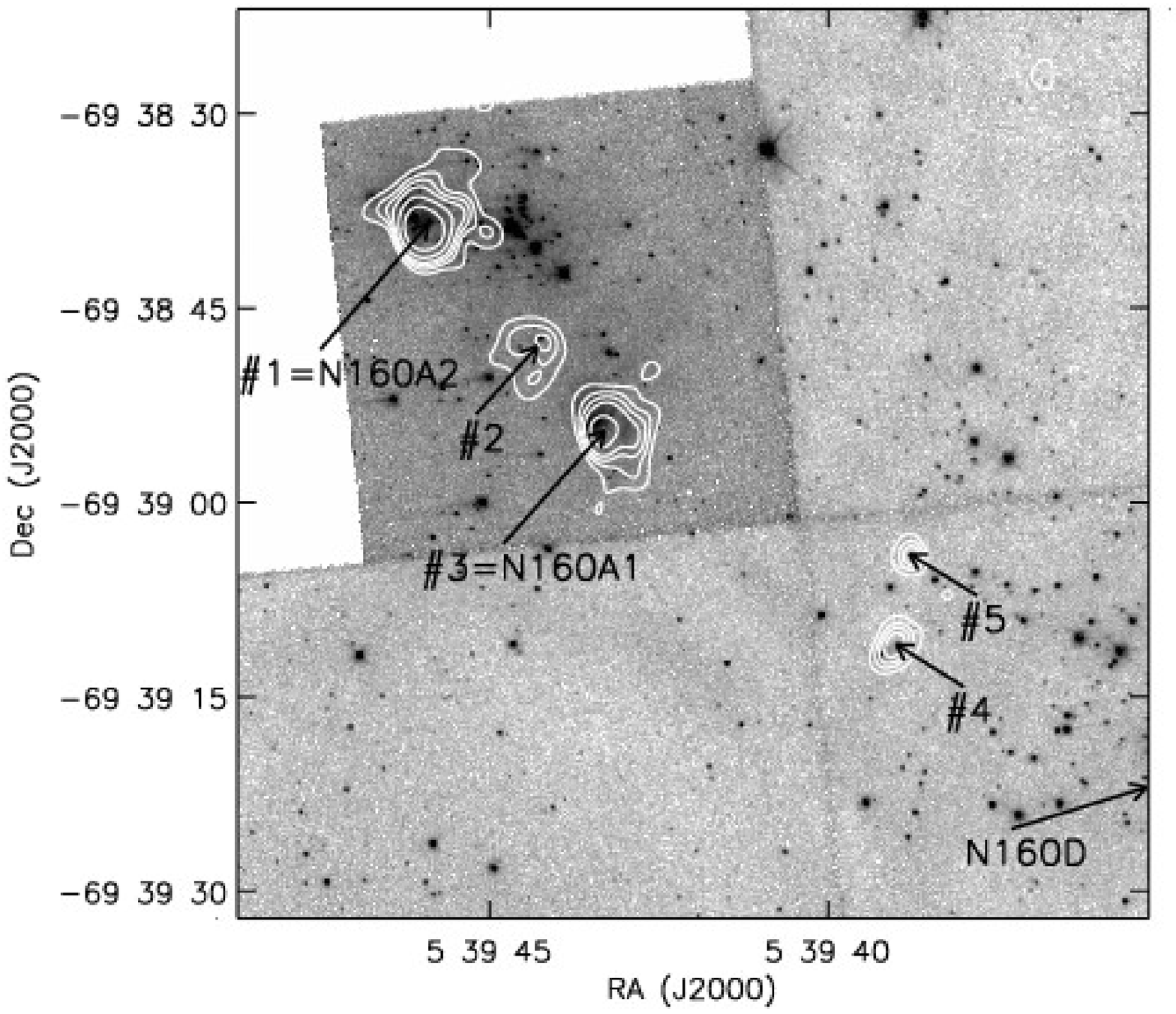}
\caption{\label{n160} N160, imaged at 6cm with the highest-resolution
configuration of the ATCA (contours) and in Str\"omgren $y$
\citep[greyscale from {\it HST},][]{heydari-n160a}.}
\end{figure}

In the more evolved N160 region, we detect 5 compact radio sources
near the optical \hii region N160A.  Figure~\ref{n160} shows our 6~cm
contours on a Str\"omgren $y$ image taken with {\it HST} by
\citet{heydari-n160a}. Our source \#3 is associated with their \hii
region N160A1, and they detect a single bright star (their \#22) in
the vicinity.  Similarly, our source \#1 is associated with what they
denote N160A2, in which they find a small cluster of stars.  The
authors derive that 8.5\up{48}~photons~s\ts{-1} and
1.2\up{48}~photons~s\ts{-1} are required to explain the observed
H$\beta$ fluxes of A1 and A2, respectively, and note that some of the
ionizing radiation could come from the cluster between the two, and
not just from the apparently associated stars.  The ionizing fluxes
required to explain our 3~cm flux densities are
1.4$\pm$0.2\up{49}~photons~s\ts{-1} and
3.4$\pm$0.2\up{49}~photons~s\ts{-1} for source \#3 and \#1 (A1 and
A2), respectively.  These are likely string lower limits to the
ionizing flux present in the region, since there is clearly an
extended halo of lower-density ionized gas around the two radio
sources which is not accounted for in our 3~cm synthesis image flux
density.  Each of our sources is also detected in the near-infrared by
2MASS -- the full infrared and multiwavelength properties of this
region are beyond the scope of this survey paper and will be discussed
in more detail in a future paper.

\begin{figure}
\plottwo{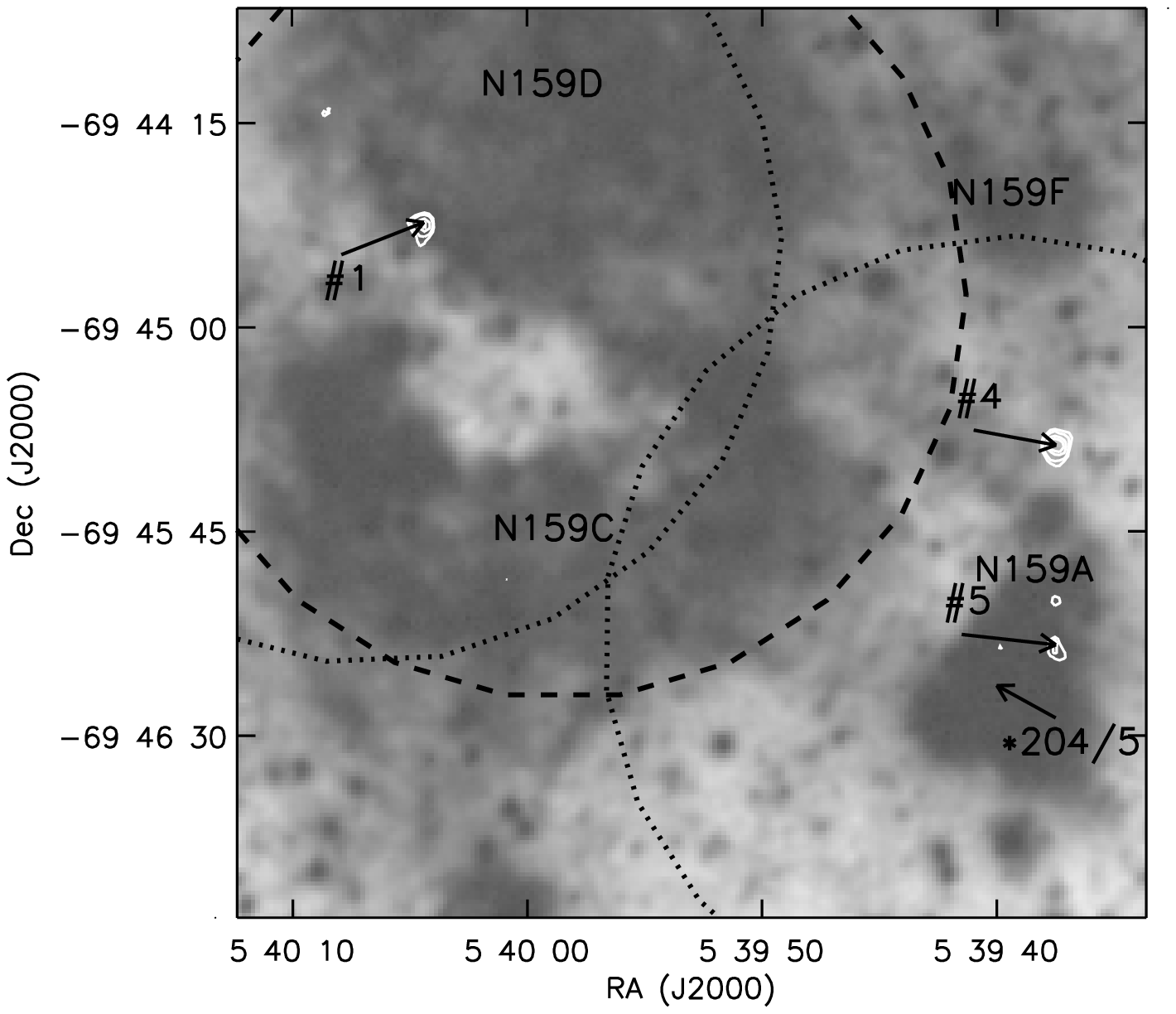}{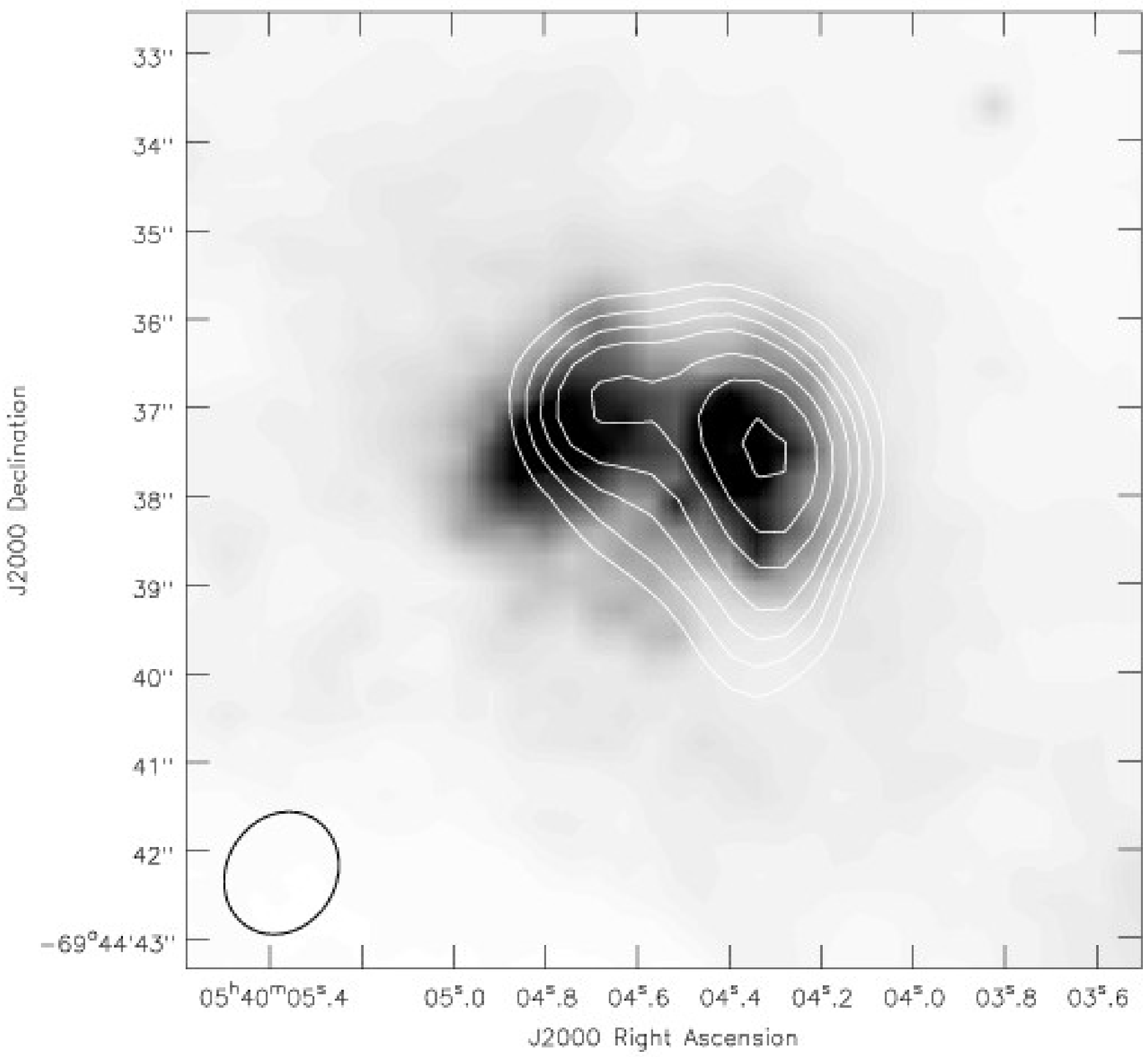}
\caption{\label{n159-5} The first panel shows a POSS-2 NIR gray-scale
image of the N159 region, with our 6~cm radio contours overlaid,
showing detections of \citet{huntwhit94} sources \#1, \#4, \#5.  Also
noted are optically identified \hii regions, and IRAS and Parkes
sources (large circles as in Fig.~\ref{n66fig}).  The position of
\citet{deharveng92} stars \#204 and \#205, which the authors say are
likely major ionizing sources in the region, are labeled ``*204/5''.
The second panel shows a close-up of source \#1, the ``Papillon''
nebula \citep{heydari-n159}.  3~cm contours (2, 3, 4, 5, 6.5, 8, and
10 mJy) are overlaid on a narrow-band H$\alpha$ ({\it HST} WFPC2 F656N)
image.  Since the \hii region was on the edge of our beam, radio
contours are from archival (C868) data.}
\end{figure}

The \hii region N159 lies to the south; Figure~\ref{n159-5} shows a
zoomed-in POSS-2 NIR image with our 6~cm contours and important
previously identified objects labeled.  We detect the compact sources
\#1, \#4, and \#5 imaged by \citet{huntwhit94} using the ATCA at lower
resolution ($\sim$4$\times$ shorter longest baseline, their Figure~1).
Their sources \#2 and \#3 are clearly extended even at their lower
resolution, so it is not surprising that there are no compact sources
detected at those locations.  \citet{huntwhit94} quote peak
intensities at 6~cm for components 1, 4, and 5 as 66, 115, and 58~mJy,
respectively.  We detect compact sources of 70, 120, and 30~mJy
(integrated, though since the sources are at best marginally resolved,
the peak flux density/beam is very similar).  These numbers indicate
that sources \#1 and \#4 are consistent with unresolved point sources,
but that \#5 is more diffuse (a fact that is fairly evident in their
radio map).

Figure~\ref{n159-5} also compares the radio continuum and H$\alpha$
emission from source \#1.  The reader should take careful note of
nomenclature, as this object is referred to as N159-5 in
\citet{heydari-n159} and \citet{heydari-papillonnew}, N159 Blob in
\citet{heydari-testor}, and source \#1 in \citet{huntwhit94}.  The
H$\alpha$ emission imaged at high resolution with {\it HST}
\citep{heydari-n159} shows two lobes, and the new high-resolution
radio imaging reveals that most of the radio continuum is associated
with the western lobe.  \citet{heydari-n159} quote an ionizing flux of
4$\times$10\ts{48}~photons~s\ts{-1} if the region is
ionization-bounded, but do not postulate any ionizing sources.  Our
3~cm flux density corresponds to an ionizing flux of
9$\times$10\ts{48}~photons~s\ts{-1}, which could be provided by a
single O7-8V star, or more likely a small cluster.  A radio-determined
ionizing flux will usually exceed an optically-determined one because
of extinction, and both are lower limits if ionizing radiation is
escaping the \hii region or absorbed by dust. \citet{deharveng92}
account for the ionization of the region with the 5 massive stars
which they observe in the region, in particular their stars \#204
(O5-6V) and \#205 (O7-8V), but these stars are at least 10\arcsec\ to
the SE of the brightest H$\alpha$ and radio region.

\subsubsection{N11}

\begin{figure}
\plotone{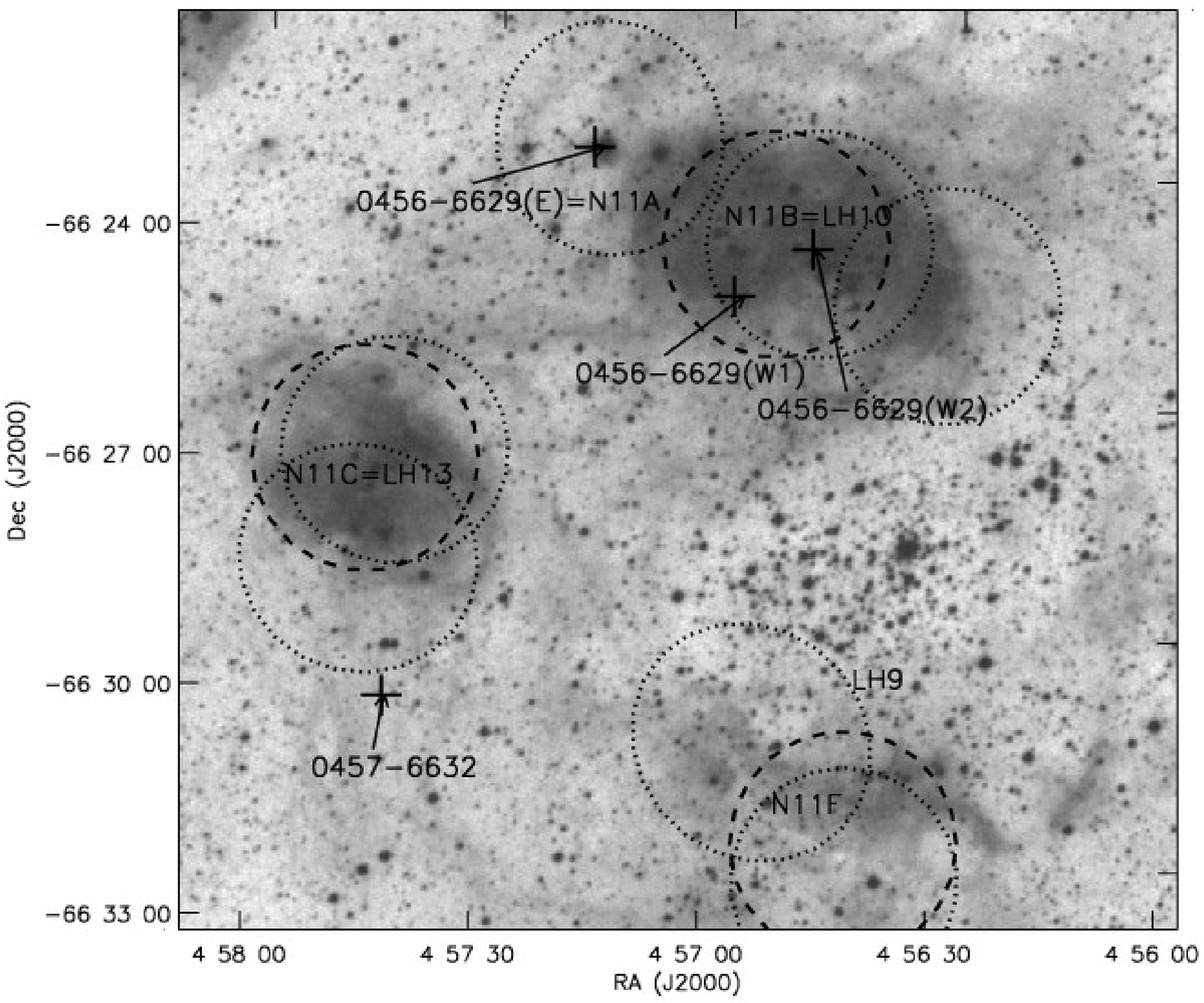}
\caption{\label{n11fig} POSS-II R image of the N11 \hii complex, with
optically identified subregions labeled.  Parkes 3 and 6~cm sources
\citep{filip98a} are indicated by dashed lines, and IRAS point sources
by dotted lines (with the size of the circle indicating the spatial
resolution of those surveys). The compact radio sources detected in
this survey are marked with crosses.}
\end{figure}

The giant \hii complex N11 has the second brightest H$\alpha$
luminosity in the LMC, after 30~Dor \citep{kh86}.  The complex
consists of numerous \hii regions roughly on the edge of a ring or
bubble \citep[][provide a good overview of the region and label the
optical subcomponents in their Figure~1]{rosado96}.  The morphology
probably results from (at least) two generations of star formation,
the older \citep[$\sim$5\up{6}~yr][]{walborn92} being the central OB
association LH9, with younger associations including LH10 (1--2\up{6}
years old) on the periphery.  Figure~\ref{n11fig} shows the region
with optical \hii regions labeled, as well as Filipovic and our radio
sources and IRAS point sources.  The older cluster LH9 is clearly
visible in the center of the ring of \hii regions delineated by
N11A,B,C,F, and H (just off the image to the west).

\begin{figure}
\plottwo{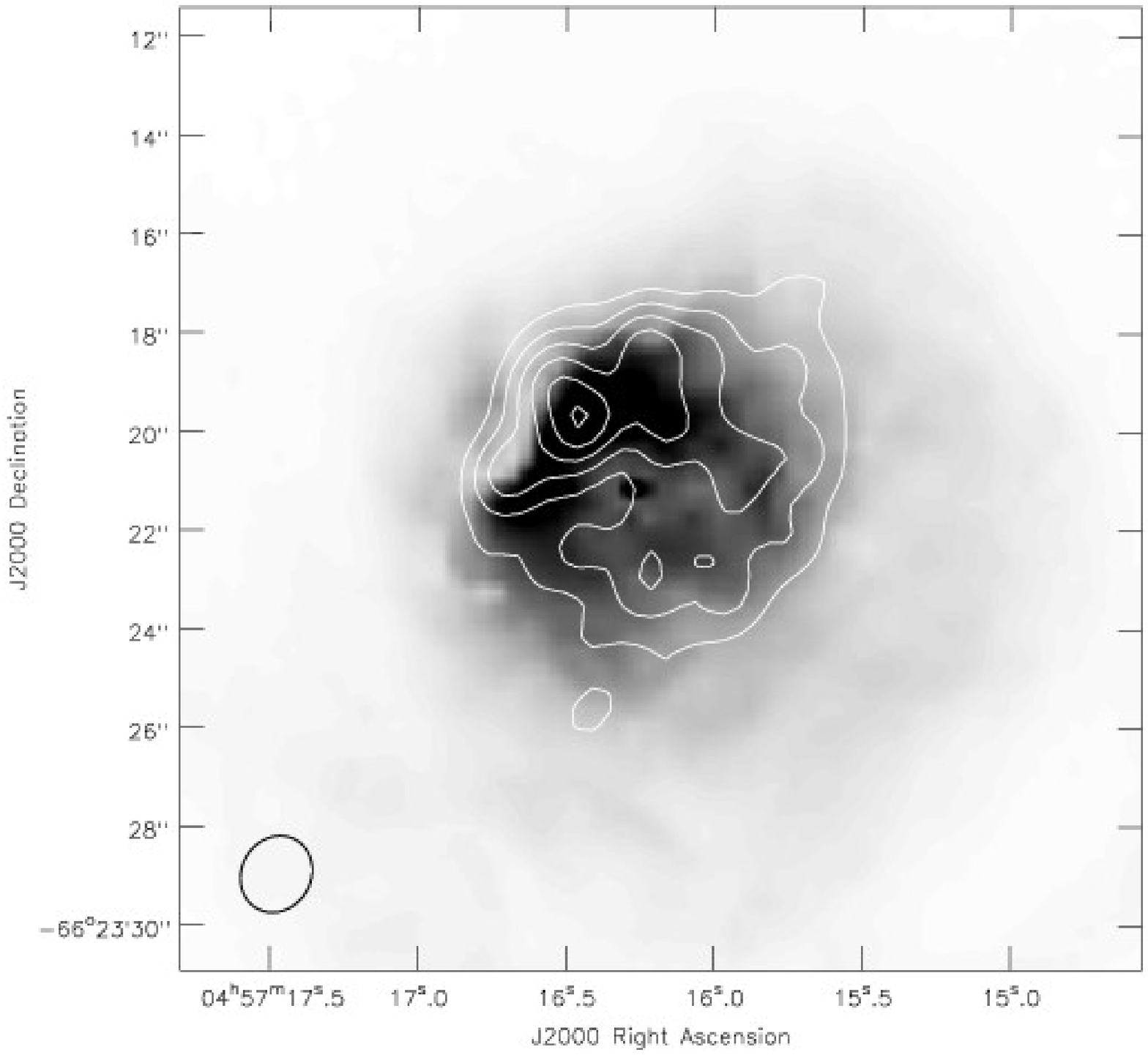}{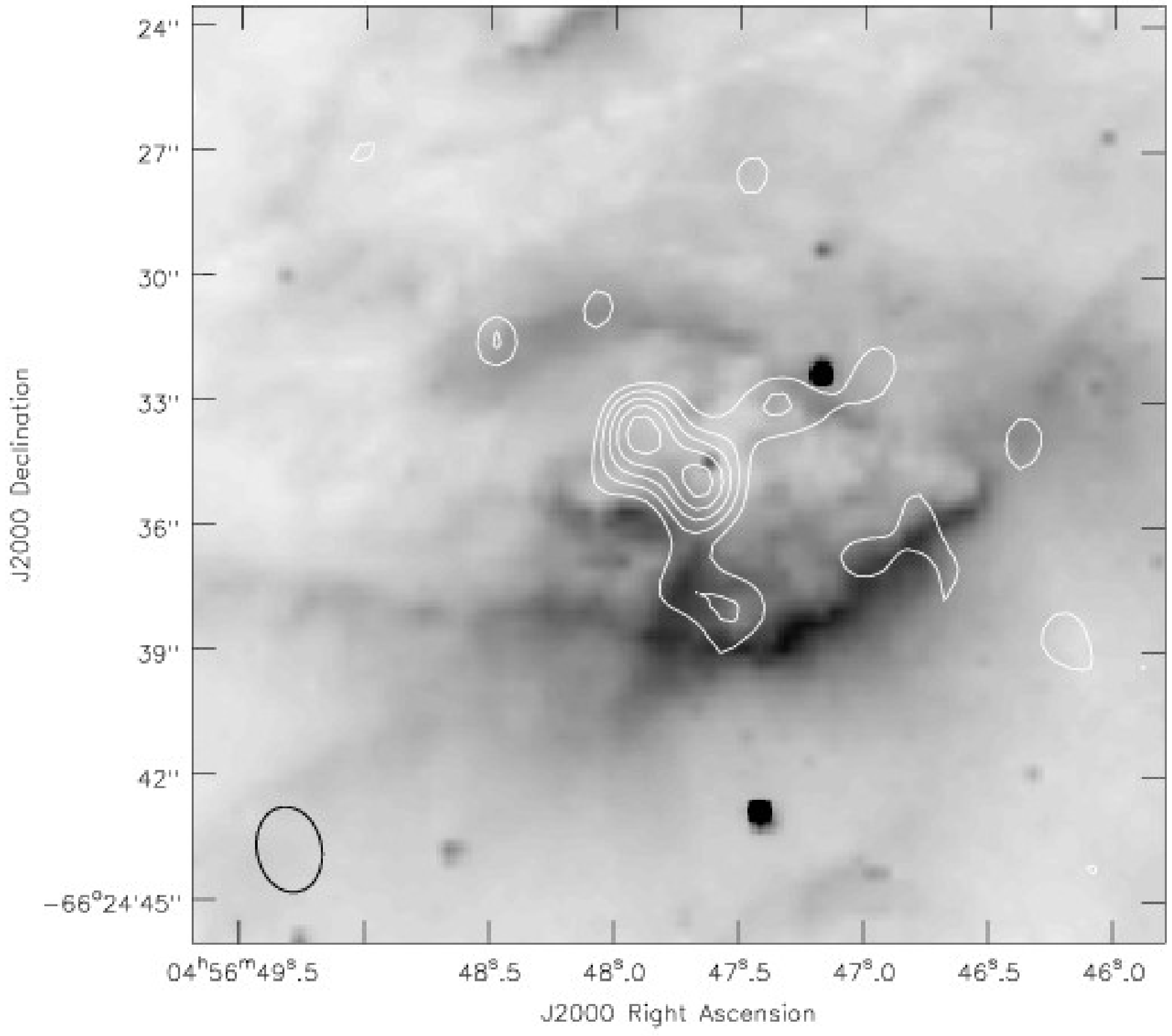}
\caption{\label{n11ab} The first panel shows 3cm contours (0.7, 1.2,
1.6, 2.1, 2.5 mJy) of N11A, and the second panel shows 6cm contours
(0.6, 0.8, 1.0, 1.2, 1.4 mJy) of N11B.  Both are overlaid on
narrow-band H$\alpha$ ({\it HST} WFPC2 F656N) images.  Since the \ion H2
region was on the edge of our 3~cm beam, the 3~cm contours of N11A are
from archival (C868) data. }
\end{figure}

We detect compact sources in the two highest excitation regions A and
B \citep[``excitation'' refers to the ratio of
{[\ion{O}{3}]}$\lambda$5007/H$\beta$ and is probably related to the
hardness of the powering radiation field:][and references
therein]{heydari-n11a}.  Figure~\ref{n11ab} shows 3~cm high-resolution
radio contours of 0456-6629(E) superimposed on {\it HST} H$\alpha$ of N11A.
The radio and H$\alpha$ morphologies of N11A agree very well,
supporting the interpretation that the \hii region is less extended
from the ionizing stars to the north-east because of the presence of a
dense molecular cloud \citep{heydari-n11a}.  The H$\alpha$ morphology
could have been highly affected by extincting material, but the
unextincted radio emission confirms the lack of ionized gas to the NE.
The second panel of Figure~\ref{n11ab} shows the western of two
sources detected in N11B, 0456-6629(W2).  The source is in the region
of brightest H$\alpha$ emission, on the northeast side of what appears
to be the powering star cluster.  Interestingly, the radio continuum
source is located in the middle of ring-like H$\alpha$ emission.  The
complex H$\alpha$ and radio morphology indicates that the nearby
material is probably inhomogeneous and clumpy, leading not only to
variable extinction of the optical light but an irregularly-shaped
\hii region.
A second source, 0456-6629(W1) was detected to the east of (W2), but
still in N11B.  The radio source is point-like or unresolved, and is
located very close to stars \#3189 and \#3193 of the extensive study
of \citet{parker92}.  The authors unfortunately do not have spectra of
these two stars, but they are fairly blue (U-B = -0.9).
 
\subsubsection{N79, N4, N5, N83, N23, N105, N119, N44, N138, N206, N55, N213, N158, N214, \& N168}

\begin{figure}
\plottwo{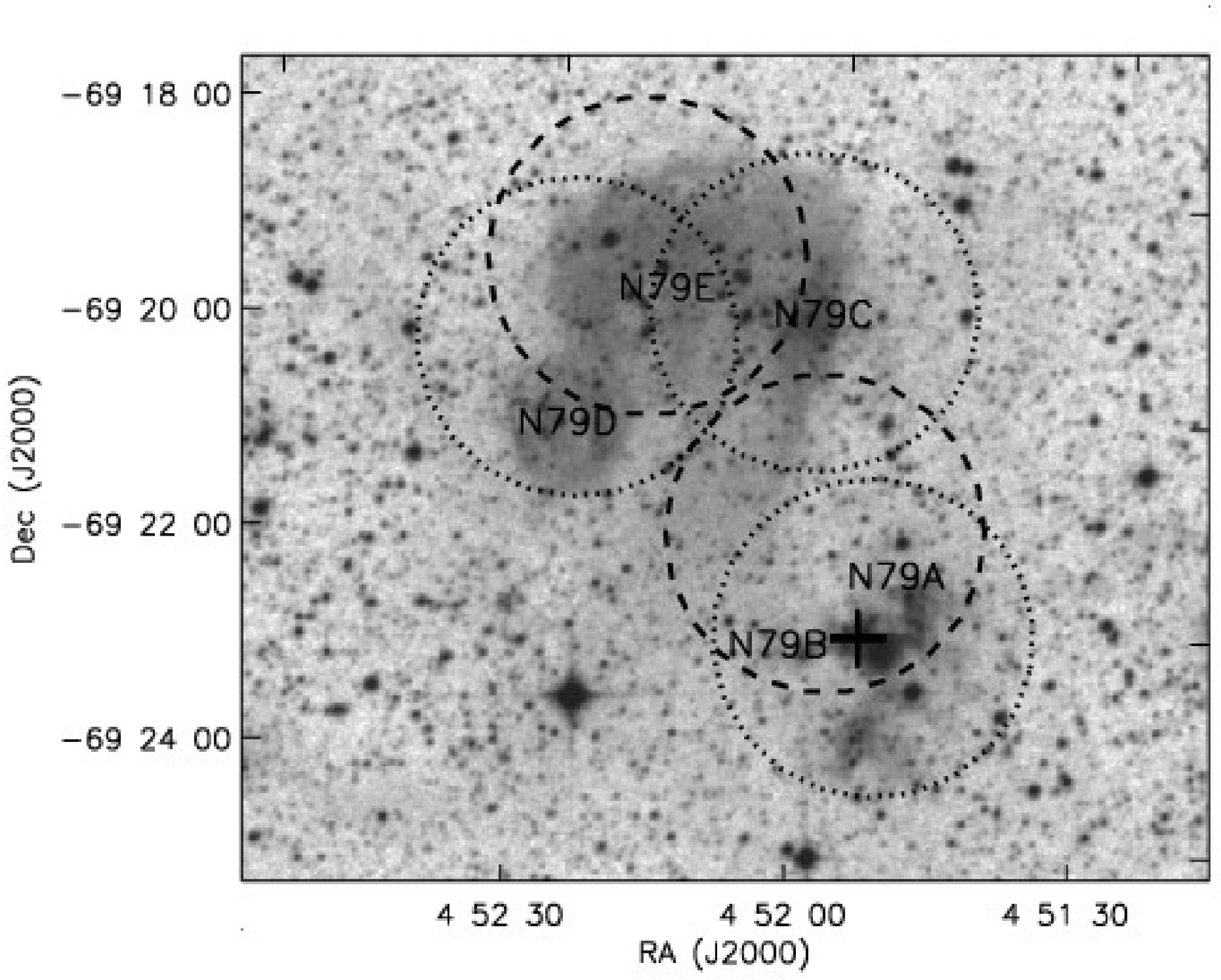}{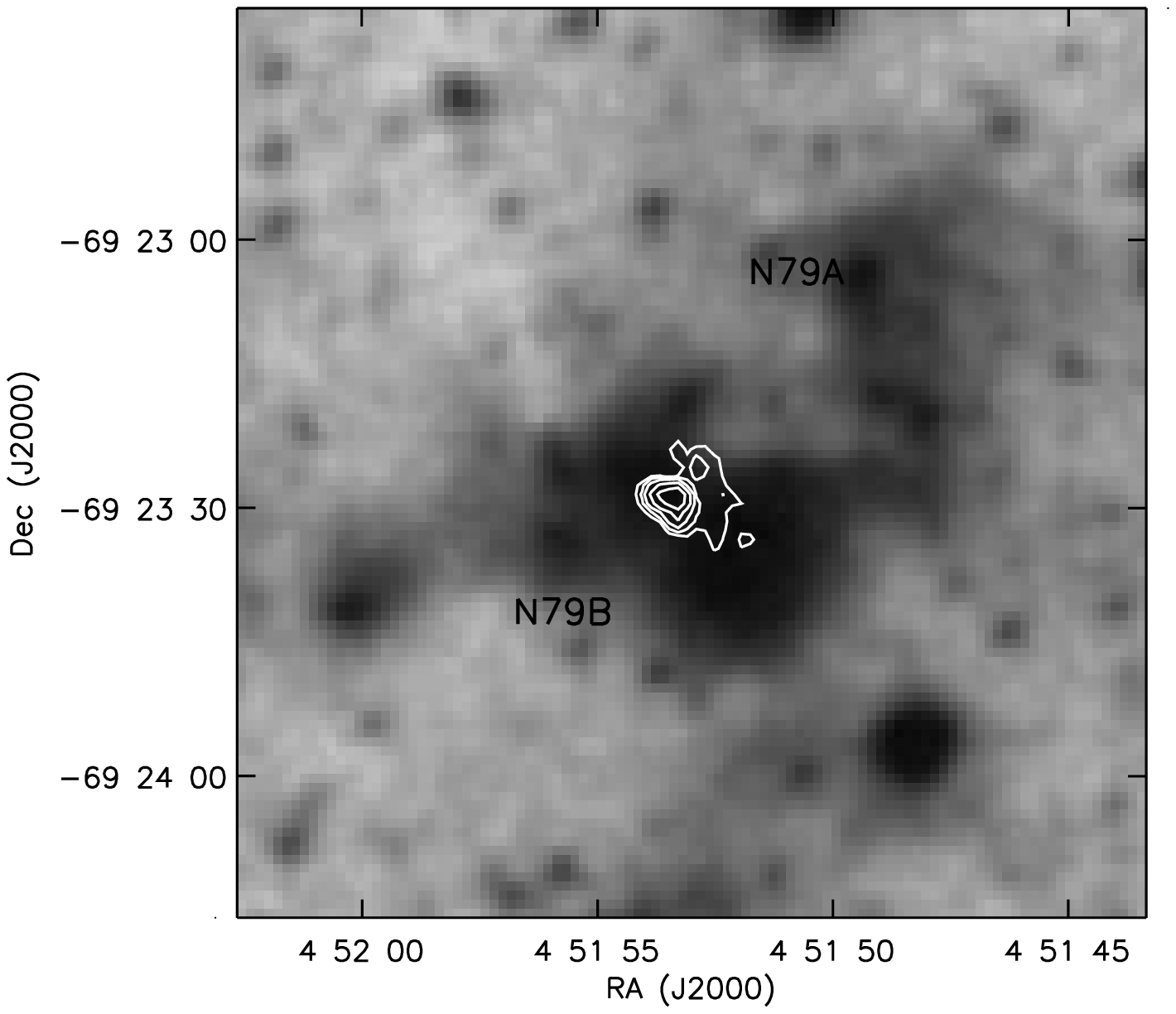}
\caption{\label{n79} Compact sources related to Parkes source
B0452-6927 and optical \hii region N79 in the LMC.  In this and the
following figures, each panel is a gray-scale POSS-II R image.  In the
first of each pair of panels corresponding to a particular region, the
locations of the radio sources detected by this study are marked with
crosses, as well as important optical nebulae, Parkes radio sources
(dashed lines) and IRAS sources (dotted lines), following the
convention of previous figures.  The second panel of each pair is a
zoomed-in version showing our 6~cm radio contours. }
\end{figure}

\begin{figure}
\plottwo{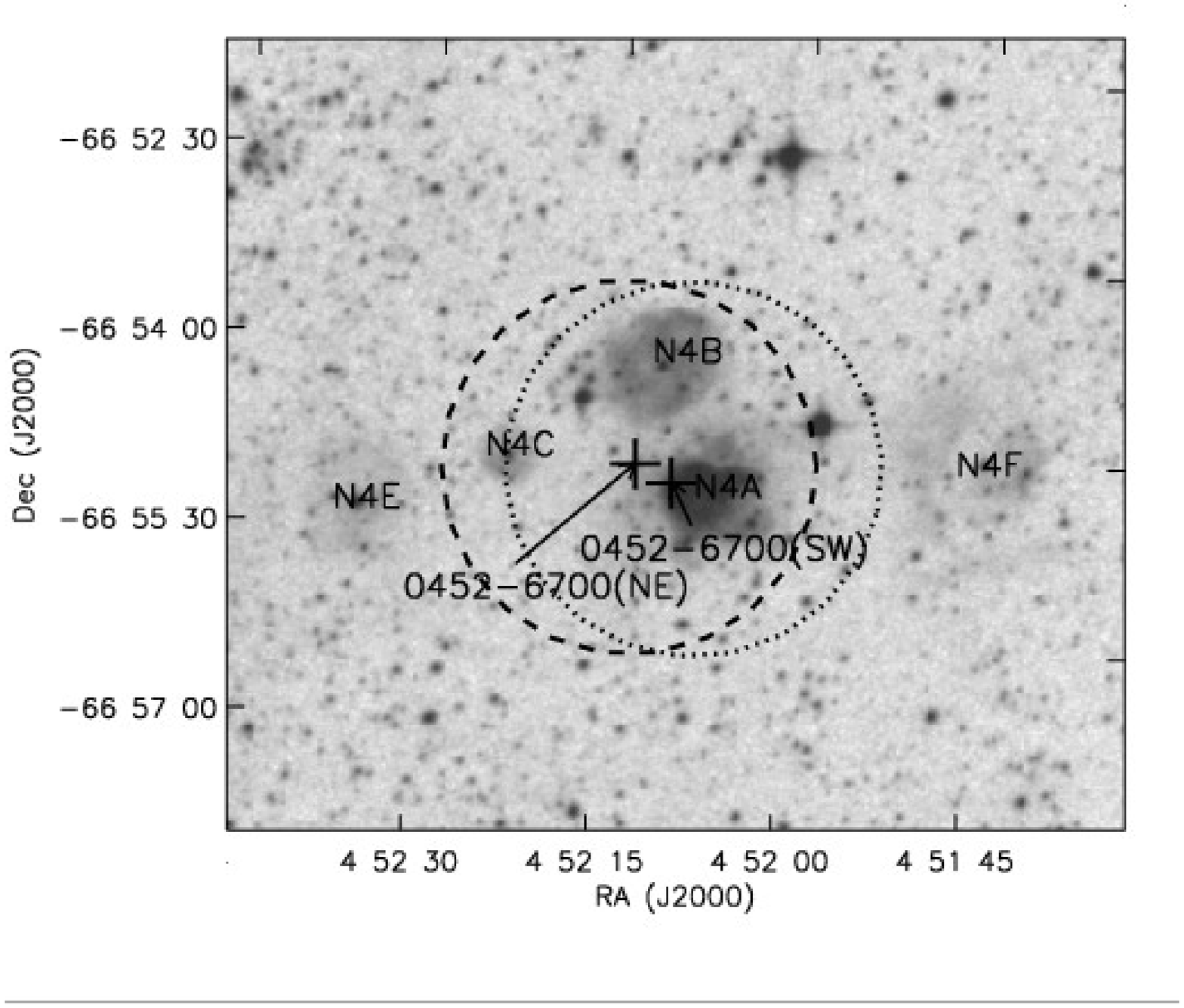}{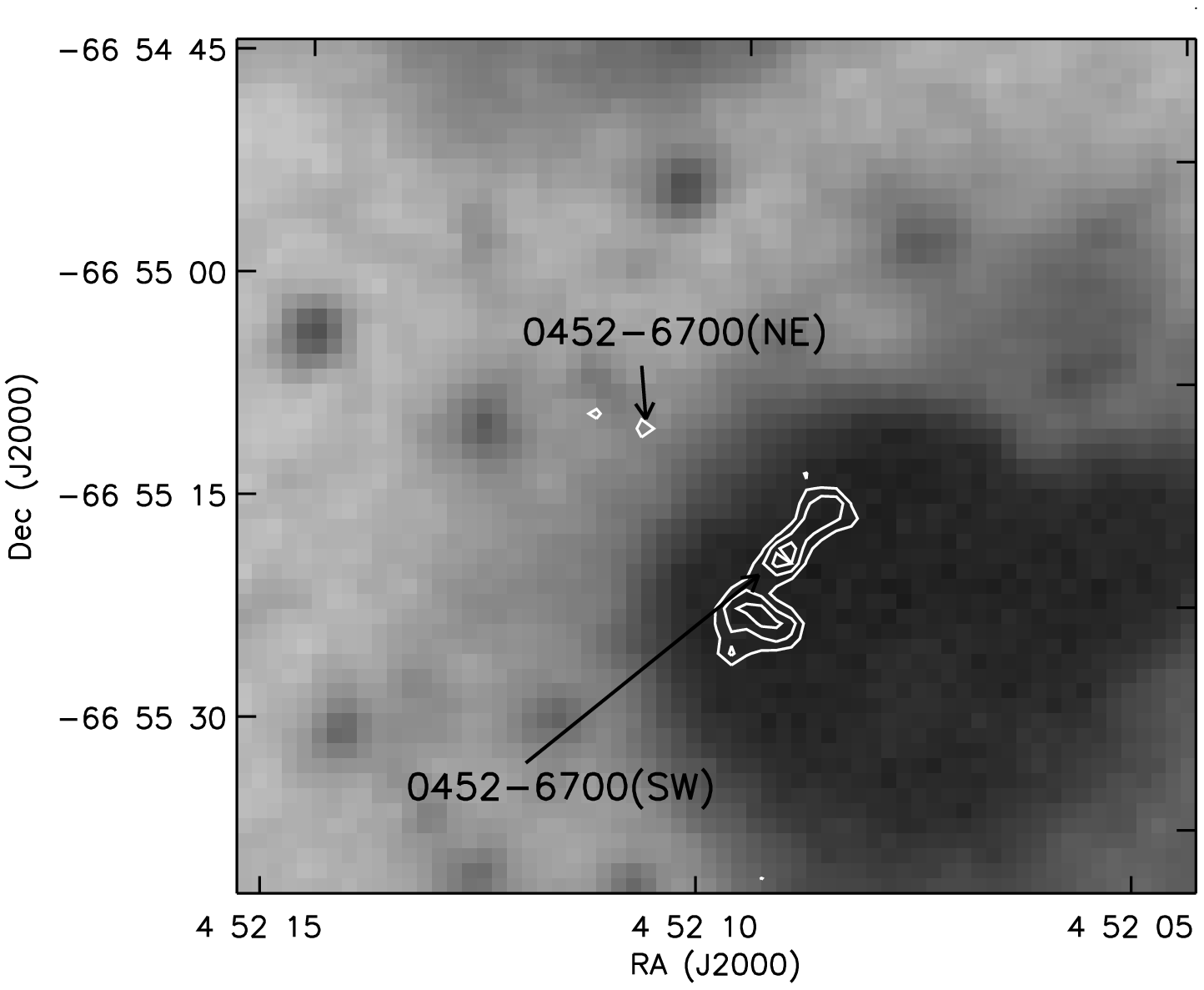}
\caption{\label{n4} Compact sources related to Parkes source
B0452-6700 and optical \hii region N4, labelled as Figure~\ref{n79}.}
\end{figure}

\begin{figure}
\plottwo{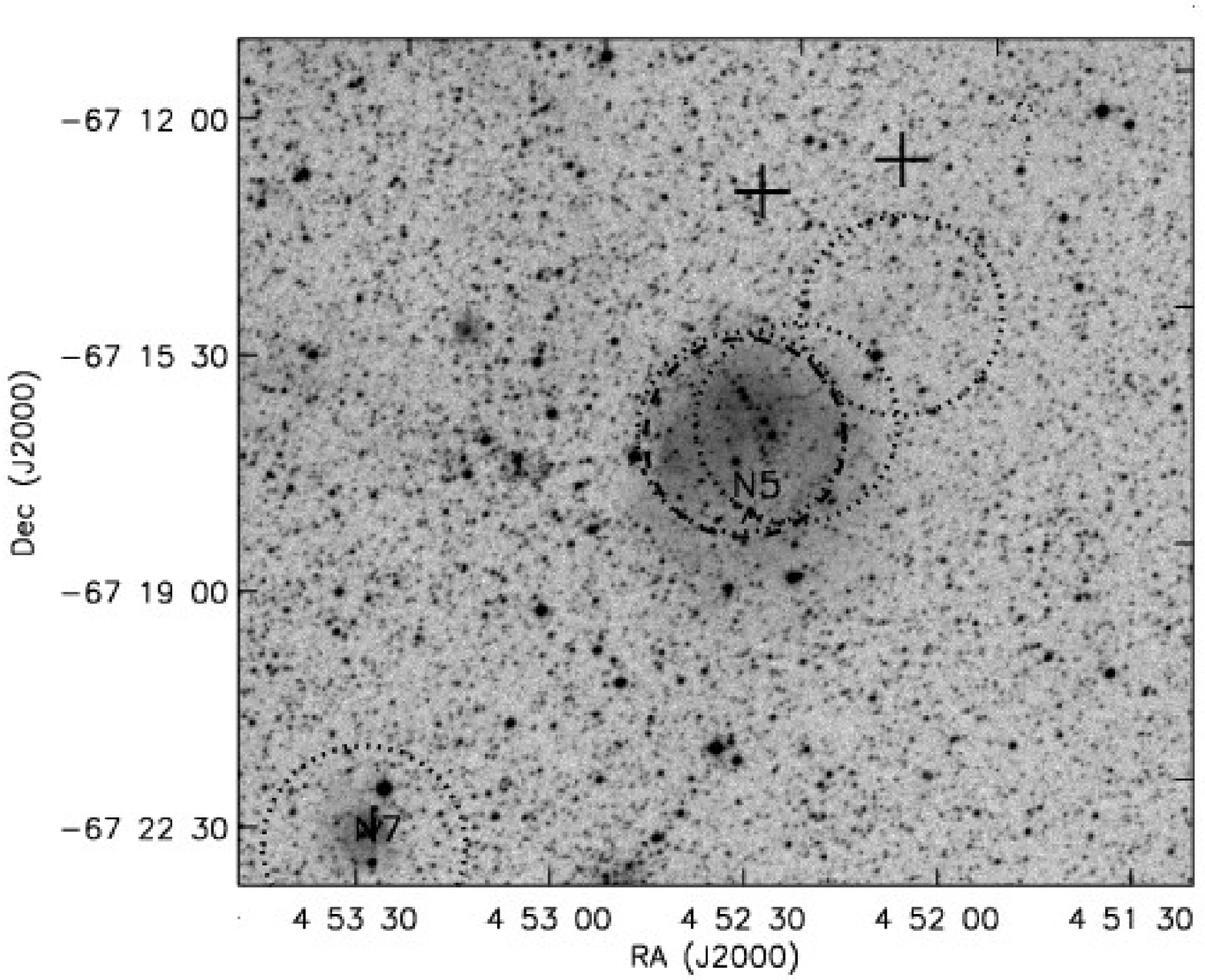}{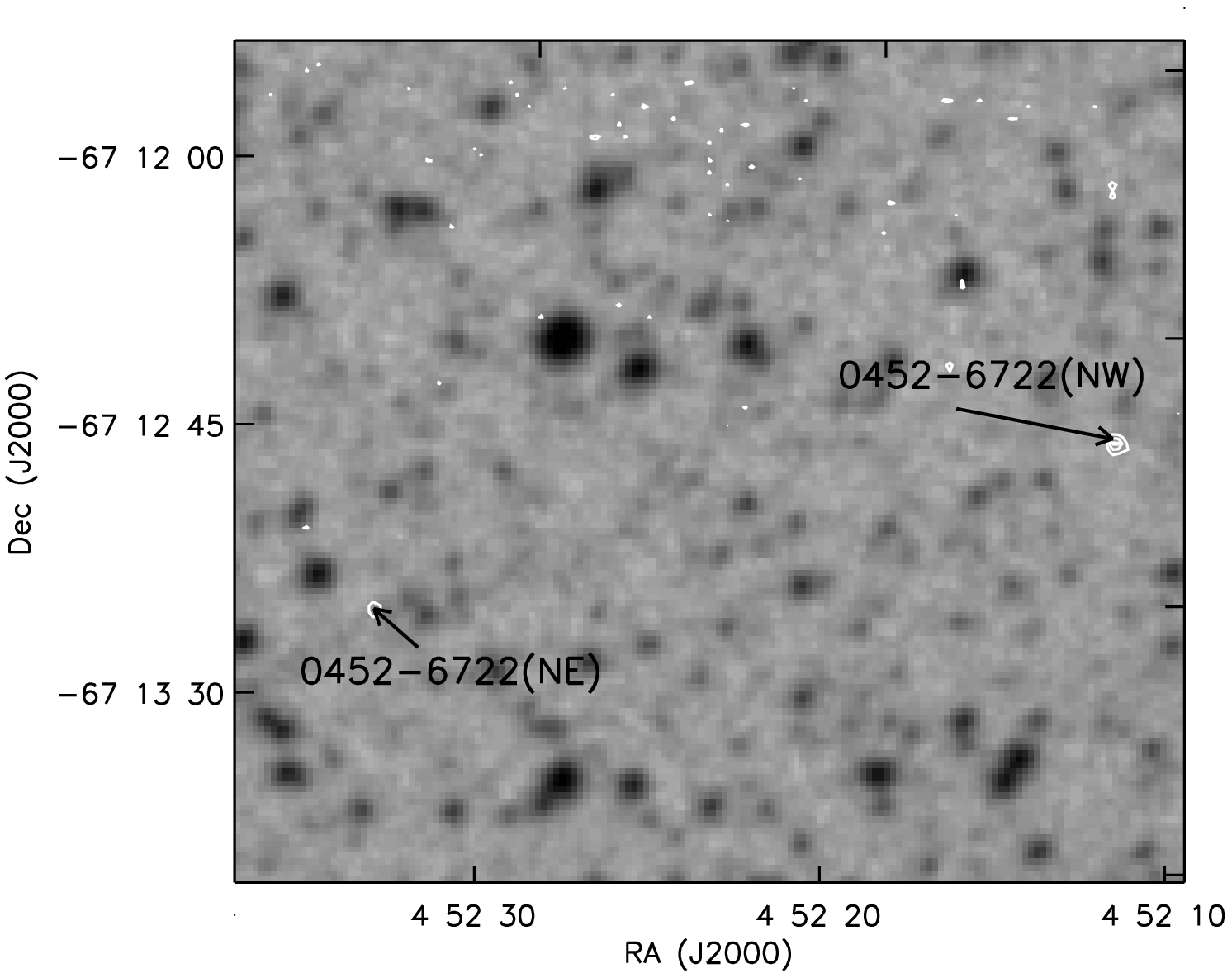}
\caption{\label{n5} Compact sources related to Parkes source
B0452-6722 and optical \hii region N5, labelled as Figure~\ref{n79}.}
\end{figure}

\begin{figure}
\plottwo{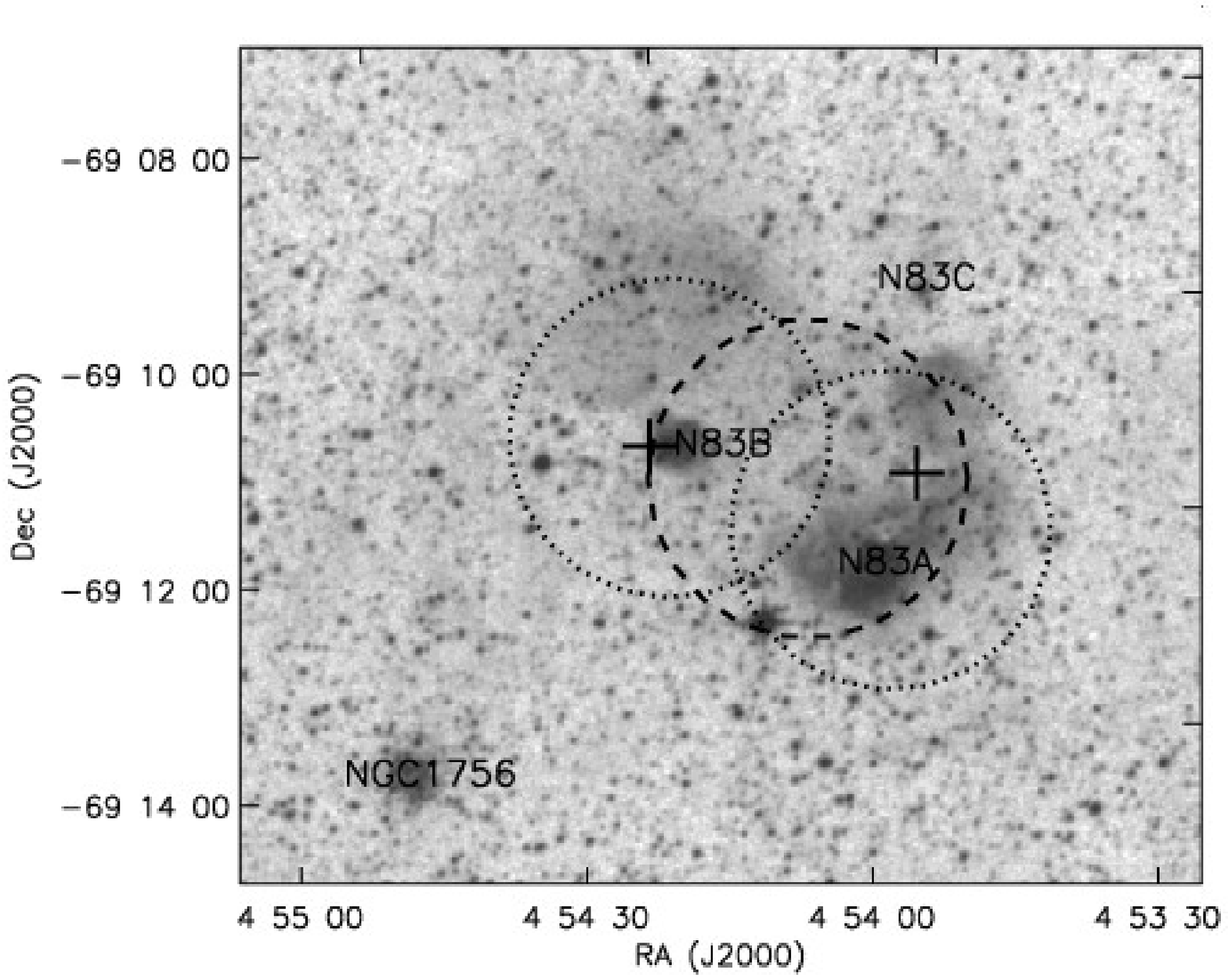}{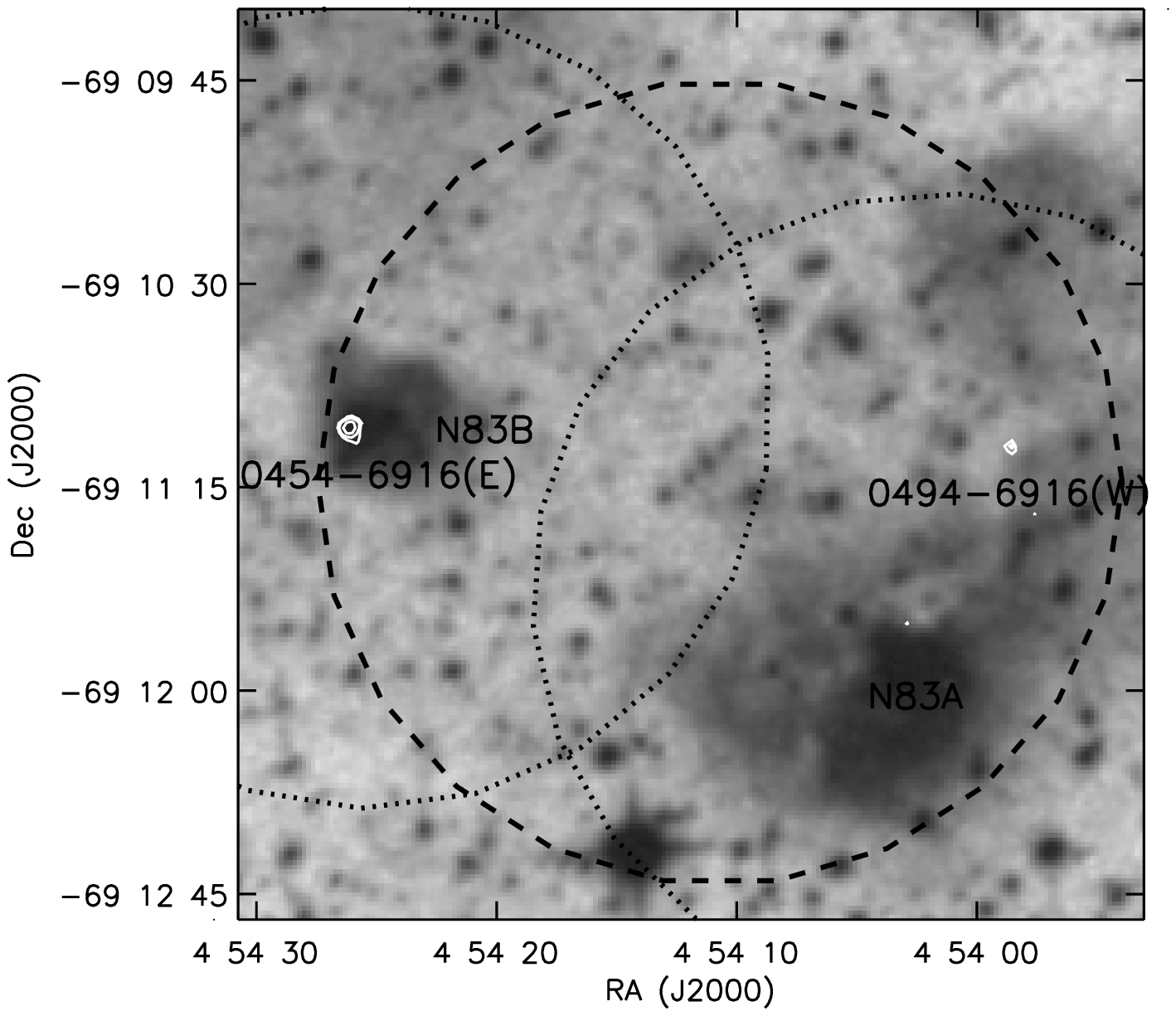}
\caption{\label{n83} Compact sources related to Parkes source
B0454-6916 and optical \hii region N83, labelled as Figure~\ref{n79}.}
\end{figure}

\begin{figure}
\plottwo{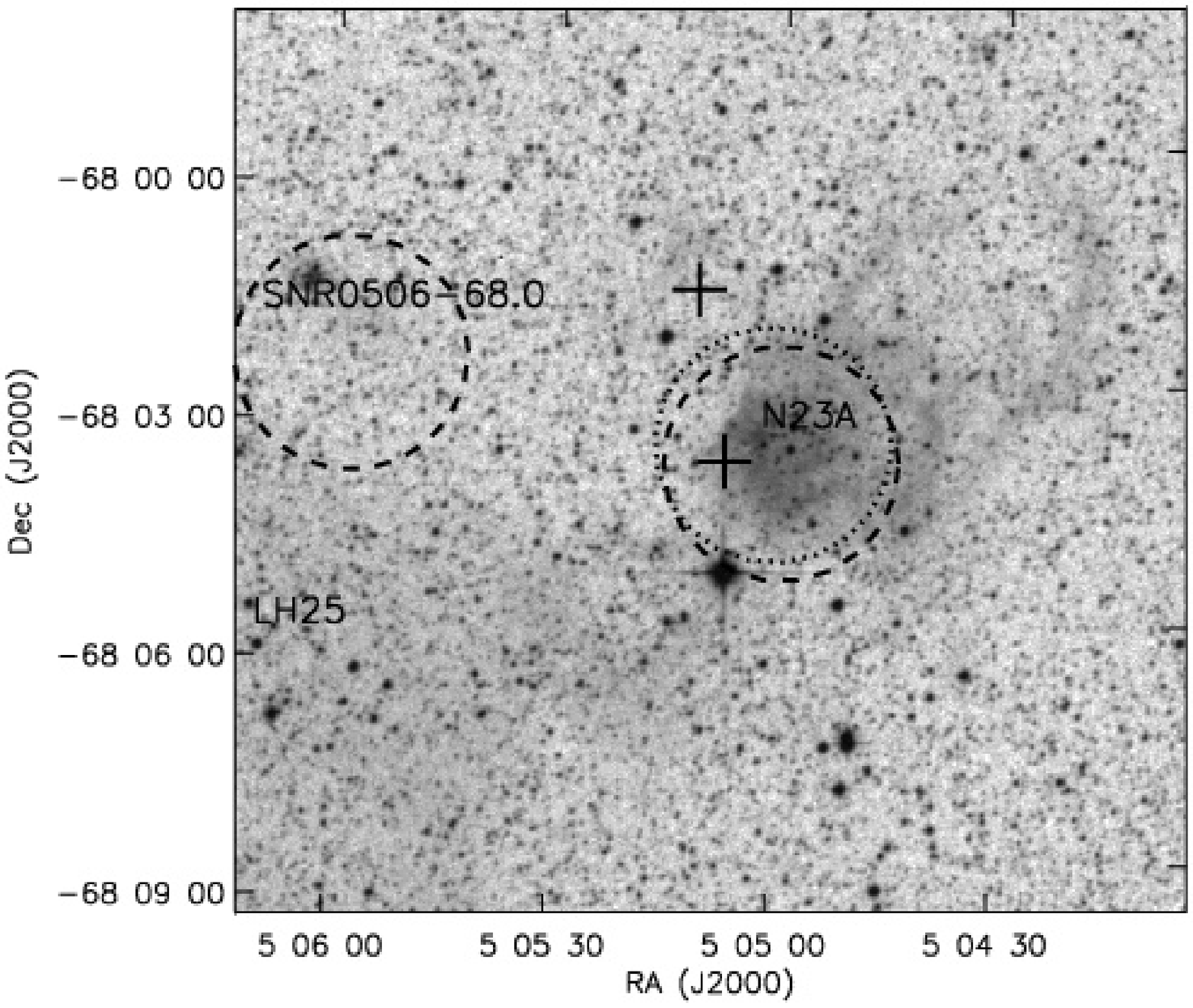}{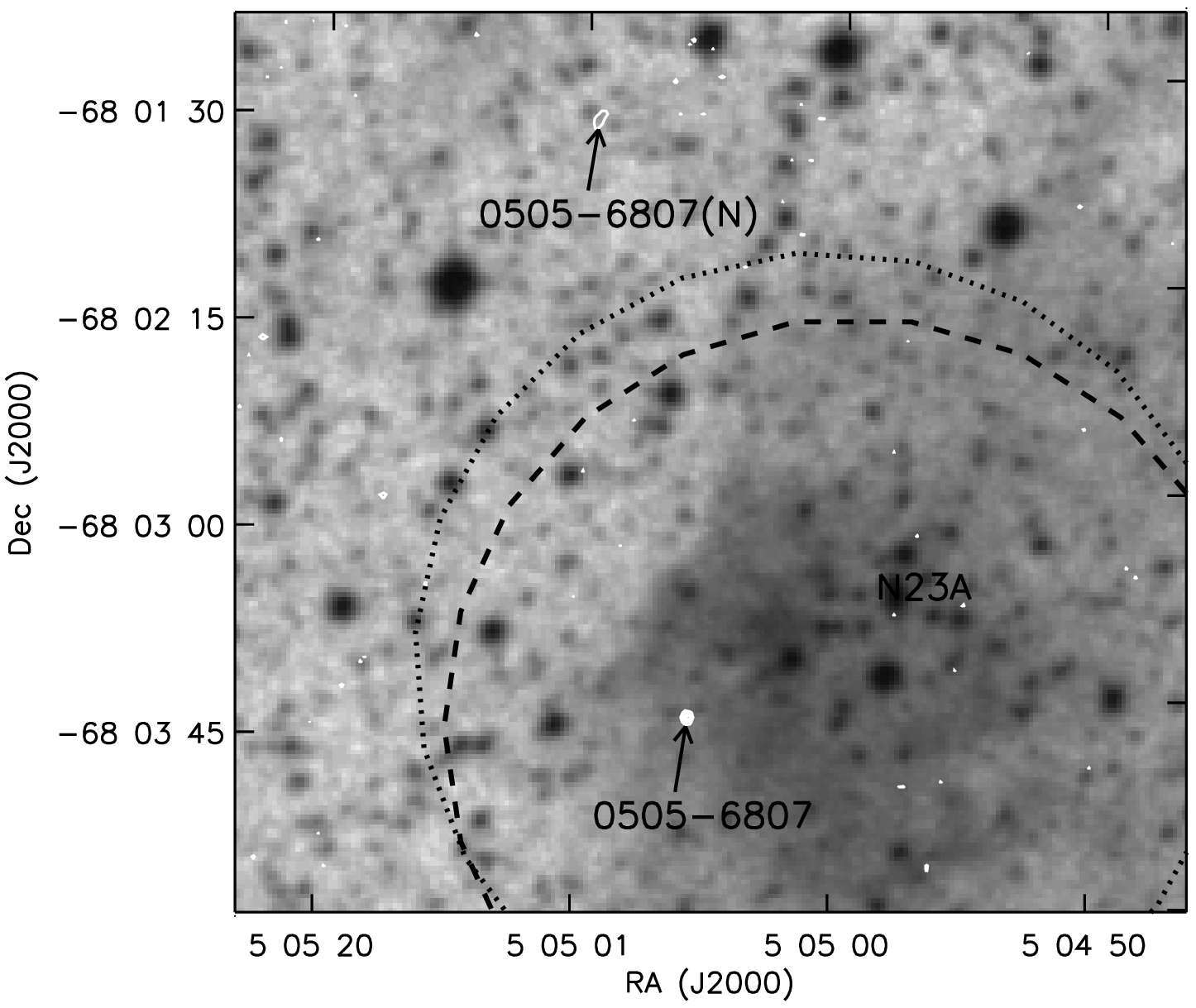}
\caption{\label{n23} Compact sources related to Parkes source
B0505-6807 and optical \hii region N23, labelled as Figure~\ref{n79}.}
\end{figure}

\begin{figure}
\plottwo{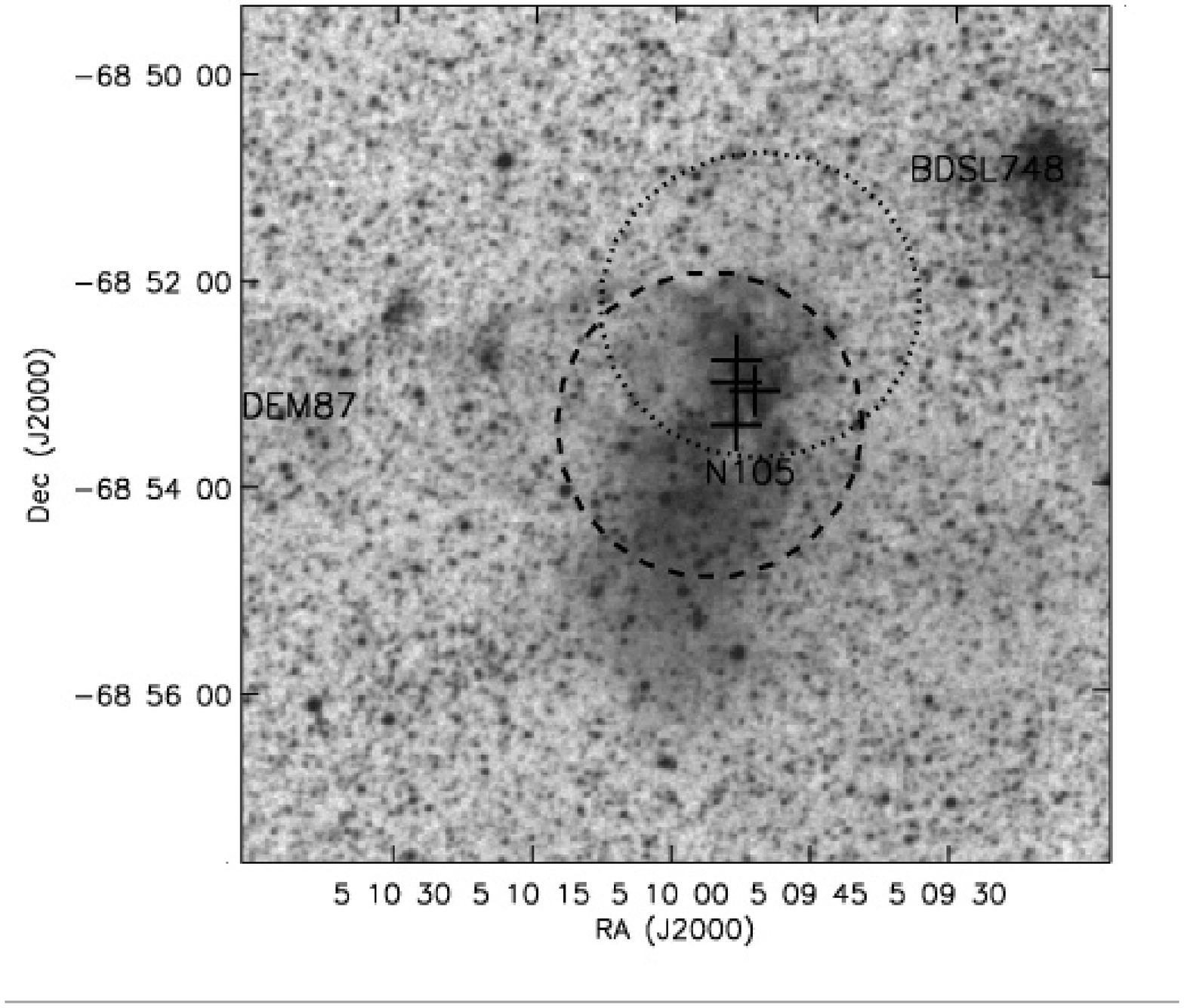}{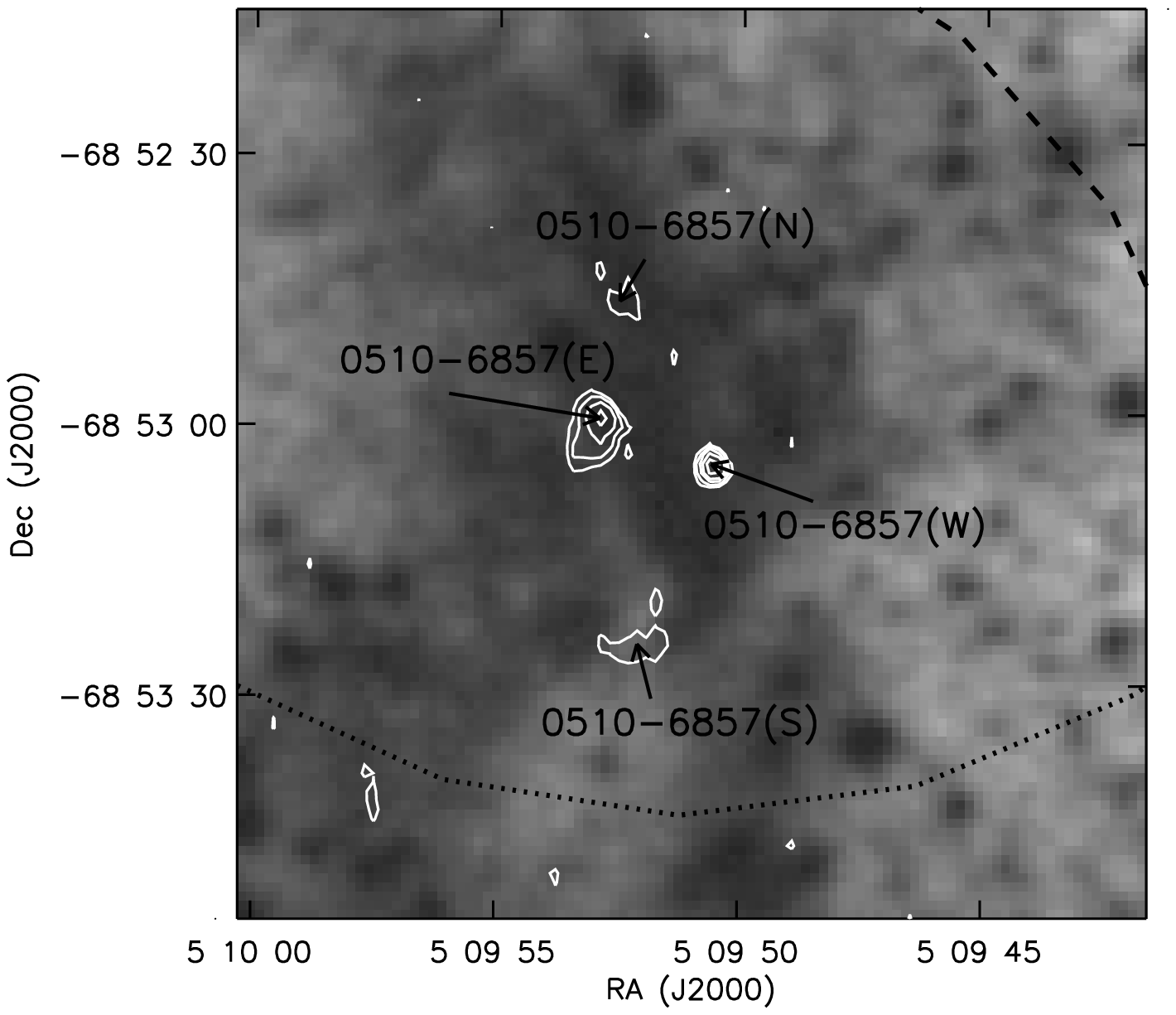}
\caption{\label{n105} Compact sources related to Parkes source
B0510-6857 and optical \hii region N105, labelled as Figure~\ref{n79}.}
\end{figure}

\begin{figure}
\plottwo{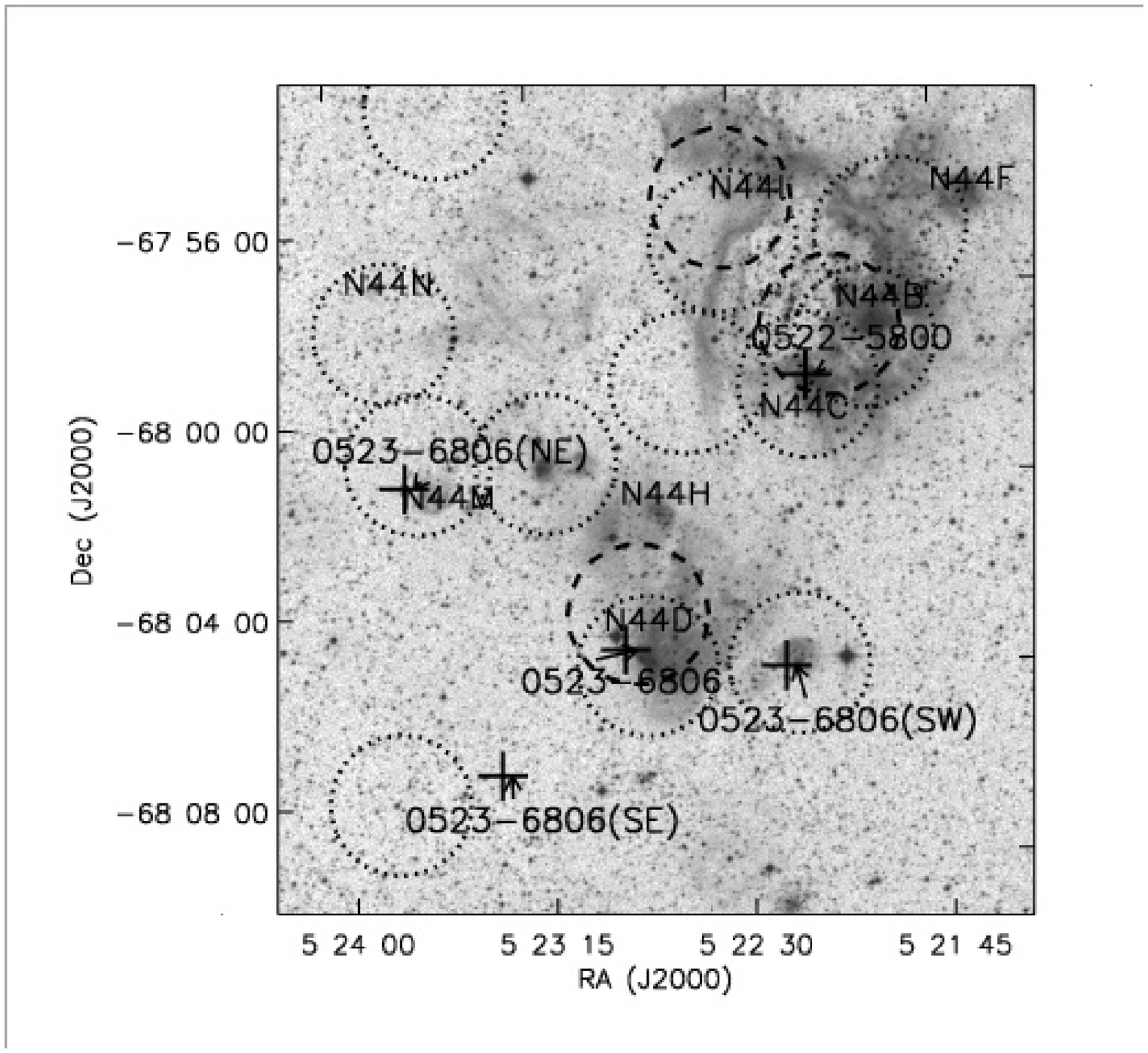}{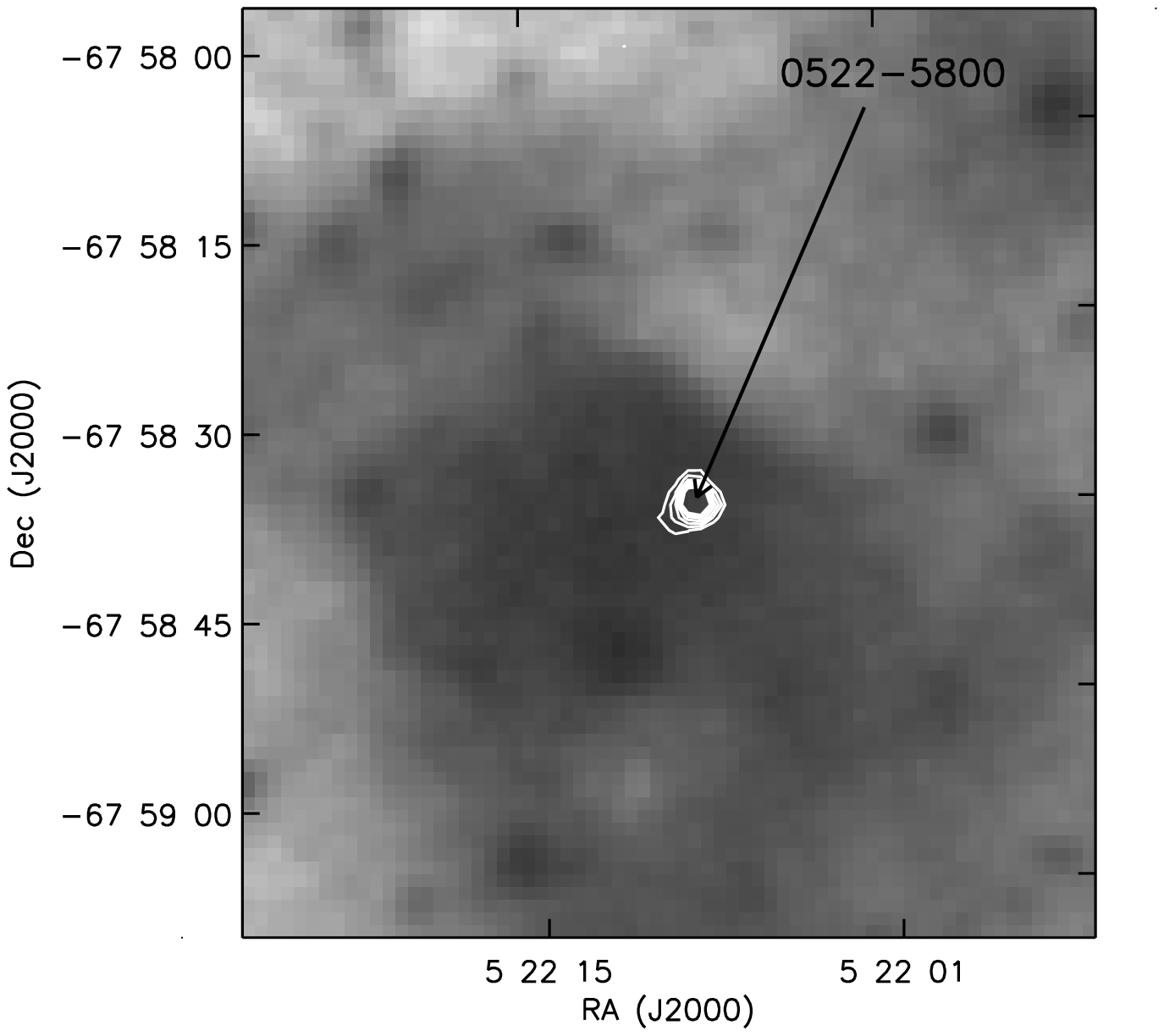}
\plottwo{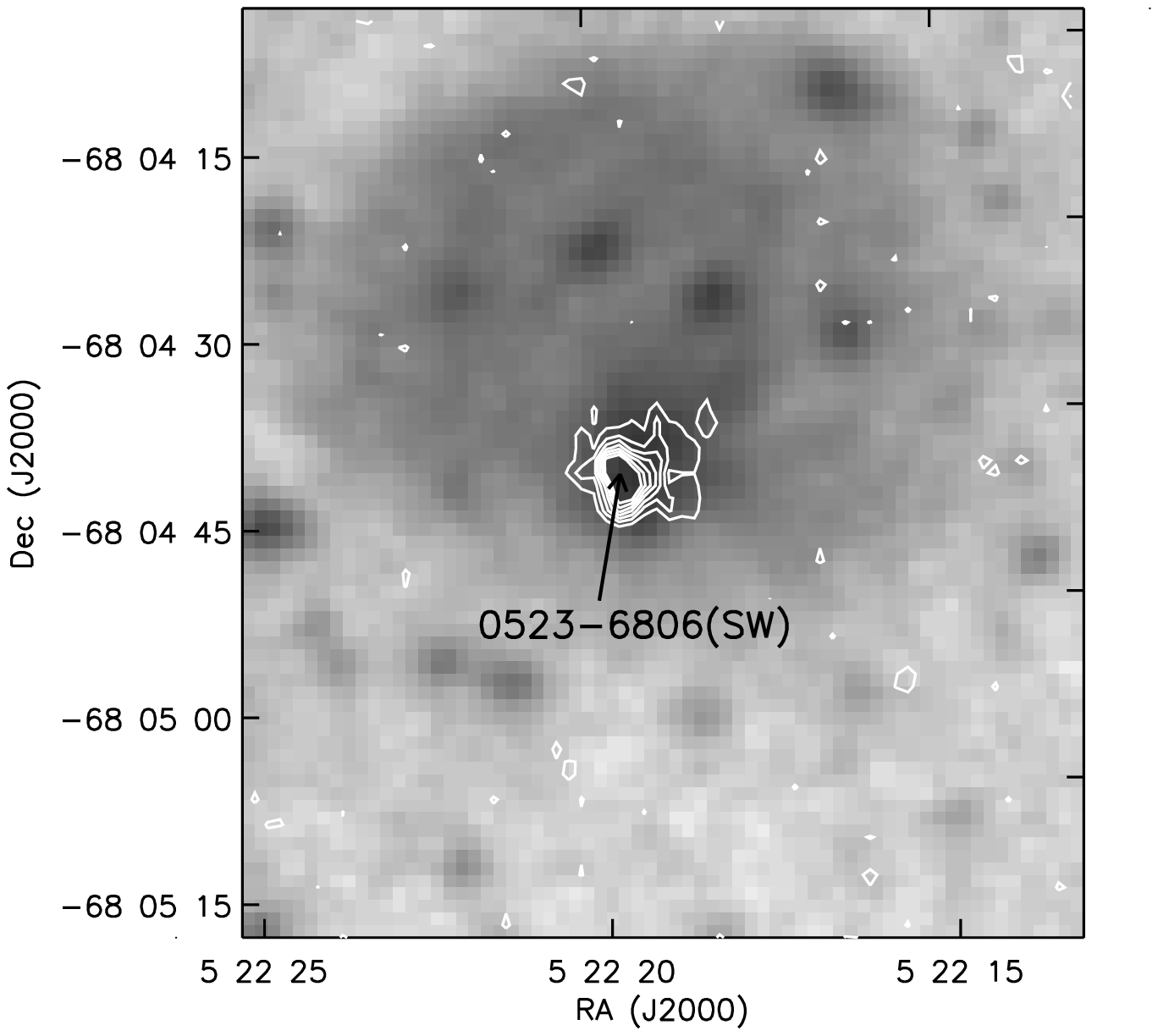}{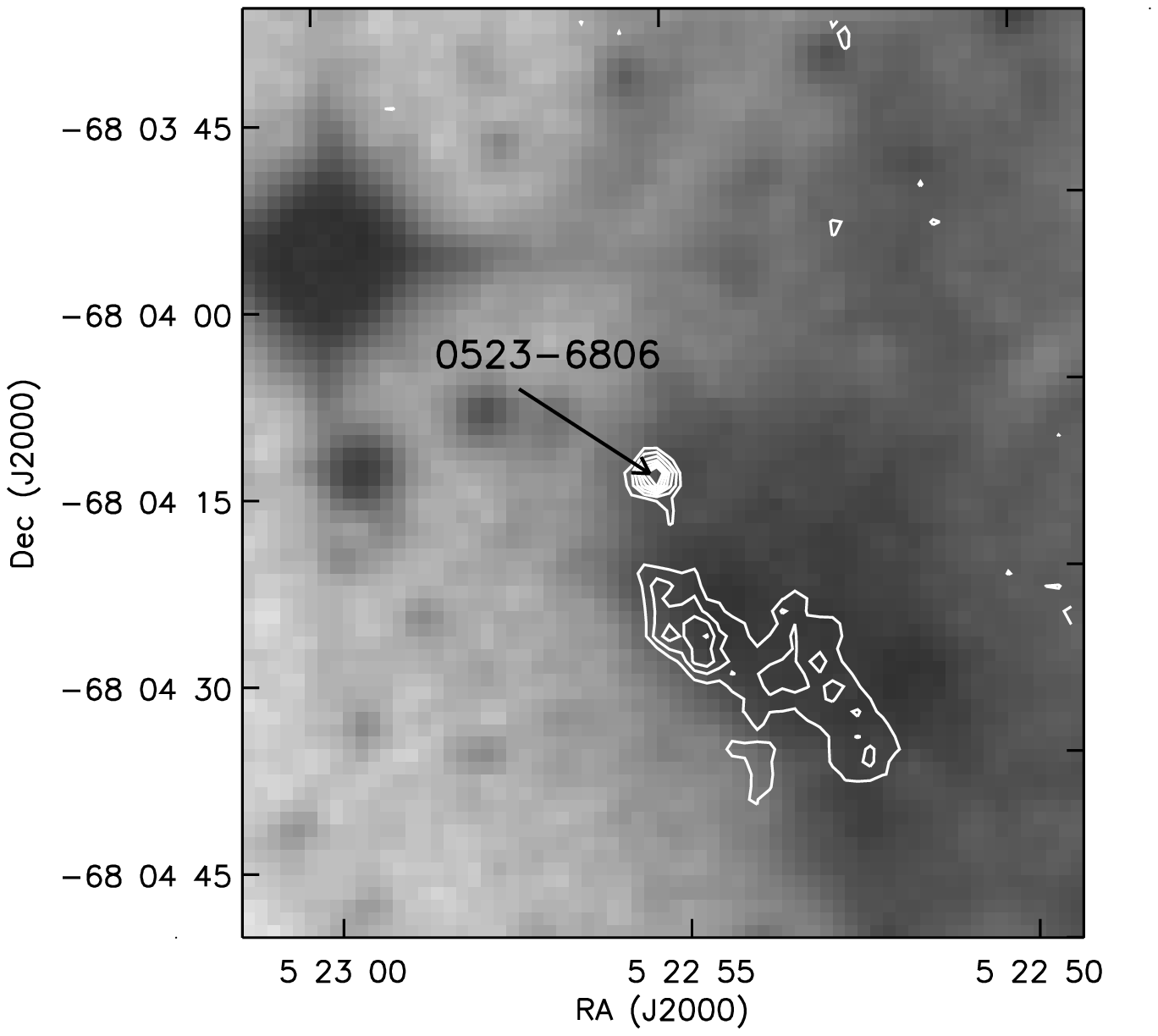}
\caption{\label{n44} Compact sources related to Parkes source
B0523-6808 and optical \hii region N44, labelled as Figure~\ref{n79}.
Zoomed-in images of three different sources in N44 are shown.}
\end{figure}

\begin{figure}
\plottwo{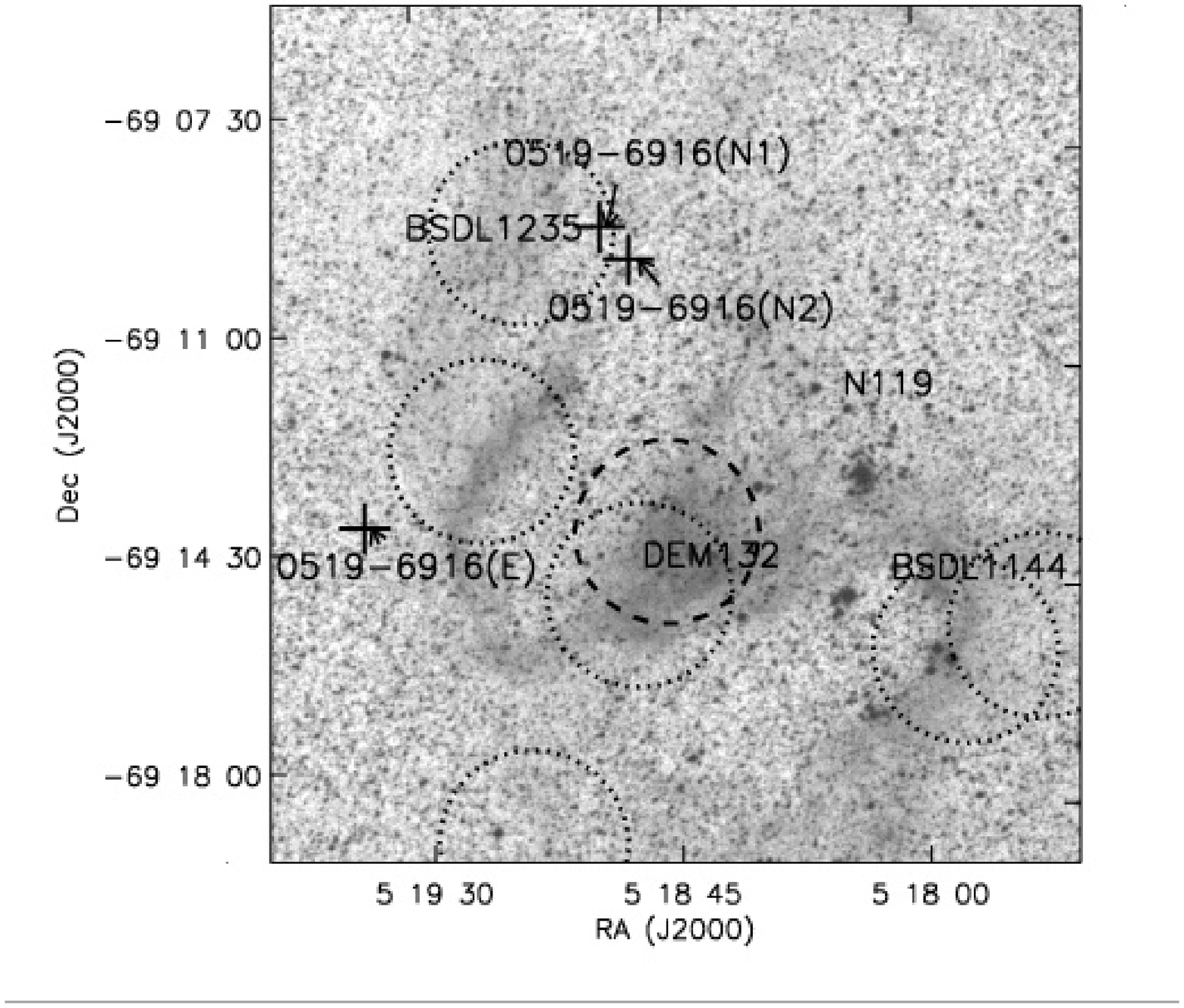}{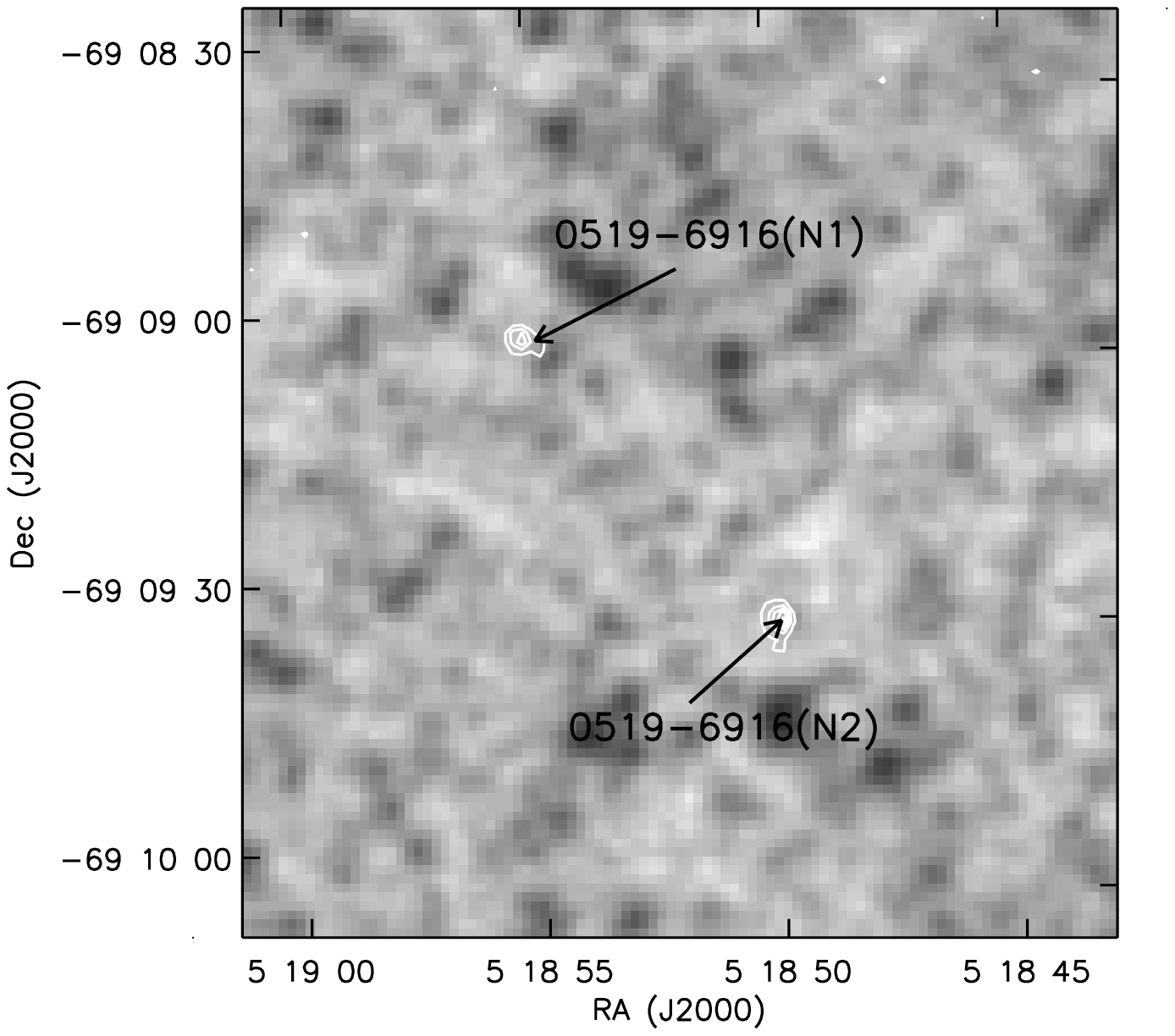}
\plottwo{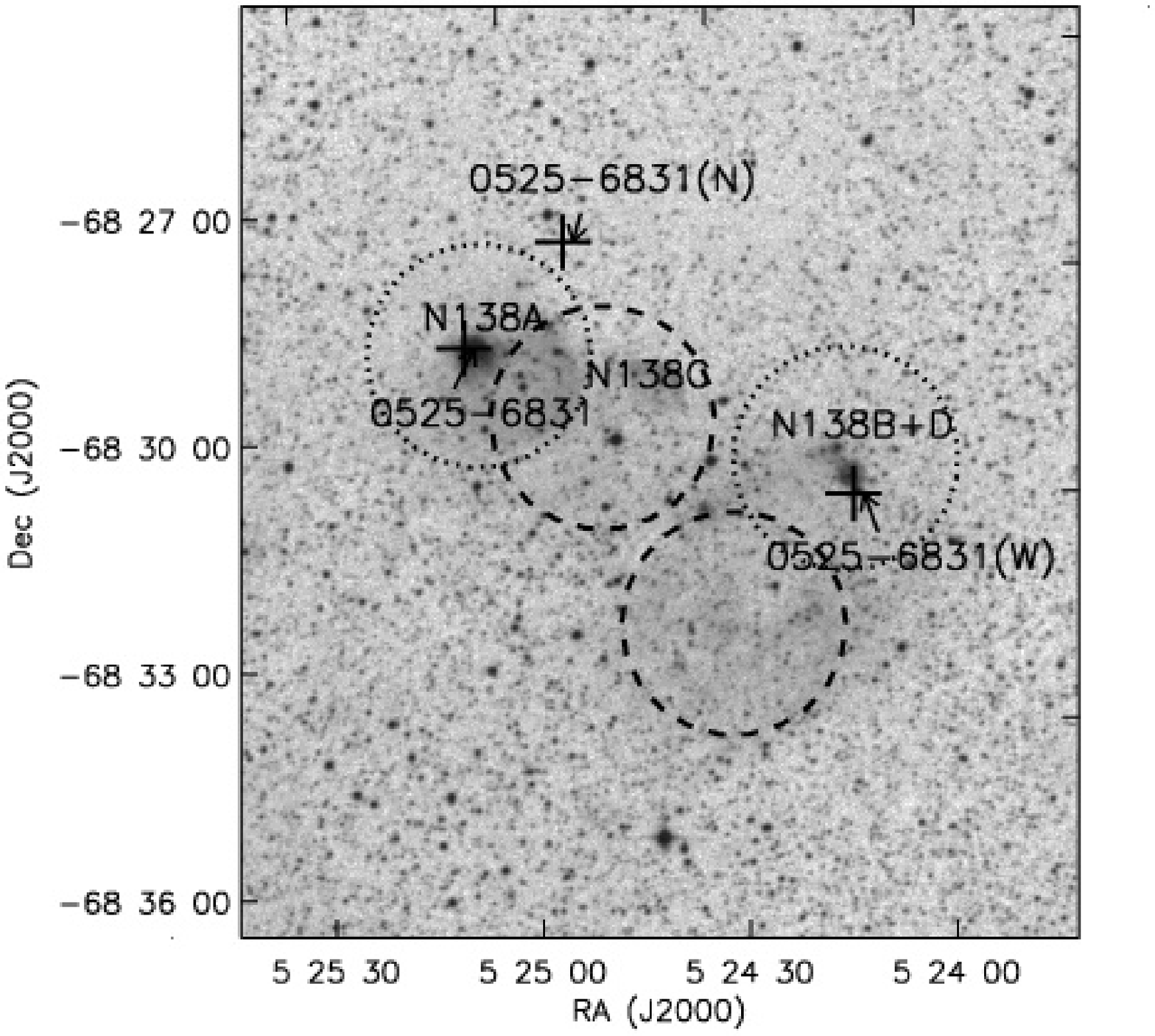}{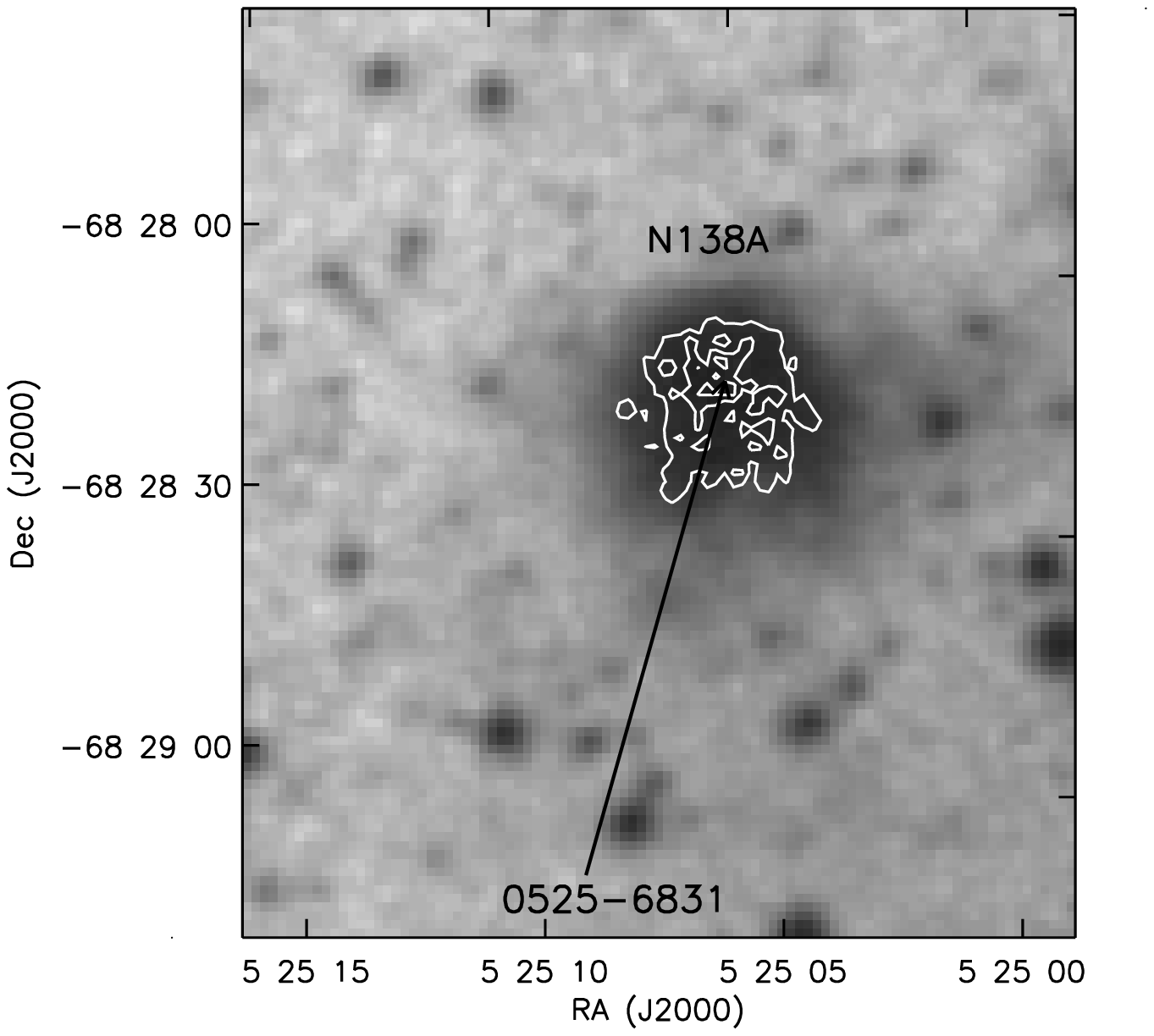}
\caption{\label{n119n138} Compact sources near Parkes source 0519-6916
and optical \hii region N119, and near Parkes source 0525-6831 and
optical \hii region N138.}
\end{figure}

\begin{figure}
\plottwo{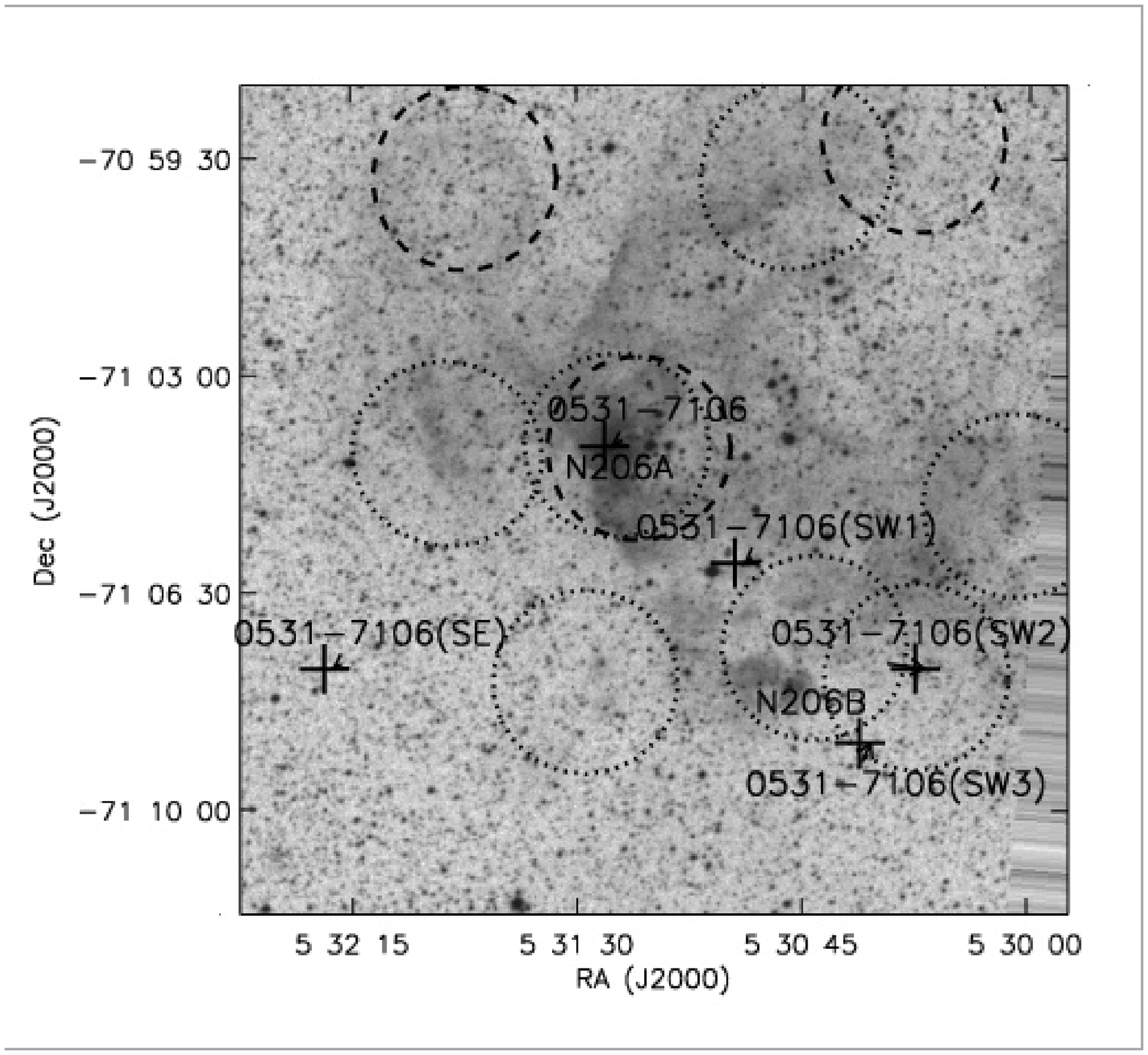}{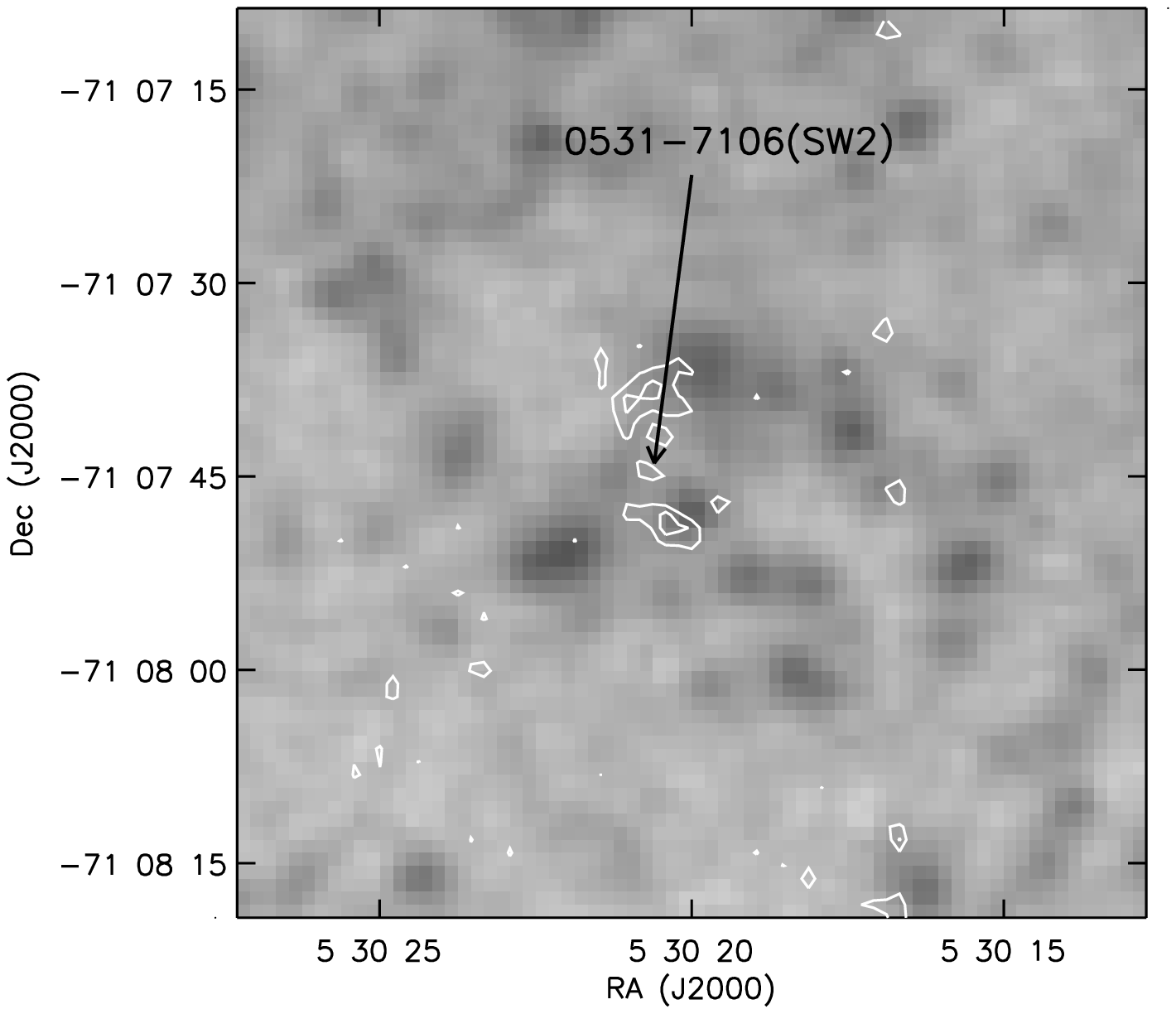}
\plottwo{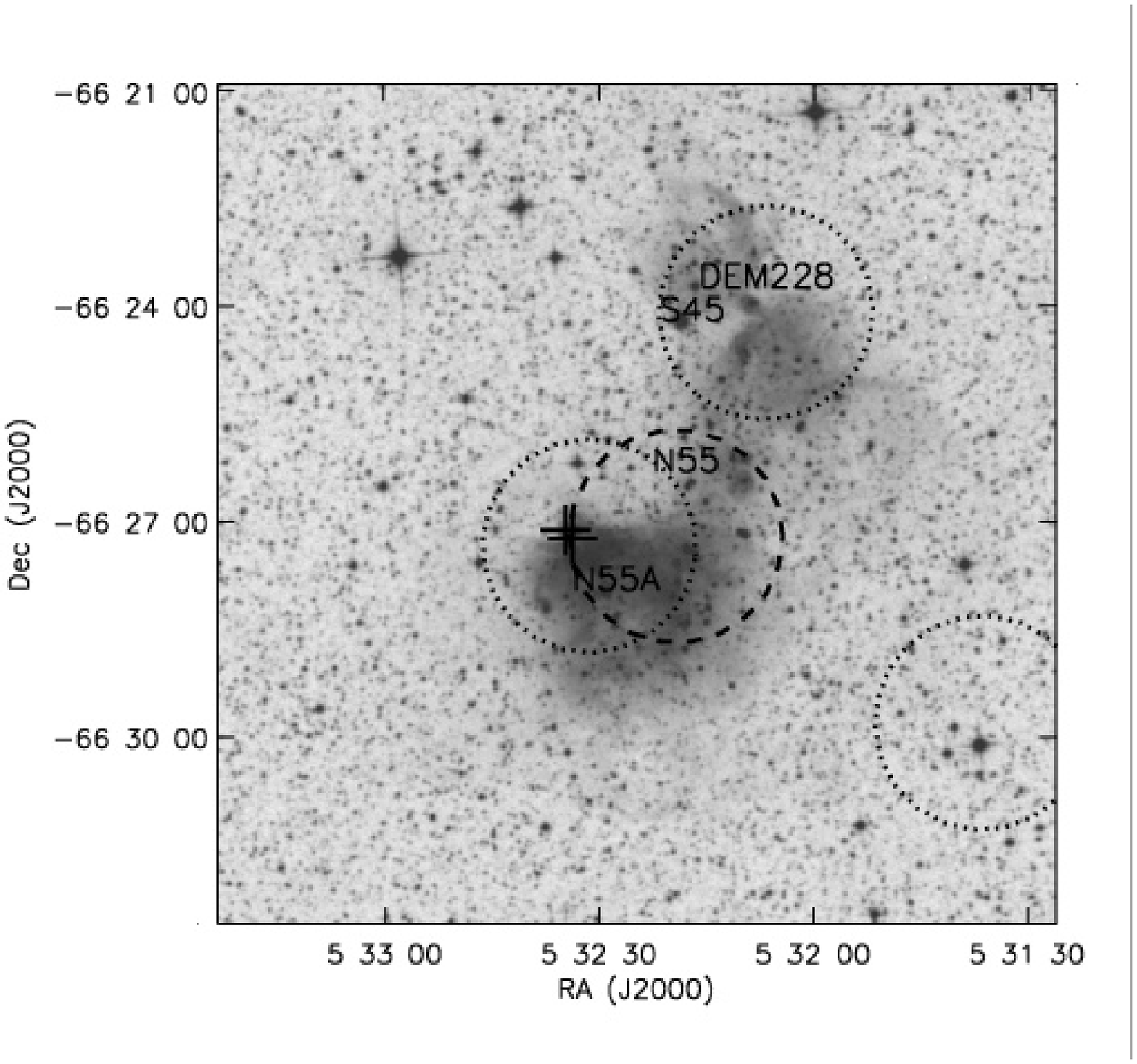}{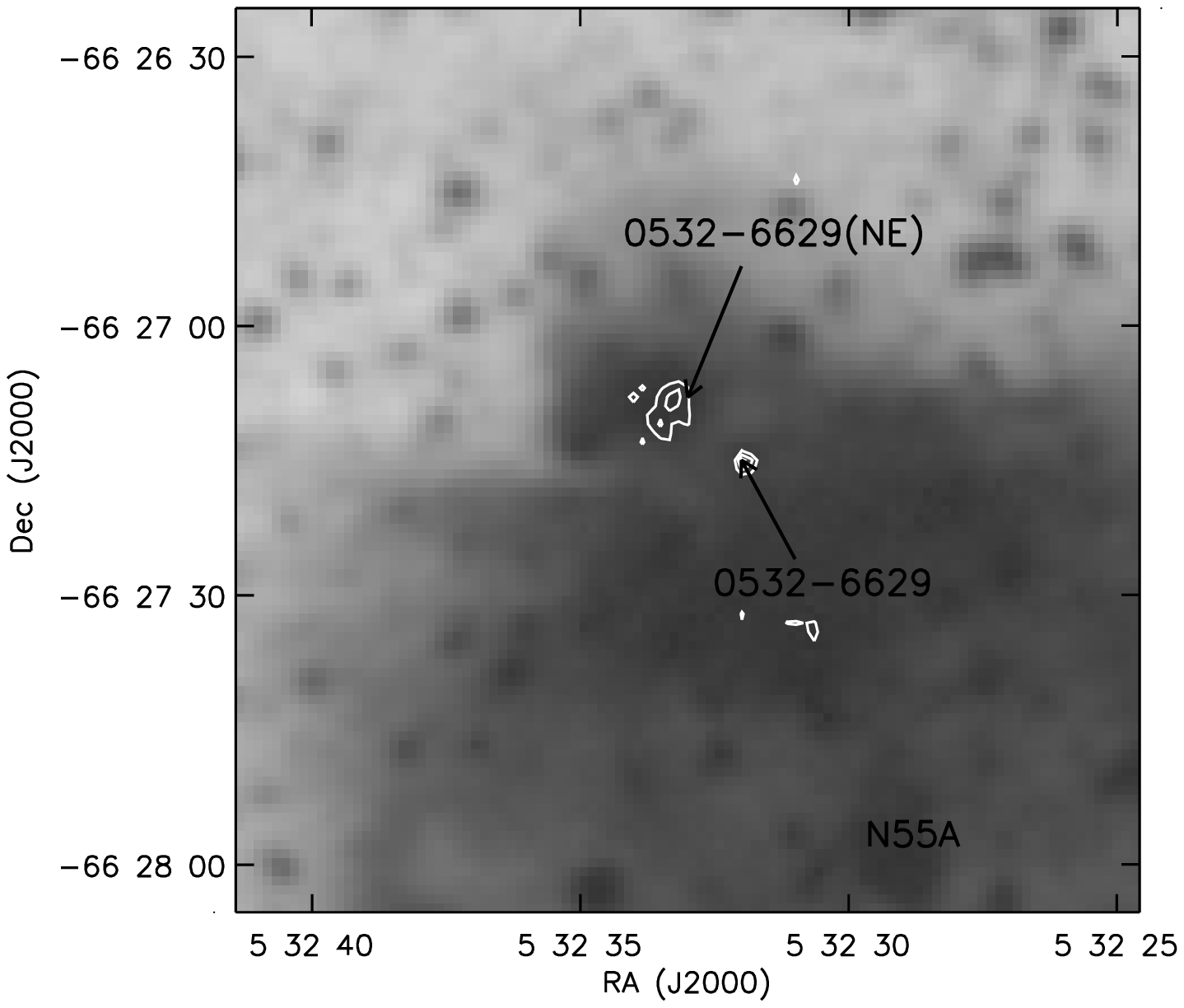}
\plottwo{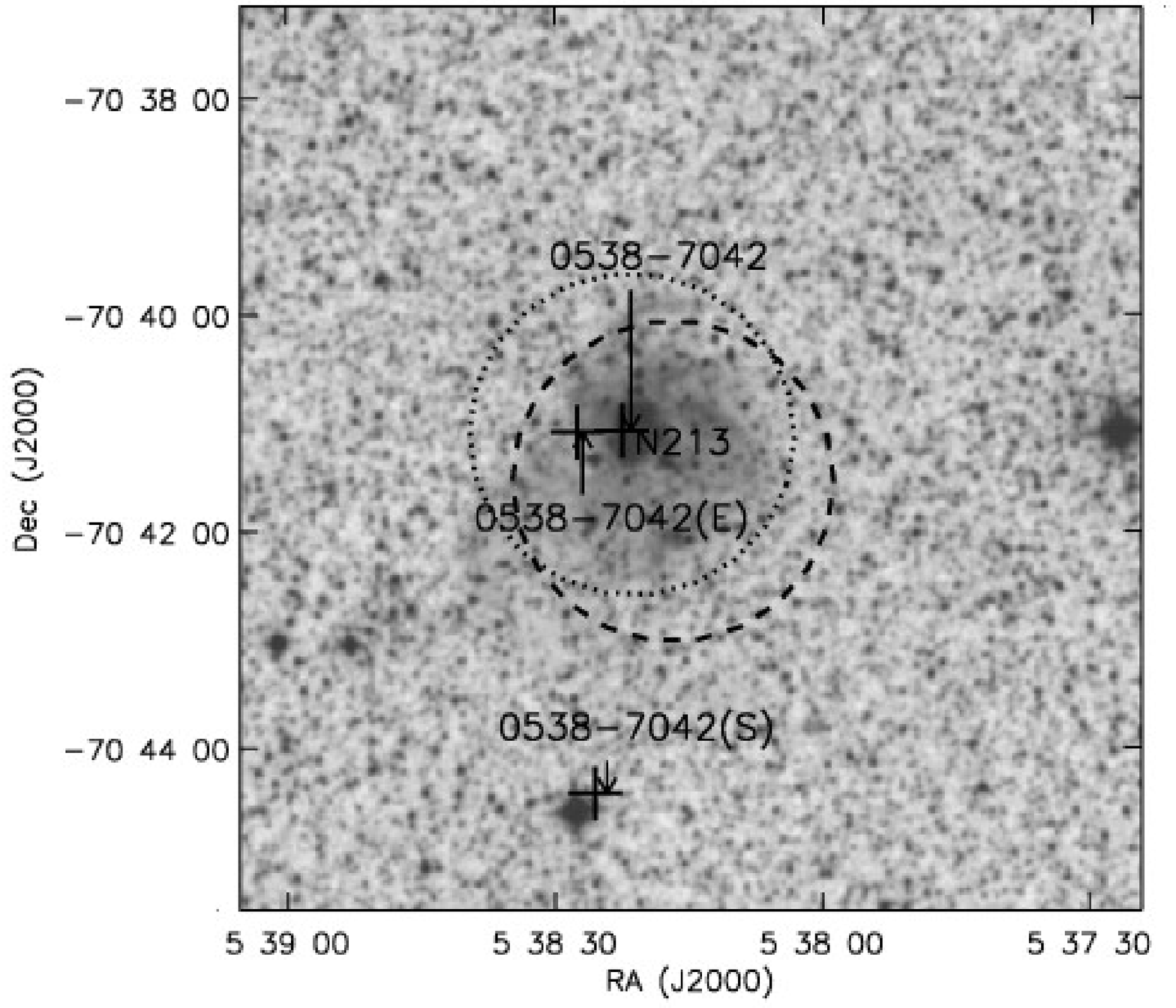}{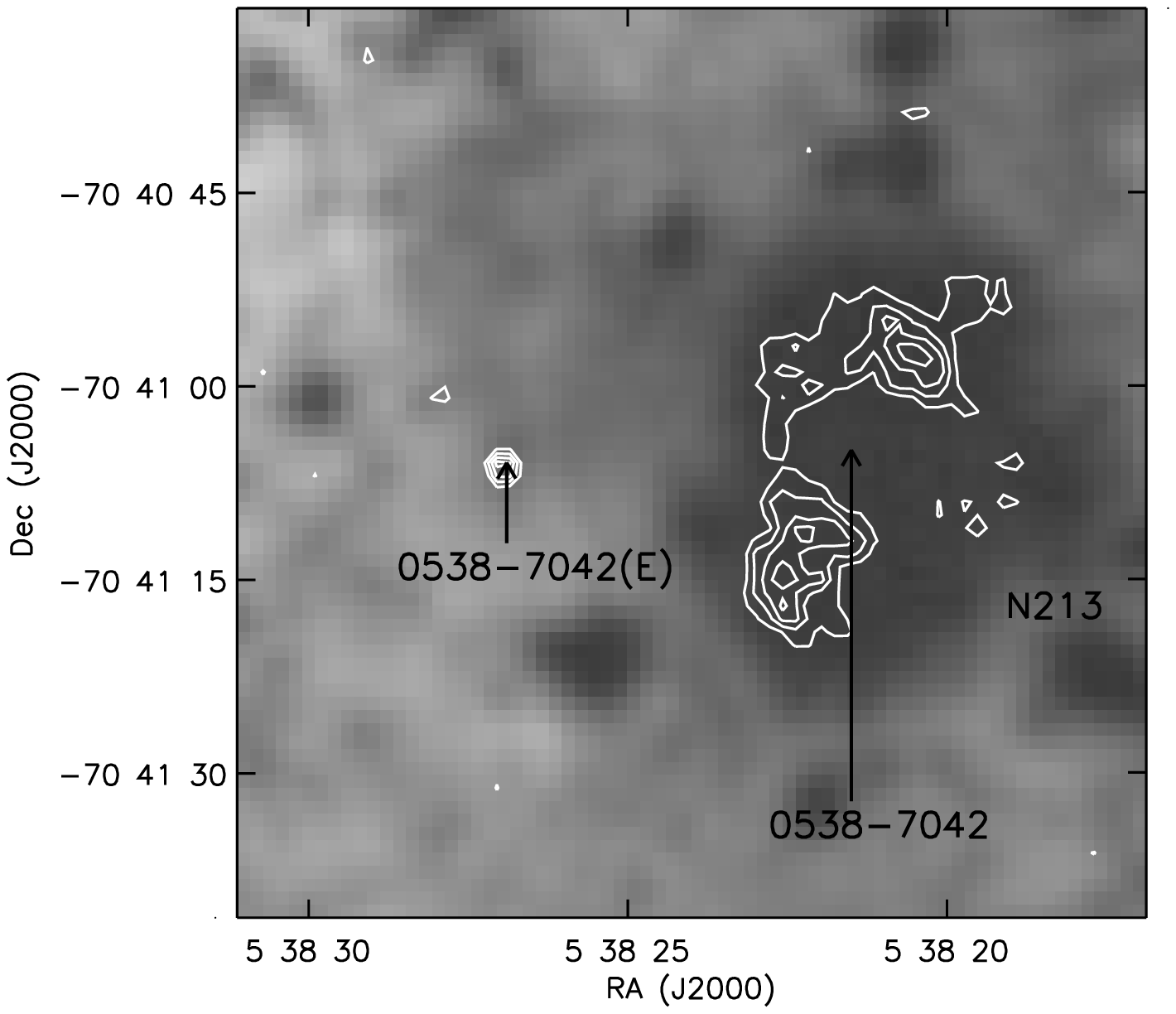}
\caption{\label{n206n55n213} Compact sources near Parkes source
0531-7106 and optical \hii region N206, near Parkes source 0532-6629
and optical \hii region N55, and near Parkes source 0538-7042 and
optical \hii region N213. Labels as Fig.~\ref{n79}.}
\end{figure}

\begin{figure}
\plottwo{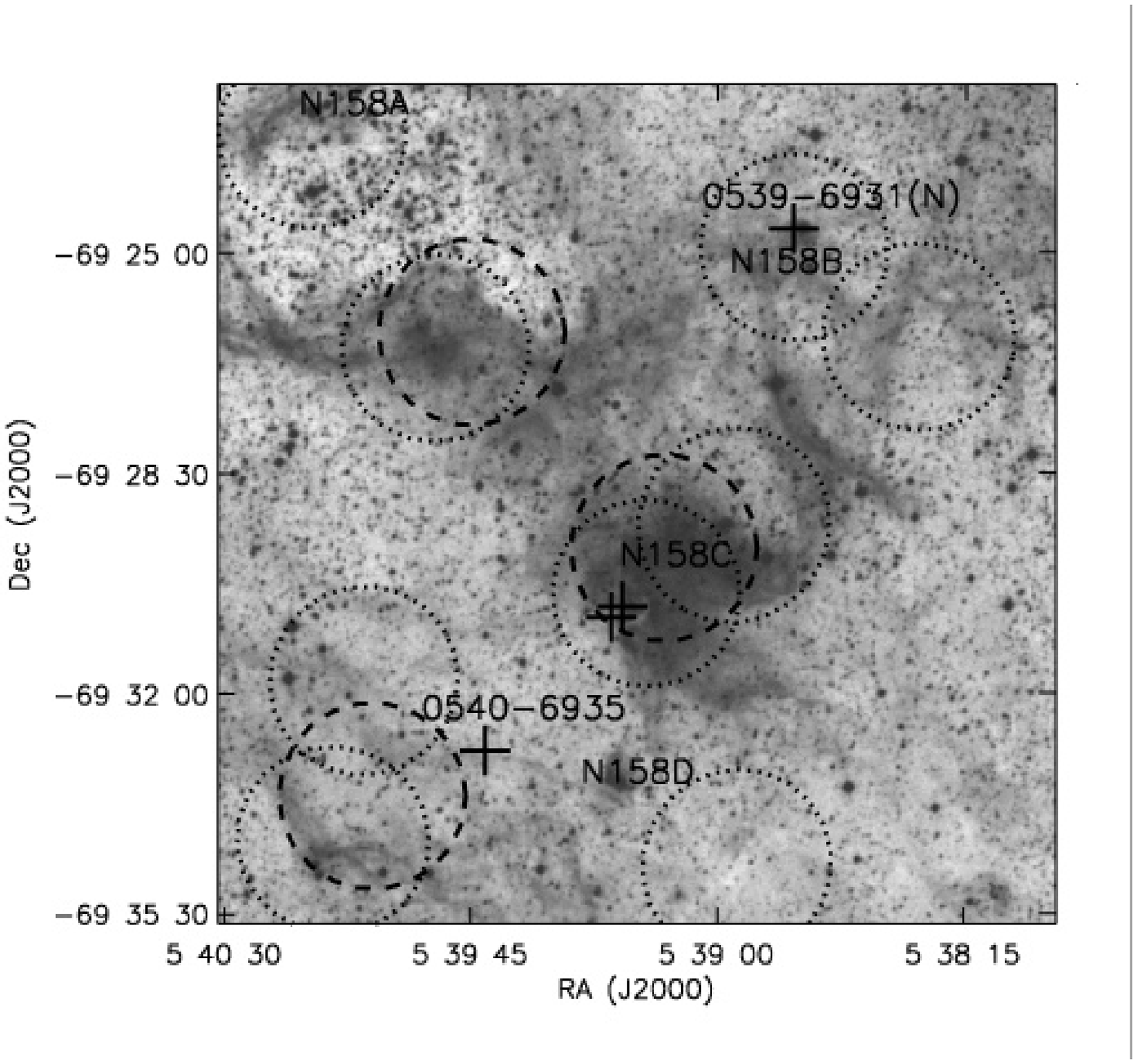}{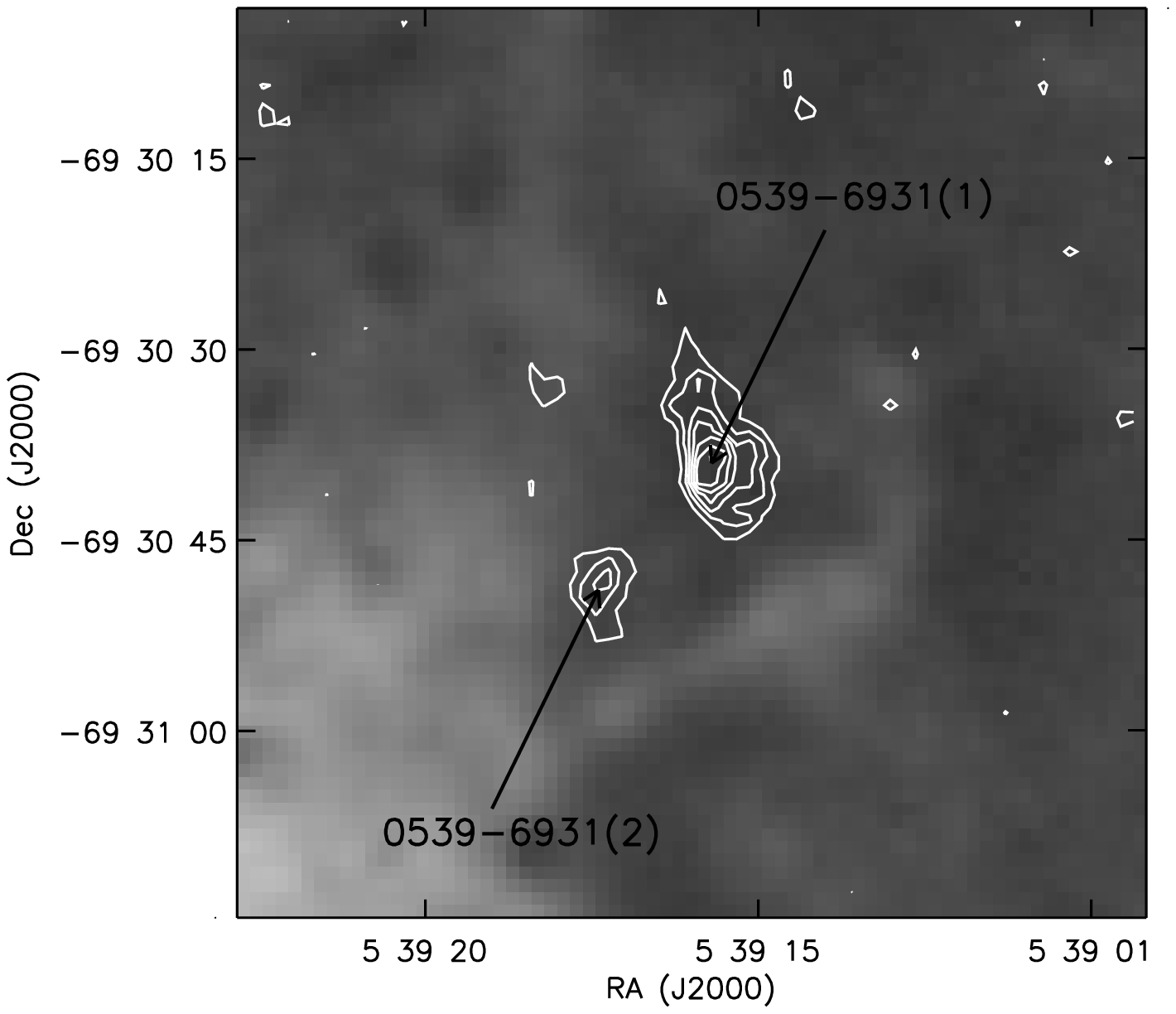}
\caption{\label{n158} Compact sources related to Parkes source
B0539-6931 and optical \hii region N158, labelled as Figure~\ref{n79}.}
\end{figure}

\begin{figure}
\plottwo{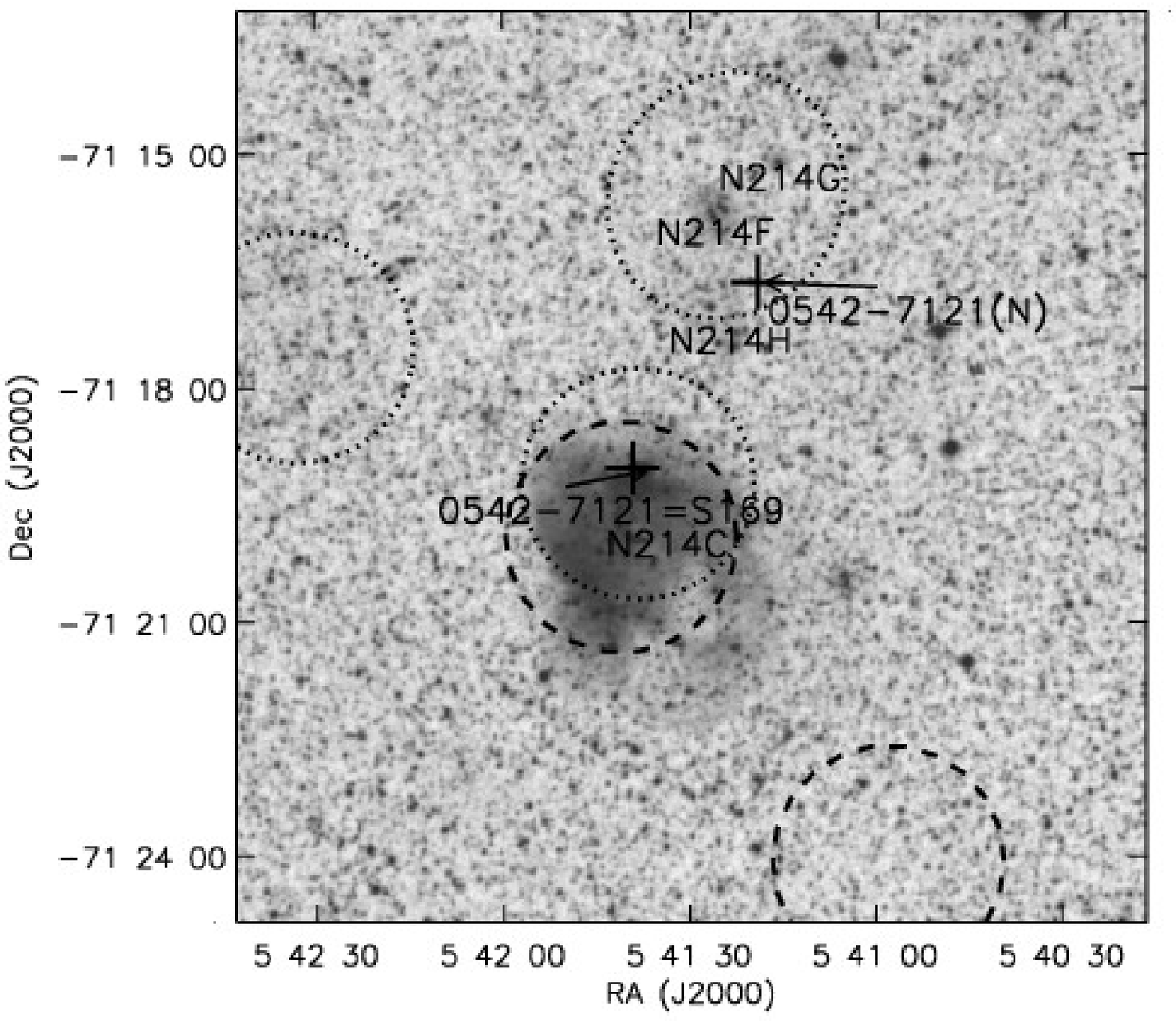}{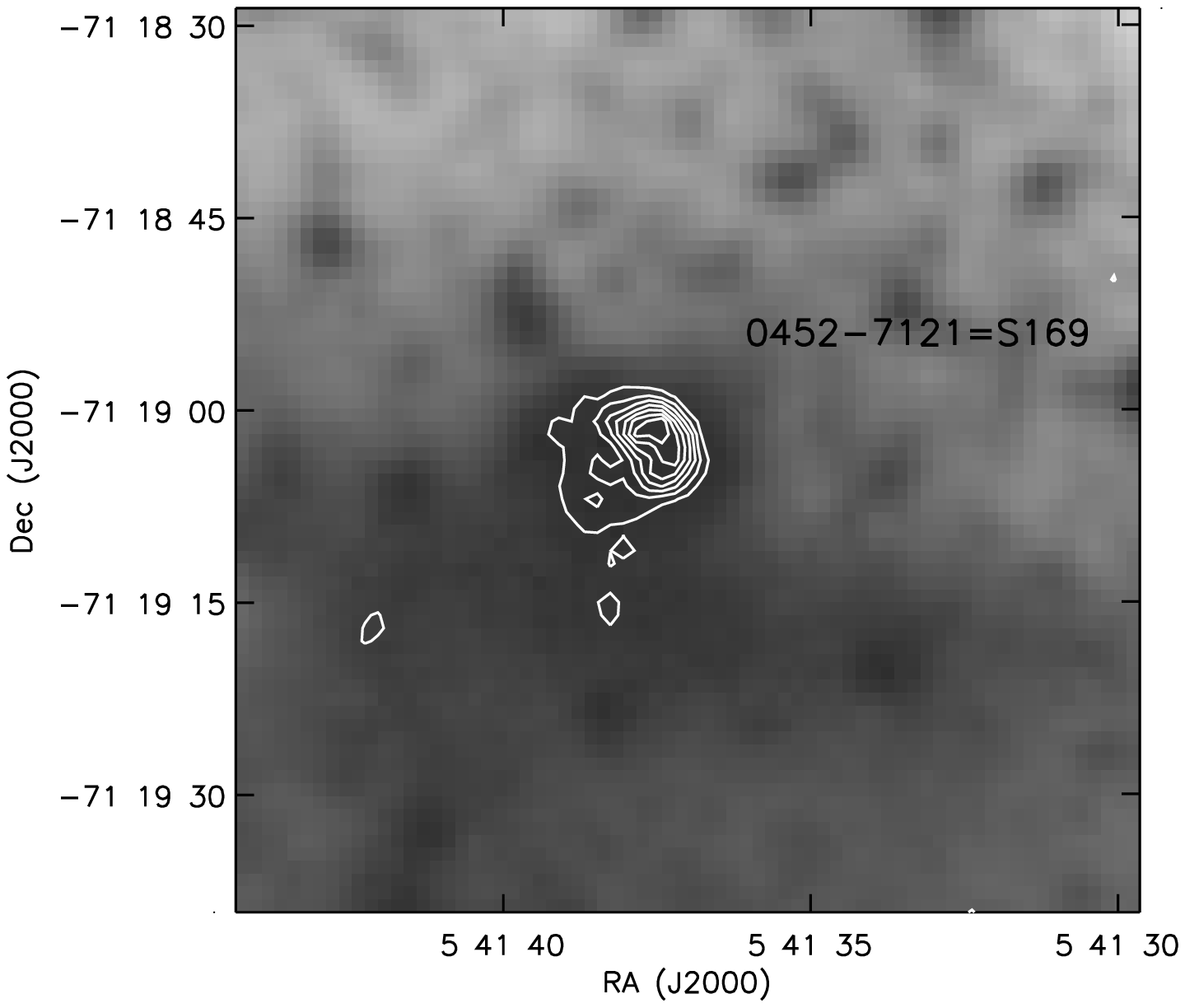}
\plottwo{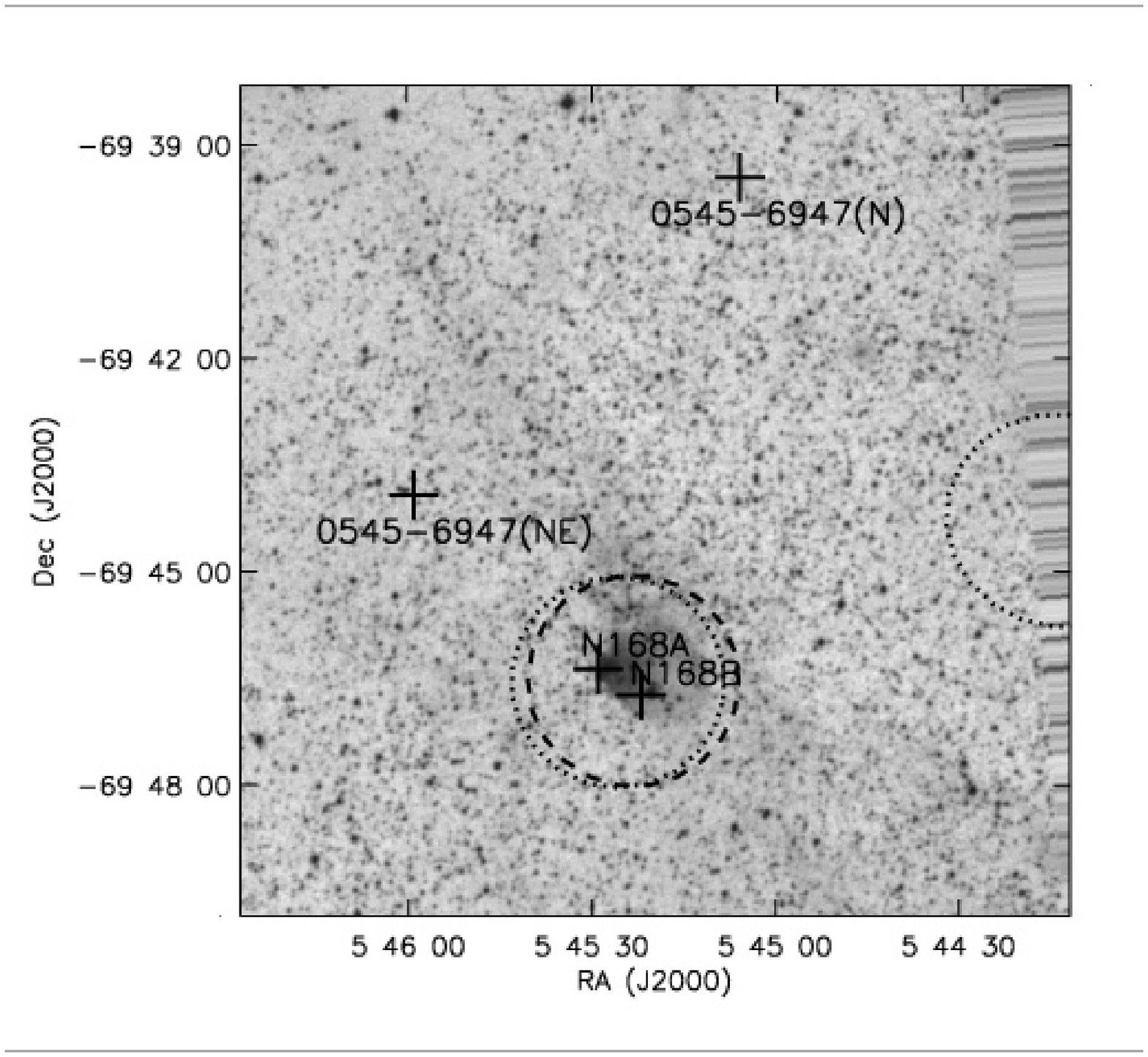}{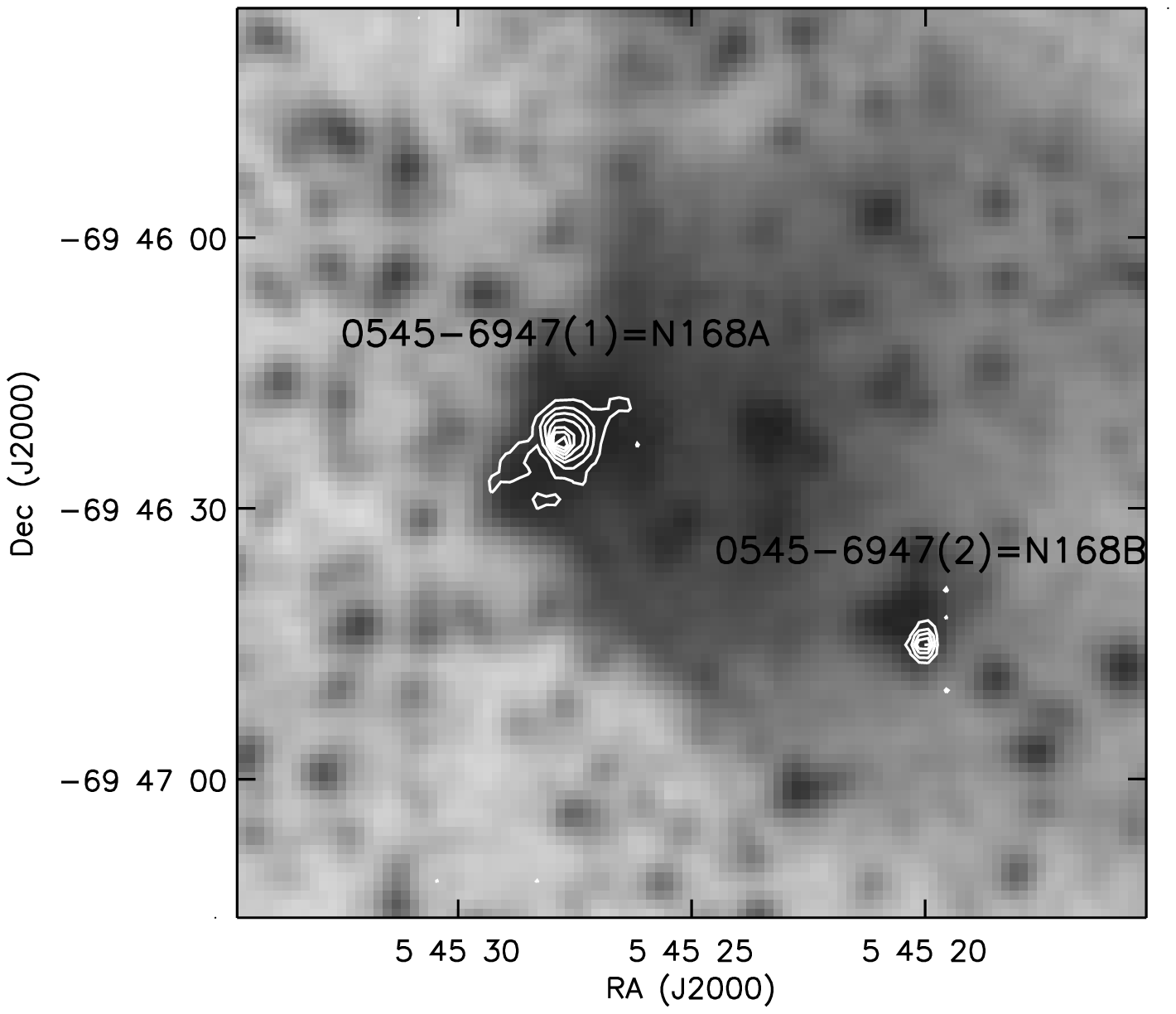}
\caption{\label{n214n168} Compact sources near Parkes source 0542-7121
and optical \hii region N214, and near Parkes source 0545-6947 and
optical \hii region N168.  Labels as Figure~\ref{n79}.}
\end{figure}

Figures \ref{n79}-\ref{n214n168} show the locations of the rest of the
LMC compact radio sources relative to nearby optically identified \hii
regions.  In many cases, the compact radio source is not associated
with the brightest optical emission.  In some but not all cases, the
radio emission is more similar to infrared.  Many of these regions are
relatively unstudied, but we describe the radio sources in as much
context as possible with a literature search:

The compact radio emission from Parkes source 0452-6927 is probably
related to optical \hii region N79B.  Figure~\ref{n79} shows the main
compact source, and a fringe of fainter emission to the east.

The N4 region in the north-west part of the LMC consists of two main
optical \hii regions, with a molecular cloud detected in CO between
them \citep{heydari-n4a}.  The southern and brighter region (A) has
higher excitation (H$\beta$/H$\alpha$) and higher H$\alpha$ surface
brightness on the NE side, toward the molecular cloud, supporting a
classical ``champagne flow'' model of the \hii region.  We detect two
compact radio sources on that more embedded side of the optical
emission (Fig.~\ref{n4}).  The brighter radio source 0452-5700(SW) is
in the optical emission, and apparently represents the densest ionized
gas near the two ionizing sources. The fainter source 0452-6700(NE) is
off the edge of the optical emission, and could be a young deeply
embedded source in the cloud.  \citet{heydari-n4a} require
3\up{49}~photons~s\ts{-1} to account for the optical hydrogen
recombination line flux. The flux density of our our brighter radio
source requires 1.3\up{49}~photons~s\ts{-1}, but the optical emission
clearly comes from a larger region of ionized gas resolved out in our
high resolution observations.

Figure~\ref{n5} shows the two sources possibly related to Parkes
source 0452-6722 and optical \hii region N5.

We detect two compact sources associated near the optical \hii region
N83 (Fig.~\ref{n83}).  0454-6916(E) is coincident with the
``high-excitation blob'' N83B \citep{heydari-n83b}, the eastern of
several optical \hii regions.  As with N4, the molecular gas detected
in CO peaks on the NE side of the complex
\citep[][Figure~1]{bolatto03}, so the higher excitation optical \hii
region is associated with denser molecular gas {\it and} a compact
radio continuum source.  \citet{heydari-n83b} calculate a required
Lyman continuum flux of 0.75\up{49}~photons~s\ts{-1} to power the
H$\beta$ nebula, and we require 1.1\up{49}~photons~s\ts{-1} to account
for the 3~cm radio continuum flux density.  The second, much fainter
compact radio source 0454-6916(W) is located in a region with little
optical emission between two optical \hii regions.

Figure~\ref{n23} shows the two sources possibly related to Parkes
source 0505-6807 and optical \hii region N23.  Nearby objects include
a supernova remnant, which is its own Parkes radio source, and a Lucke
\& Hodge star cluster.

We detect a cluster of 4 compact radio sources near the center of N105
(Fig.~\ref{n105}), an \hii region associated with at least 18 OB
stars, 2 WR stars, and bright [\ion{O}{3}] emission \citep[][and
references therein]{ambrocio98}.  The compact radio sources are in the
highest excitation part of the optical nebula, just to the west of the
WRC5+O star Brey~16a, in a region associated with masers and a
proposed IR protostar.

We detect compact radio sources associated with several of the
brighter optical \hii regions in the N44 complex (C, D, G and M),
around the edges of an \ion{H}{1} shell possibly associated with a
superbubble \citep{kim98}.  In N44M, \citet{naze} quote an ionizing
flux of 1.5--2.1\up{48}~photons~s\ts{-1} to account for the H$\alpha$
flux, and we need 2.8\up{48}~photons~s\ts{-1} to account for the 3~cm
flux density of 0523-6806(NE).  \citep{oey95} find many OB stars in the N44C
region, but without coordinates no particular association with radio
source 0522-6800 can be made.

Figure~\ref{n119n138} shows the sources near Parkes source 0519-6916
and optical \hii region N119, and near Parkes source 0525-6831 and
optical \hii region N138.  Figure~\ref{n206n55n213} shows the sources
near Parkes source 0531-7106 and optical \hii region N206, near Parkes
source 0532-6629 and optical \hii region N55, and near Parkes source
0538-7042 and optical \hii region N213.

The sources 0539-6931(1) and 0539-6931(2) (Fig.~\ref{n158}) are
located in the optical \hii region N158C, very near several identified
OB stars \citep{testor98}.  Source \#1 is near identified O7V and
O9V stars, and source \#2 is near an identified B0V star.  The Lyman
continuum fluxes required by the 3~cm measurements are 3\up{48} and
1\up{48}~photons~s\ts{-1} for sources \#1 and \#2, respectively,
consistent with single O7.5V--O8V and O9V stars.

Finally, Figure~\ref{n214n168} shows the sources near Parkes source
0542-7121 and optical \hii region N214, and near Parkes source
0545-6947 and optical \hii region N168.

\end{document}